\documentclass[a4paper,onecolumn,11pt,accepted=2022-05-24]{quantumarticle}
\pdfoutput=1
\usepackage[utf8]{inputenc}
\usepackage[english]{babel}
\usepackage[T1]{fontenc}
\usepackage{amsmath}
\usepackage{hyperref}

\usepackage{tikz}
\usepackage{lipsum}

\usepackage{amsthm}
\theoremstyle{definition}
\newtheorem{definition}{Definition}[section]

\newtheorem{theorem}{Theorem}[section]

\usepackage{amsfonts, amssymb, amsmath, amsthm, mathrsfs, braket, xcolor, longtable}
\usepackage{graphicx}

\usepackage{algcompatible}

\newcommand{\R}{\mathbb{R}}
\newcommand{\C}{\mathbb{C}}
\newcommand{\N}{\mathbb{N}}

\newcommand{\ID}{\mathbb{I}}

\newcommand{\real}{\operatorname{Re}}
\newcommand{\imag}{\operatorname{Im}}

\newcommand{\sumtwo}{\operatorname*{\sum\sum}}

\usepackage{tikz}
\usepackage[linesnumbered,ruled,vlined]{algorithm2e}

\SetCommentSty{mycommfont}

\SetKwInput{KwInput}{Input}
\SetKwInput{KwOutput}{Output}
\usetikzlibrary{quantikz}
\theoremstyle{definition}
\allowdisplaybreaks[4]



\begin{document}

\title{A variational quantum algorithm for the Feynman-Kac formula}

\author{Hedayat Alghassi}
\email{hedayat.alghassi@ibm.com}
\affiliation{%
IBM Quantum, Yorktown Heights, NY, US}
\orcid{0000-0001-7553-6975}

\author{Amol Deshmukh}
\email{amol.deshmukh@ibm.com}
\affiliation{%
IBM Quantum, Yorktown Heights, NY, US}
\orcid{0000-0001-8591-7085}

\author{Noelle Ibrahim}
\email{noel.ibrahim@ibm.com}
\affiliation{%
IBM Quantum, Yorktown Heights, NY, US}
\orcid{0000-0002-3056-8700}

\author{Nicolas Robles}
\email{nicolas.robles@ibm.com}
\affiliation{%
IBM Quantum, Yorktown Heights, NY, US}
\orcid{0000-0002-8369-4516}

\author{Stefan Woerner}
\email{wor@zurich.ibm.com}
\affiliation{%
IBM Quantum, IBM Research Europe -- Zurich, Switzerland}
\orcid{0000-0002-5945-4707}

\author{Christa Zoufal}
\email{ouf@zurich.ibm.com}
\affiliation{%
IBM Quantum, IBM Research Europe -- Zurich, Switzerland}%
\affiliation{Institute for Theoretical Physics, ETH Zurich, Switzerland}
\orcid{0000-0003-4126-3141}

\maketitle

\begin{abstract}
We propose an algorithm based on variational quantum imaginary time evolution for solving the Feynman-Kac partial differential equation resulting from a multidimensional system of stochastic differential equations. We utilize the correspondence between the Feynman-Kac partial differential equation (PDE) and the Wick-rotated Schr\"{o}dinger equation for this purpose. The results for a $(2+1)$ dimensional Feynman-Kac system obtained through the variational quantum algorithm are then compared against classical ODE solvers and Monte Carlo simulation. We see a remarkable agreement between the classical methods and the quantum variational method for an illustrative example on six and eight qubits.
In the non-trivial case of PDEs which are preserving probability distributions -- rather than preserving the $\ell_2$-norm -- we introduce a proxy norm which is efficient in keeping the solution approximately normalized throughout the evolution. The algorithmic complexity and costs associated to this methodology, in particular for the extraction of properties of the solution, are investigated. Future research topics in the areas of quantitative finance and other types of PDEs are also discussed. 
\end{abstract}

\maketitle
\section{\label{sec:introduction}Introduction}
Differential equations govern the evolution of complex systems arising in diverse fields such as physical and natural sciences as well as in economics and finance. The Feynman-Kac formula connects stochastic differential equations (SDEs), which describe the evolution of random variables, to parabolic (second-order, linear) partial differential equations (PDEs)  \cite{feynman2005principle, kac}. This kind of equation is similar and can formally be identified to the Schr\"{o}dinger equation \cite[$\mathsection$1]{kac}. This remarkable connection provides a methodology for linking the heat equation to Brownian motion and yields a technique for solving PDEs by Monte Carlo simulations. Conversely, many stochastic processes are amenable to be studied using deterministic methods. A perennial application of these notions is the Black-Scholes equation and geometric Brownian motion which enables pricing of financial derivatives \cite{blackscholes}. This connection also plays a crucial role in the rigorous and constructive formulation of quantum field theories, which are essential for the description of nature at the most fundamental level \cite{functional_integrals, LorincziHiroshimaBetz+2011}. On a more practical level, this formula is routinely utilized for the numerical evaluation of the quantum properties of atomic and subatomic systems in quantum chemistry as well as in atomic and particle physics \cite{fk_path_int1, doi:10.1063/1.454227, fk_path_int3}. 

Recent research on the suitability of quantum computers for solving differential equations has grown substantially due to the promise of efficiency and speed-up brought by solving these problems quantum mechanically. In \cite[$\mathsection$I]{gonzalezconde2021pricing}, the link between the Feynman-Kac formula and stochastic processes is pointed out but not further developed. Similarly, on \cite[$\mathsection$II]{santosh}, a remark about a `deep' connection between the heat equation and stochastic processes is made. In \cite[$\mathsection$2.1]{fontanela2021quantum}, Feynman-Kac formula is invoked to transform the price of a European call option from a conditional expectation value to the solution of the Wick-rotated Schr\"{o}dinger equation through a particular Hamiltonian. Furthermore, in \cite{stochastic}, the connection between trinomial trees for random walks, SDEs, and the Fokker-Planck equation, is established but the full picture of generalizing to the Feynman-Kac formula is not developed. In this paper, we will explore the advantages of entirely \textsl{transitioning} from an SDE to a PDE by not relying on a random-walk tree or other kinds of discrete probabilistic structures.

An alternative approach to estimate properties of stochastic processes is by leveraging Quantum Amplitude Estimation (QAE) \cite{brassard_qae} and its variants \cite{montanaromontecarlo, suzukiqae, 2021npjQI...7...52G}, which can achieve a quadratic speed-up over classical Monte Carlo simulation.
In contrast to classical Monte Carlo simulation, all possible paths of a (discretized) stochastic process are modelled in a single quantum state in superposition. In other words, a set of qubits is prepared such that measuring their state would result in sampling from the distribution of the stochastic process of interest. Although preparing such states is in principle always possible for reasonable stochastic processes \cite{PhysRevApplied.15.034027}, efficient realization of this method demands a careful analysis \cite{ibmoption, woerner_2019_quantum_risk, 9259208, Chakrabarti2021thresholdquantum} and may not always result in a practical quantum advantage. 
Once such a state is prepared, QAE can be used to evaluate different (path-dependent) objective functions. In \cite{An2021quantumaccelerated}, a similar connection between SDEs and PDEs is developed focused on Monte Carlo methods rather than variational methods associated to PDEs.

In most of the existing literature on option pricing for equities using quantum computers, e.g. \cite{ibmoption, montecarloxanadu}, however, an SDE is tacitly solved, namely the so-called Geometric Brownian motion $dX_t = \mu X_t dt + \sigma X_t dW_t$, where $\mu$ and $\sigma$ are constants known as drift and volatility respectively, and $W_t$ is the Brownian motion. Once this SDE is solved and shown to be $X_t = X_0 e^{(\mu-\sigma^2/2)t + \sigma(W_t-W_0)}$ with $X_0>0$, which produces a log-normal distribution, then the pricing of a particular security begins by applying QAE. However, this SDE is one of the simplest and one of the few that can be solved analytically. 
For the most part SDEs rarely admit analytical solutions \cite{oksendal, mitsde} and are usually dealt with numerically with large amounts of classical resources. 
Here we bypass this problem by not solving the SDE and instead simulating an associated -- and deterministic -- PDE for the calculations of the conditional expectations of the stochastic process $X_t$. 

We now propose to synthesize these ideas into a unified framework and bring economy of thought by describing how the heat, Schr\"{o}dinger, Black-Scholes, Hamilton-Jacobi, Fokker-Planck equations can be thought of as different manifestations of the Feynman-Kac formulation. Once this framework is established, we then employ recent techniques from variational quantum time evolution \cite{PhysRevLett.125.010501, theoryvariational, zoufal_error_bounds}
to introduce an algorithm for solving expectations of SDEs on quantum computers by \textsl{first} linking them to PDEs and \textsl{then} by establishing a differential operator, an infinitesimal generator, that can be evolved on a quantum computer.

The paper is structured as follows. 
In Section \ref{sec:fk}, we introduce the Feynman-Kac formula.
The algorithm along with its complexity and advantages is developed in Section \ref{sec:algorithm} and numerically illustrated in Section \ref{sec:numerical}. Future exploratory items are brought forward in Appendix \ref{sec:futurework}.

\section{\label{sec:fk}The Feynman-Kac formula}
In this section we first introduce the definition and existence of Brownian motion and immediately use its properties to derive a partial differential equation whose solution will be a conditional expectation of the Brownian motion. We formalize these results in one and several dimensions. A table of formal identifications of the Feynman-Kac formula and other types of second order PDEs is also provided. Moreover, we establish the discretization process that will be needed in Section \ref{sec:numerical} for our quantum and classical algorithms with periodic boundary conditions.
\subsection{Brownian motion and differential equations}

We start by reminding the reader about Brownian motion and link it to the heat equation. We shall proceed axiomatically in order to provide all the tools needed throughout the paper. Suppose $\Omega$ is a sample space, $\omega \in \Omega$ is an event of this space and $\mathcal{N}(\mu, \sigma)$ is the normal distribution with mean $\mu$ and variance $\sigma^2$, respectively.

\begin{definition} \label{defBM}
Brownian motion $W_t=W(t)$ with $t \ge 0$ is a continuous stochastic process that satisfies the following characteristics \cite[$\mathsection$5.3]{riskneutral}:
\begin{itemize}
\item[1)] $W(0) =0$ almost surely;
\item[2)] $W(t)$ has independent increments: $W(t+u)-W(t)$ is independent of $\sigma(W(s):s \le t)$ for $u \ge 0$, where $\sigma$ denotes the sigma algebra generated by $W(s)$;
\item[3)] $W(t)$ has Gaussian increments: $W(t+u)-W(t)$ is normally distributed with mean $0$ and variance $u$, i.e. $W(t+u)-W(t) \sim \mathcal{N}(0,u)$;
\item[4)] $W(t)$ has continuous paths: $W(t)$ is a continuous function of $t$, i.e. $t \to W(t,\omega)$ is continuous in $t$ for all $\omega \in \Omega$.
\end{itemize}
\end{definition}

The probability measure under which $W_t$ is a Brownian motion is denoted by $\mathbb{P}$ and the filtered probability space will be denoted by $(\Omega, \mathcal{F}, \mathbb{P})$. The existence of Brownian motion -- often overlooked in the literature of quantum computing, e.g. \cite[$\mathsection$II.A]{stochastic}, \cite[$\mathsection$II.A]{unary} or \cite[$\mathsection$II]{montecarloxanadu} -- was established by Wiener by proving that indeed the first three items were compatible with the continuity of the paths of $W_t$.

\begin{theorem}[Wiener]
Brownian motion exists.
\end{theorem}

\begin{figure}
	\centering
	\includegraphics[scale=0.34]{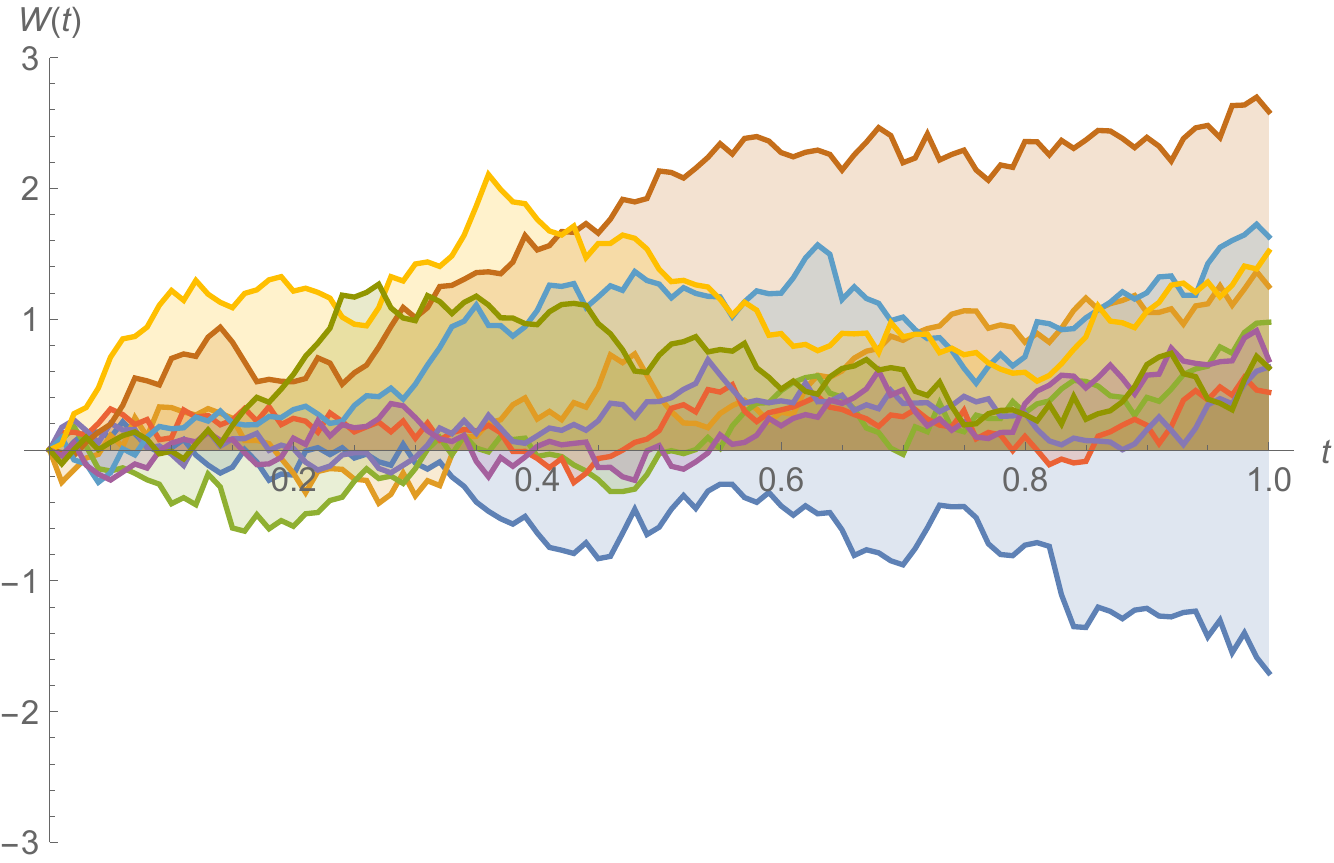}
	\includegraphics[scale=0.31]{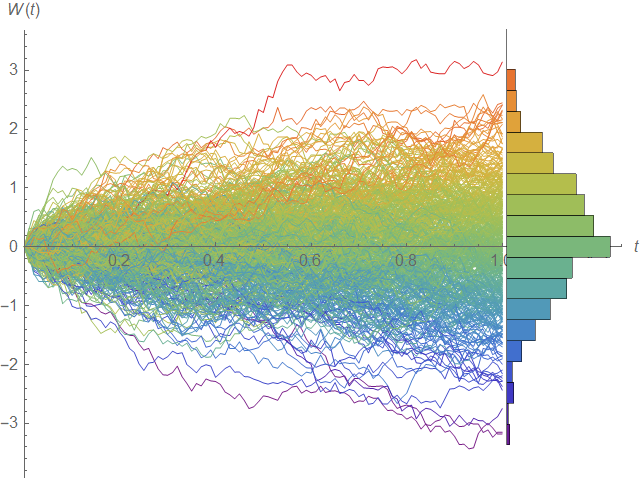}
	\caption{\underline{Left}: 10 realizations of Brownian motion. \underline{Right}: 500 realizations of Brownian motion and their resulting distribution.}
	\label{fig:brownian_motion}
\end{figure}

In Figure \ref{fig:brownian_motion} we display some possible paths of Brownian motion and their distribution. A key result, which is best motivated informally, is due to It\^{o} and it states the following \cite[$\mathsection$3.1]{karatzas} and \cite[Appendix B]{martingale}, see also \cite{baxterrennie, riskneutral, stochasticbook}.

\begin{theorem}[It\^{o}'s lemma] 
For infinitesimal Brownian motion increments, one has that
\begin{align} \label{ito}
dW_t^2 = dt.
\end{align}
\end{theorem}

This result states that the variance of the Brownian increment is proportional to the time interval of the increment. In other words, Brownian motion accumulates quadratic variation at rate one per unit time.

Furthermore, Brownian motion has some unusual properties. For instance while it is continuous everywhere it is (with probability one) nowhere differentiable and it scales like a fractal. Moreover, Brownian motion will eventually attain any and every value no matter how large or negative and once Brownian motion attains a value, it immediately attains it again infinitely often, and then after that again from time to time in the future, see \cite[$\mathsection$3.1]{baxterrennie} for other items in the `bestiary' of Brownian motion. Globally, the law of large numbers states that $W_t/t \to 0$ almost surely as $t \to \infty$, \cite[Proposition 2.8]{LorincziHiroshimaBetz+2011}.

Next, we move on to the backward heat equation \cite[$\mathsection$3.1.9]{martingale} as an illustrative example.
Suppose that $(W_t)_{t \ge 0}$ is a Brownian motion defined on the filtered probability space $(\Omega, \mathcal{F}, \mathbb{P})$, and consider the simplest type of SDE
\begin{align} \label{babysde}
dX_t = dW_t.
\end{align}
Let $\psi: \mathbb{R} \to \mathbb{R}$ be a Borel-measurable function such that
\begin{align} \label{babycondition}
\int_{- \infty}^{\infty} e^{-ax^2} |\psi(x)|dx < \infty
\end{align}
for some $a>0$. Set the conditional expectation
\begin{align} \label{babyexpectation}
u(x,t) = \mathbb{E} [\psi(X_T) \;| \; X_t=x], \quad \forall (x,t) \in \mathbb{R} \times [0,T].
\end{align}
From the properties of Brownian motion we can write $u$ as
\begin{align} \label{babyintegral}
u(x,t) = \frac{1}{\sqrt{2 \pi (T-t)}} \int_{- \infty}^{\infty} \psi(y) \exp\bigg(-\frac{(y-x)^2}{2(T-t)}\bigg) dy
\end{align}
for every $(x,y,t) \in \mathbb{R}^2 \times [0,T]$. Taking the partial derivative with respect to $t$ and the second order partial derivative with respect to $x$, one can easily verify that $u$ satisfies the PDE
\begin{align} \label{babypde}
-\frac{\partial u}{\partial t} = \frac{1}{2} \frac{\partial^2 u}{\partial x^2}, \quad \forall (x,t) \in \mathbb{R} \times [0,T)
\end{align}
with the terminal condition $u(x,T)=\psi(x)$ for $x \in \mathbb{R}$. This result contains the essence of Feynman-Kac formulation as it expresses the solution to the parabolic PDE \eqref{babypde} as the expected value of a functional of Brownian motion \eqref{babyexpectation} coming from an SDE \eqref{babysde} with a terminal condition that is subjected to \eqref{babycondition}. While \eqref{babypde} is indeed the heat equation, some guesswork would be involved in generalizing this to a full SDE of the type $dX_t= \mu(X_t,t)dt + \sigma(X_t,t)dW_t$, let alone to multi-dimensional generalizations. In order to overcome this, it is worth investing in seeing how It\^{o}'s lemma plays a critical role in the derivation of the Feynman-Kac formula. This investment will also explain how \eqref{babyintegral} was arrived at. The details are in Appendices \ref{sec:details} and \ref{sec:fullfk}.

\subsection{One-dimensional Feynman-Kac formula}

As we have seen, the Feynman-Kac technique is a strategy for solving a certain type of partial differential equations by using Brownian motion and Monte Carlo methods. The converse can also be quite useful as it implies that expectations of Wiener processes such as \eqref{babysde} or more general, which are stochastic, can be translated to and computed by purely deterministic methods. Let us now formalize these notions.

Let $(W_t)_{t \ge 0}$ be a one-dimensional Brownian motion defined on a filtered probability space $(\Omega, \mathcal{F}, \mathbb{P})$. We shall consider the Wiener process $W_t$ and the stochastic differential equation
\begin{align} \label{processX}
dX_t = \mathfrak{b}(X_t,t)dt + \sqrt{2\mathfrak{a}(X_t,t)}dW_t,
\end{align}
where $\mathfrak{a}$ and $\mathfrak{b}$ are suitable analytic functions with $\mathfrak{a} > 0$ and with condition $X_t=x$, see \cite[$\mathsection$4]{karatzas}. Note how \eqref{processX} now incorporates a drift term $\mathfrak{b}(X_t,t)$, unlike \eqref{babysde}. Let $\mathfrak{f}: \R \to \R$ be an analytic function. For a Borel-measurable function $\psi: \R \to \R$, we define the function $u: \R \times [0,T] \to \R$ by setting
\begin{align} \label{longcondexp}
u(x,t) &= \mathbb{E} \bigg[  \int_t^T \exp\bigg(-\int_t^s \mathfrak{c}(X_\tau, \tau) d \tau\bigg)  \mathfrak{f}(X_s, s)ds  \nonumber \\
&\quad +  \exp\bigg(-\int_t^T \mathfrak{c}(X_\tau, \tau)d\tau \bigg) \psi(X_T) \; | \; X_t=x \bigg],
\end{align}
where $\mathfrak{c}$ is a suitable real function called the discount function. Moreover, suppose that 
\begin{align}
\int_{-\infty}^\infty e^{-\delta x^2} |\psi(x)|dx < \infty, \nonumber
\end{align}
for some $\delta > 0$. The function $u$ is defined for $0<T-t<\tfrac{1}{2}\delta$ and $x \in \R$, and has derivatives of all orders. In particular, it belongs to the class $\mathcal{C}^{2,1}(\R \times (0,T))$. The Feynman-Kac result is the fact that $u$ satisfies the following partial differential equation
\begin{align} \label{longPDE}
\begin{dcases}
\frac{\partial u}{\partial t} + \mathfrak{a}(x,t)\frac{\partial^2 u}{\partial x^2} +\mathfrak{b}(x,t) \frac{\partial u}{\partial x} - \mathfrak{c}(x,t)u + \mathfrak{f}(x,t)=0, \quad \textnormal{for} \quad t<T  \\
~  \
u(x,T) = \psi(x).
\end{dcases}
\end{align}
From this point, we shall drop the term involving $\mathfrak{f}$ by setting it to zero as it will not be very useful in our formulation. In this case, equation \eqref{longPDE} becomes
\begin{align} \label{shortPDE}
\begin{dcases}
\frac{\partial u}{\partial t} + \mathfrak{a}(x,t)\frac{\partial^2 u}{\partial x^2} +\mathfrak{b}(x,t) \frac{\partial u}{\partial x} = \mathfrak{c}(x,t)u,  \quad \textnormal{for} \quad t<T \\
~  \
u(x,T) = \psi(x).
\end{dcases}
\end{align}
and its associated conditional expectation \eqref{longcondexp} reduces to 
\begin{align} \label{shotcondexp}
u(x,t) = \mathbb{E} \bigg[\exp\bigg(-\int_t^T \mathfrak{c}(X_\tau, \tau)d\tau \bigg) \psi(X_T) \; | \; X_t=x \bigg].
\end{align}
Shortly it will become clear that the term $\frac{\partial u}{\partial x}$ is not always helpful for our objectives and it is advantageous to remove it now by applying changes of variables and transformations. First we set $u(x,t)=e^{g(x,t)}v(x,t)$ where $g$ is a function of $\mathfrak{a}$ and $\mathfrak{b}$ given \eqref{integratingfactor} in Appendix \ref{sec:fkschro}. Moreover, we switch to an initial value problem by setting $\tau = T-t$ and then perform a Wick rotation $\xi = -i \tau$, where $i$ is the imaginary unit. The last step requires analytical continuation. This transforms the problem to 
\begin{align} 
-i \frac{\partial v}{\partial \xi} = \mathfrak{a}(x,\xi) \frac{\partial^2 v}{\partial x^2}  +\mathfrak{w}(x,\xi) v,  \nonumber
\end{align}
where the new term $\mathfrak{w}$ can be written in terms of the other coefficients $\mathfrak{a}, \mathfrak{b}$ and $\mathfrak{c}$ of the PDE by the formula
\begin{align} 
\mathfrak{w}(x,\xi) = - \frac{\partial}{\partial \xi} \int \frac{\mathfrak{b}(x,\xi)}{2 \mathfrak{a}(x,\xi)}dx - \frac{\mathfrak{b}^2(x,\xi)}{4 \mathfrak{a}(x,\xi)} - \frac{1}{2} \mathfrak{a}(x,\xi) \bigg(\frac{\partial}{\partial x} \nonumber \frac{\mathfrak{b}(x,\xi)}{\mathfrak{a}(x,\xi)}\bigg) - \mathfrak{c}(x,\xi).
\end{align}
We shall now write this in Dirac notation as
\begin{align} \label{schrodingertransformed2}
-i \frac{\partial}{\partial \xi} \ket{\psi} = \hat{\mathcal{H}} \ket{\psi},
\end{align}
where the wave function is written in Dirac notation as $\psi(x,t)=\braket{x,t|\psi}$ and it takes the role of the modified solution in the setting of an infinite dimensional Hilbert space. The resulting function $\ket{\psi}$ does not necessarily correspond to an $\ell_2$ normalized wavefunction and $\hat{\mathcal{H}}$ is not necessarily Hermitian. 
The wave function $\ket{\psi}$ is not to be confused with the boundary condition $\psi$ from \eqref{shotcondexp}. In this case, the Hamiltonian is given by the expression
\begin{align} 
\hat{\mathcal{H}} := \mathfrak{a}(x,\xi) \frac{\partial^2}{\partial x^2} + \mathfrak{w}(x,\xi) \ID, \nonumber
\end{align}
where $\ID$ is the identity operator. This Hamiltonian is time-dependent as $\xi$ appears in both $\mathfrak{a}$ and $\mathfrak{w}$. The reason the term $\frac{\partial u}{\partial x}$ in \eqref{shortPDE} was undesirable now becomes more apparent. While performing a canonical quantization the momentum operator is taken to be
$
{\hat p}_x \to \frac{\hbar}{i} \frac{\partial}{\partial x}. \nonumber
$
The differential operator $\tfrac{\partial}{\partial x}$ is skew-Hermitian, i.e. $\tfrac{\partial}{\partial x} ^ \dagger = - \tfrac{\partial}{\partial x}$, whereas the second order derivative is Hermitian, $\tfrac{\partial^2}{\partial x^2} ^ \dagger = \tfrac{\partial^2}{\partial x^2}$. Therefore, when we performed the Wick rotation, the absence of the derivative of first order leaves the condition of the Hermicity of $\hat {\mathcal{H}}$ solely on the term $\mathfrak{a}(x,\xi) \frac{\partial^2}{\partial x^2}$. In general this term will not be Hermitian unless $\mathfrak{a}$ and $\frac{\partial^2}{\partial x^2}$ commute. However, for our case of interest in Section \ref{sec:numerical}, we shall set $\mathfrak{a}$ to be constant and this will result in our Hamiltonian being Hermitian. This means that the solution to the Schr\"{o}dinger-type equation \eqref{schrodingertransformed2} is
\begin{align} \label{solutiontime}
\ket{\psi(x,\xi)} = \exp \bigg(i \int_0^\xi \hat{\mathcal{H}} (\chi) d\chi \bigg) \ket{\psi(x,0)}. 
\end{align}
If there were no $\xi$ dependency in $\hat{\mathcal{H}}$, then \eqref{solutiontime} would reduce to
$
\ket{\psi(x,\xi)} = \exp (i \hat{\mathcal{H}} \xi ) \ket{\psi(x,0)}. 
$

The first order derivative in \eqref{shortPDE} is associated to the drift term $\mathfrak{b}(X_t,t)$ in \eqref{processX}. An SDE that has no $dt$ term is called a strict local martingale \cite[$\mathsection$2.3]{papanicolaou} or \cite{stochasticbook, karatzas}. Alternatively, transforming an initial SDE into a martingale also helps in obtaining a Wick rotated equation without the first order derivative. 

In the real time evolution, the evolution operator $e^{-i \hat{\mathcal{H}} t}$ must be unitary. This operator is unitary if the Hamiltonian $\hat{\mathcal{H}}$ is Hermitian. In the imaginary time evolution, however, the evolution operator $e^{-\tau \hat{\mathcal{H}}}$ is not in general unitary and thus we do not need to be concerned about the Hermiticity of $\hat{\mathcal{H}}$. The situation for non-Hermitian Hamiltonians is worth discussing. In \cite[$\mathsection$2.1]{fontanela2021quantum}, a sequence of very judicious transformations and changes of variables transformed the Black-Scholes equation into the heat equation, see also \cite[$\mathsection$6.5.1]{wilmott} as well as the Lamperti transform. The advantage is that the Wick-rotated heat equation becomes a Schr\"{o}dinger equation with a Hermitian Hamiltonian. In \cite[$\mathsection$II]{gonzalezconde2021pricing} (which does not employ variational methods), these changes were not applied and as a result the authors have to deal with a non-Hermitian Hamiltonian which gets embedded, by unitary dilation \cite{dilation}, into a larger space and then deal with probabilities of success when measuring the outcomes of quantum circuits. Moreover, in \cite[$\mathsection$II]{santosh} (which does make use variational methods), these transformation were not employed either and the author has to utilize similarity transformations motivated from supersymmetry \cite{Bagarello:2020int} in order to obtain a Hermitian Hamiltonian. These techniques are indeed useful for situations where no change of variables or transformations lead to Hermitian Hamiltonians, see also \cite{paritytime} for useful insights.

\subsection{Special cases}
According to the values of the coefficients in \eqref{shortPDE}, we can make the following formal identifications.
\subsubsection{Feynman-Kac} 
\begin{itemize}
    \item \underline{Coefficients}: Suitable $\mathfrak{a}, \mathfrak{b}, \mathfrak{c}$ with $\mathfrak{f}=0$.
    \item \underline{Transformations}: None.
    \item \underline{Interpretation}: The solution is an expectation \eqref{shotcondexp} of a Wiener process \eqref{processX} 
    involving Brownian motion.
    \item \underline{Notes}: Most general case before specializing the coefficients.
\end{itemize}
\subsubsection{Heat equation}
\begin{itemize}
    \item \underline{Coefficients}: $\mathfrak{a}=1$, $\mathfrak{b}=\mathfrak{c}=0$.
    \item \underline{Transformations}: None.
    \item \underline{Interpretation}: The solution $u$ corresponds to the temperature field and the Laplacian indicates whether the material surrounding each point $x$ is hotter or colder, on average, than the material at the point $x$.
    \item \underline{Notes}: Density and properties of materials can be taken into account by applying certain coefficients to the terms that appear in the equation. It is $\ell_1$-norm preserving.
\end{itemize}
\subsubsection{Black-Scholes} 
\begin{itemize}
    \item \underline{Coefficients}: $\mathfrak{a} =\tfrac{1}{2}\sigma^2 x^2, \mathfrak{b}=rx, \mathfrak{c}=r$.
    \item \underline{Transformations}: $t\to T-t$.
    \item \underline{Interpretation}: The solution is the price of an option whose underlying follows a geometric Brownian motion.
    \item \underline{Notes}: Type of financial derivative depends on payoff boundary condition.
\end{itemize}
\subsubsection{Schr\"{o}dinger} 
\begin{itemize}
    \item \underline{Coefficients}: $\mathfrak{a}=\tfrac{\hbar^2}{2m}, \mathfrak{b}=0, \mathfrak{c}=V(x)$, with $V(x)$ the potential.
    \item \underline{Transformations}: $t \to - i t$.
    \item \underline{Interpretation}: The solution is interpreted as the amplitude of wave function in quantum mechanics.
    \item \underline{Notes}: The resulting Hamiltonian $\hat{\mathcal{H}}$ is Hermitian and the equation is $\ell_2$-norm preserving.
\end{itemize}
\subsubsection{Fokker–Planck} 
\begin{itemize}
    \item \underline{Coefficients}: $\mathfrak{a}=-D(x,t), \mathfrak{b}=-2\tfrac{\partial D}{\partial x}+\mu, \mathfrak{c}=-\tfrac{\partial^2 D}{\partial x^2}+\tfrac{\partial \mu}{\partial x}$.
    \item \underline{Transformations}: Here $D$ is the diffusion coefficient given by $D(X,t)=\tfrac{1}{2}\sigma^2$.
    \item \underline{Interpretation}: The solution $p(x,t)$ is the probability density function of the random variable $X_t$ whose It\^{o} process is $dX_t=\mu(X_t,t)dt +\sigma(X_t,t)dW_t.$
    \item \underline{Notes}: Time evolution of the probability density function of a particle under the influence of drag forces and random forces. The equation is $\ell_1$-norm preserving.
\end{itemize}
\subsubsection{Kolmogorov} 
\begin{itemize}
    \item \underline{Coefficients}: $\mathfrak{a}=D(x,t), \mathfrak{b}=\mu(x,t)$.
    \item \underline{Transformations}: See Fokker-Planck equation above.
    \item \underline{Interpretation}: These equations are used to characterize stochastic processes by describing how they change over time.
    \item \underline{Notes}: `Forward' equation is Fokker-Planck, and `backward' equation is used to determine the probability distribution of a state at a later time $s>t$. The equation is $\ell_1$-norm preserving.
\end{itemize}
\subsubsection{Hamilton-Jacobi} 
\begin{itemize}
    \item \underline{Coefficients}: Use ansatz $\psi(x,t)=\psi_0 e^{i S /\hbar}$ where $S$ is action on Schr\"{o}dinger.
    \item \underline{Transformations}: See Schr\"{o}dinger equation above.
    \item \underline{Interpretation}: Take $\hbar \to 0$ for limiting semi-classical case and use $S=\int L dt$, where $L$ is the Lagrangian.
    \item \underline{Notes}: This is used to derive equations of motion and identifying conserved quantities of mechanical systems. It is non-linear.
\end{itemize}
\subsubsection{Ornstein - Uhlenbeck} 
\begin{itemize}
    \item \underline{Coefficients}: See Fokker-Planck above. In one dimension, Ornstein - Uhlenbeck satisfies $\tfrac{\partial D}{\partial t} = \theta \tfrac{\partial}{\partial x}(x P) + D \frac{\partial^2 P}{\partial x^2}$.
    \item \underline{Transformations}: Here $P(x,t)$ is the probability of finding the process in the state $x$ and time $t$ and diffusion $D=\tfrac{\sigma^2}{2}$.
    \item \underline{Interpretation}: The associated SDE is $dX_t = - \theta X_t dt + \sigma dW_t$ and it is frequently written as a Langevin equation of the type $\tfrac{dX_t}{dt} = - \theta X_t + \sigma \eta(t)$, where $\eta(t)$ is white noise. Note that $dW_t/dt$ does not exist, as Brownian motion is not differentiable, so this is only heuristic.
    \item \underline{Notes}: The Ornstein - Uhlenbeck process is a stationary Gauss-Markov process and it is temporarily homogeneous. Wide applications in finance (interest rate modeling), mechanical systems (overdamped processes) and biology (modeling of phenotypes).
\end{itemize}
Other kinds of formal identification in several dimensions also exist.

\subsection{Multi-dimensional Feynman-Kac formula}

Let $D \in \N$ and $N \in \N$ and consider a $D$-dimensional stochastic process \[{\bf X}_t = (X^1_t, X^2_t, \cdots, X^D_t)^T\] governed by an $N$-dimensional Brownian motion 
\begin{align} \label{DdimensionalBrownian}
{\bf W}_t=(W_t^1, W_t^2, \cdots, W_t^N)^T,
\end{align}
whose time evolution is given by the following system of SDEs:
\begin{align} \label{systemSDEs}
dX_t^i = \mu_i({\bf X}_t, t)dt + \sum_{\ell=1}^N \sigma_{i \ell}({\bf X}_t,t)dW_t^{\ell}, \quad \textnormal{for} \quad i=1,2,\cdots,D.
\end{align}
We may abbreviate this to
\begin{align}
\begin{cases}
d{\bf X}_t &= {\bf \mu}({\bf X}_t, t)dt + {\bf \Sigma} \cdot d{\bf W}_t, \nonumber \\
{\bf X}_0 &= {\bf x}_0,
\end{cases}
\end{align}
where ${\bf X}_t, {\bf x}_0$ and ${\bf \mu}$ are in $\R^D$, ${\bf W}_t$ is in $\R^N$, and ${\bf \Sigma}$ is the $\R^{D \times N}$ matrix with elements $\sigma_{ik}({\bf X}_t, t)$. The infinitesimal generator $\mathcal{G}$ of the diffusion process ${\bf X}_t$ is given by the differential operator (see e.g. \cite[$\mathsection$4.1]{nolen} as well as \cite[$\mathsection$4.2]{lecture12})
\begin{align} \label{generator}
\mathcal{G} = \frac{1}{2} \sum_{i=1}^D \sum_{j=1}^D \sum_{\ell=1}^N \sigma_{i \ell}({\bf X}_t, t) \sigma_{j \ell}({\bf X}_t, t) \frac{\partial^2}{\partial x_i \partial x_j} + \sum_{i=1}^D \mu_i({\bf X}_t, t) \frac{\partial}{\partial x_i}.
\end{align}
We may further set $({\bf\Sigma \Sigma^T})_{ij} = \sum_{\ell=1}^N \sigma_{i \ell}({\bf X}_t, t) \sigma_{j \ell}({\bf X}_t, t)$. The terms in $\mathcal{G}$ correspond to the terms that appear in the multidimensional It\^{o} formula
\begin{align} \label{dudetstoc}
du({\bf X}_t, t) = \frac{\partial u}{\partial t}({\bf X}_t, t)dt + \mathcal{G}u({\bf X}_t, t) dt + R(dW_t^j),
\end{align}
where $R$ stands for the various stochastic terms involving the different $dW_t^j$. Thus, $\mathcal{G}$ represents the deterministic, non-time dependent, part of \eqref{dudetstoc}. Now set
\begin{align} \label{physicalexpectation}
u({\bf x}, t) = \mathbb{E} \bigg[\exp\bigg(-\int_0^t r({\bf X}_s, s) ds \bigg) \; \psi({\bf X}_t) \; | \; {\bf X}_t = {\bf x}\bigg],
\end{align}
for some function $\psi$ satisfying appropriate Lipschitz and growth hypotheses \cite[Theorem 6]{lecture12}. The function $u$ defined by \eqref{physicalexpectation} satisfies the diffusion equation
\begin{align}\label{generalizedOp}
\frac{\partial u}{\partial t} = \mathcal{G}u - ru, \quad t>0
\end{align}
as well as the initial condition $u({\bf x}, 0)=\psi({\bf x})$. This problem could be mapped to the corresponding Schr\"odinger-type equation
\begin{align} 
\frac{\partial}{\partial t}\ket{\psi(t)} = -i\tilde{\mathcal{G}} \ket{\psi(t)} \quad \textnormal{where} \quad \tilde{\mathcal{G}} = \mathcal{G} - r\mathbb{I}, \nonumber
\end{align}
by introducing an appropriate Wick rotation, however $\tilde{\mathcal{G}}$ might not necessarily be Hermitian.

There is no need to impose correlations between the elements of the $N$-dimensional Brownian motion as we shall see in Section \ref{sec:anisotropic}. Not only does this approach bypass the need to set individual correlations between Brownian motions, it also allows for substantial flexibility when modeling different particles whose behavior is governed by \textsl{several} Brownian motions $n=1,2,\cdots, N$ per dimension $d=1,2,\cdots, D$ as in \eqref{DdimensionalBrownian}. We further note that generalization we present here, in addition to multi Brownian motions, also contains the discount function $r$. None of these features appear in the previous literature of VarQITE and stochastic differential equations \cite{santosh, fontanela2021quantum, stochastic}. 

\subsection{\label{sec:finitedifference}Finite difference approximation}
We briefly explain the discretization and the embedding schemes needed to solve the Feynman-Kac system using a Schr\"odinger equation solver, such as VarQITE which will be described in Section \ref{sec:algorithm}. Let $n_q$ be the total number of qubits. We demonstrate the discretization procedure for a simple problem of a two dimensional Feynman-Kac generator, as specified in \eqref{hamiltonianpaper} below which is a special case of \eqref{generator}, involving the second order derivatives $\frac{\partial^2}{\partial x^2}, \frac{\partial^2}{\partial y^2}$ and $\frac{\partial^2}{\partial x \partial y}$.
To solve the system using $n= 4$ qubits (i.e. $2^4=16$ computational basis states: $\ket{0}, \ket{1}, \dots, \ket{15}$), we embed the solution $u_{ij}$ in the quantum state as shown in the diagram below. This type of diagram in fact can be used to elucidate the embedding of a multidimensional differential operator on a spatial grid. The mesh is in general a hypercube in $d$ dimensions, but if $n_q$ is not divisible by $D$, then the mesh is a $D$ dimensional hypercube with the sum of its dimensions equal to $n_q$.

\begin{center}
\begin{tikzpicture}
\label{theMaindiag}
\draw[step=2cm, black, thick] (0,0) grid (6,6);
\foreach \x in {0, 2, 4, 6}
    \foreach \y in {0, 2, 4, 6}
        \fill (\x, \y) circle [radius=2pt];
\draw(0.35, 5.65) node{$|0\rangle$};
\draw(2.35, 5.65) node{$|1\rangle$};
\draw(4.35, 5.65) node{$|2\rangle$};
\draw(6.35, 5.65) node{$|3\rangle$};
\draw(0.35, 3.65) node{$|4\rangle$};
\draw(2.35, 3.65) node{$|5\rangle$};
\draw(4.35, 3.65) node{$|6\rangle$};
\draw(6.35, 3.65) node{$|7\rangle$};
\draw(0.35, 1.65) node{$|8\rangle$};
\draw(2.35, 1.65) node{$|9\rangle$};
\draw(4.35, 1.65) node{$|10\rangle$};
\draw(6.35, 1.65) node{$|11\rangle$};
\draw(0.35,-0.35) node{$|12\rangle$};
\draw(2.35,-0.35) node{$|13\rangle$};
\draw(4.35,-0.35) node{$|14\rangle$};
\draw(6.35,-0.35) node{$|15\rangle$};
\end{tikzpicture}
\end{center}
Note that the size of the square mesh is $2^\frac{n_q}{D} = 2^2 = 4$, where $n_q$ was the number of qubits and $D$ is the number of dimensions. Considering the above mesh, the second order derivatives from \eqref{hamiltonianpaper} can be approximated by using the stencil method \cite{num_recipe} as
\begin{align}
\frac{\partial^2 u}{\partial x^2} &\approx \frac{u_{i+1, j} - 2u_{i, j} + u_{i-1, j}}{\Delta x^2}, \nonumber\\
\frac{\partial^2 u}{\partial y^2} &\approx \frac{u_{i, j+1} - 2u_{i, j} + u_{i, j-1}}{\Delta y^2}, \nonumber\\
\frac{\partial^2 u}{\partial x\partial y} &\approx 
\frac{u_{i+1, j+1} - u_{i+1, j-1} - u_{i-1, j+1} + u_{i-1, j-1} }{4\Delta x \Delta y}. \nonumber
\end{align}
Here $\Delta x$ and $\Delta y$ refer to the lattice spacing in $x$ and $y$ directions respectively. We additionally employ periodic boundary conditions on the grid in both directions.

The second order derivatives appearing in the general case of $D$ dimensional Feynman-Kac operator \eqref{generator} can be approximated as
\begin{align} \label{process}
\frac{\partial u(x_{i_1}, x_{i_2}, \cdots, x_{i_D}, t)}{\partial x_{i_d}} &\approx \frac{u_{{\bf i}+ {\bf e}_d} - u_{{\bf i}- {\bf e}_d}}{2\Delta x_{i_d}}, \nonumber \\
\frac{\partial^2 u(x_{i_1}, x_{i_2}, \cdots, x_{i_D}, t)}{\partial x_{i_d}^2} &\approx \frac{u_{{\bf i}+ {\bf e}_d} - 2u_{{\bf i}} + u_{{\bf i}- {\bf e}_d}}{\Delta x_{i_d}^2}, \nonumber \\
\frac{\partial^2 u(x_{i_1}, x_{i_2}, \cdots, x_{i_D}, t)}{\partial x_{i_d}\partial x_{i_{d'}}} &\approx 
\frac{u_{{\bf i}+ {\bf e}_d+ {\bf e}_{d'}} - u_{{\bf i}+ {\bf e}_d- {\bf e}_{d'}} - u_{{\bf i}- {\bf e}_d+ {\bf e}_{d'}} + u_{{\bf i}- {\bf e}_d- {\bf e}_{d'}} }{4\Delta x_{i_d} \Delta x_{i_{d'}}},
\end{align}
where ${\bf i}= \{{i_1, i_2, \cdots, i_D}\}$ and ${\bf i} \pm {\bf e}_d = \{i_1, \dots, i_{d-1}, i_d \pm1, i_{d+1}, \dots, i_D\}$ and ${\bf i} \pm {\bf e}_d\pm {\bf e}_{d'} = \{ i_1, \dots, i_{d-1}, i_d \pm1, i_{d+1}, \dots, i_{d'-1}, i_{d'} \pm1, i_{d'+1}, \dots, i_D\}$.
Similarly if one intends to consider periodic boundary conditions then appropriate modular operations should be taken into account. Using the approximations above we obtain the discretized form for the operator $\tilde{\mathcal{G}}$ as
\begin{align} \label{discreteG1}
\tilde{\mathcal{G}} = \sum_{i_1=0}^{n_m-1}\sum_{i_2=0}^{n_m-1} \cdots \sum_{i_D=0}^{n_m-1}
\sum_{j_1=0}^{n_m-1}\sum_{j_2=0}^{n_m-1} \cdots \sum_{j_D=0}^{n_m-1}
[\hat{\mathcal{G}}]_{i_1, i_2, \cdots, i_D, j_1, j_2, \cdots, j_D} \ket{{\bf i}} \bra{{\bf j}},
\end{align}
where $n_m = 2^{n_q/D}$. The elements of the matrix are given by
\begin{align} \label{discreteG2}
 [\tilde{\mathcal{G}}]_{{\bf i, j}} = 
\begin{cases}
\big(\sum_{d=1}^{D}\frac{({\bf\Sigma\Sigma^T})_{i{_d}i{_d}}}{\Delta x_{i_d}^2}\big) - r\quad ({\bf i} = {\bf j}), \\
\frac{1}{2} \frac{({\bf\Sigma\Sigma^T})_{i{_d}i{_d}}}{(\Delta x_{i_d})^2} \mp \frac{1}{2}\frac{\mu_{i_d}}{\Delta x_{i_d}} \quad ({\bf i} = {\bf j}\pm{\bf e}_d),  \\
\frac{1}{4}\frac{({\bf\Sigma\Sigma^T})_{i_{d}i_{d'}}}{(\Delta x_{i_{d}}\Delta x_{i_{d'}})} \quad ({\bf i} = {\bf j} \pm {\bf e}_d \pm {\bf e}_{d'}),\\
-\frac{1}{4}\frac{({\bf\Sigma\Sigma^T})_{i_{d}i_{d'}}}{(\Delta x_{i_{d}}\Delta x_{i_{d'}})} \quad ({\bf i} = {\bf j} \pm {\bf e}_d \mp {\bf e}_{d'}),\\
0 \quad \text{otherwise}.
\end{cases}
\end{align}

\section{\label{sec:algorithm}Quantum algorithm for solving the Feynman-Kac formula}

In this section, we describe a quantum algorithm for solving the PDE associated to the SDE of interest as prescribed within the Feynman-Kac formulation. We first discuss variational quantum imaginary time evolution (VarQITE), i.e., the fundamental algorithm used, then how to embed the problems into quantum states and the efficient decomposition of the differential operators of the PDE into unitaries. After that, we discuss how to extract properties of the solution, and lastly how to (approximately) enforce the $\ell_1$ norm preservation during the evolution.

\subsection{\label{sec:VarQITEembedding}VarQITE and state embedding}

Suppose we have a time-independent differential operator $\mathcal{E}$ acting on $n$ qubits with associated propagator $\exp(-i \mathcal{E} t)$, evolving for real values of $t$.
Provided the operator $\mathcal{E}$ is Hermitian, the quantum real time evolution (QRTE) is unitary and can be implemented, e.g., by a suitable Trotter decomposition, solving the Schr\"{o}dinger equation.

If we apply the Wick rotation $\tau = i t$, i.e., we translate the problem to the realm of quantum imaginary time evolution (QITE), the evolution operator $\exp(-\mathcal{E} \tau)$ will not be a unitary operation anymore and the underlying equation is more complicated to (approximately) solve on a quantum computer. If we are given an initial state $\ket{\psi(0)}$, then the normalized imaginary time evolution is defined by
$
\ket{\psi(\tau)} = \alpha(\tau) e^{-\mathcal{E}\tau} \ket{\psi(0)},
$
where $\alpha(\tau) = 1 / \sqrt{\bra{\psi(0)} e^{-2 \mathcal{E} \tau} \ket{\psi(0)}}$ is the normalization factor with respect to the $\ell_2$ norm.
The corresponding Wick-rotated Schr\"{o}dinger equation is
\begin{align} \label{wickrotatednormalized}
\frac{\partial \ket{\psi(\tau)}}{\partial \tau} = -(\mathcal{E}-\lambda_\tau \mathbb{I})  \ket{\psi(\tau)},
\end{align}
where $\lambda_\tau = \bra{\psi(\tau)} \mathcal{E} \ket{\psi(\tau)}$ ensures the normalization.

VarQITE is an algorithm to approximate QITE, see \cite{PhysRevX.7.021050, stochastic, PhysRevLett.125.010501, fontanela2021quantum, variationalansatz, theoryvariational}. Instead of evolving quantum states in the complete exponential state space, a parameterized ansatz is used and the evolution is approximated by mapping it to the ansatz parameters via McLachlan's variational principle \cite{McLachlan64}. The main advantage is the possibility of implementing it using shallow quantum circuits, suitable for near-term quantum devices \cite[p. 1]{variationalansatz}. More precisely, instead of considering a general state $\ket{\psi(\tau)}$, we encode the trial state $\ket{\phi(\boldsymbol \theta(\tau))}$, with 
$
\boldsymbol \theta(\tau) = (\theta_1 (\tau), \theta_2 (\tau), \cdots, \theta_N (\tau)),
$
where $N$ is an integer depending on the ansatz circuit of our choice.
The choice of the ansatz is crucial for the performance of the algorithm and we discuss this for our concrete example in Sec.~\ref{sec:ansatz}.

Instead of introducing standard VarQITE for quantum states, we now discuss the evolution for states that are possibly not $\ell_2$ normalized \cite{PhysRevLett.125.010501}.
In the situation of an unnormalized state $\ket{\tilde \psi({\bf x}, t)}$ governed by the equation
\begin{align} \label{masterequation}
\frac{\partial}{\partial t} \ket{\tilde \psi({\bf x}, t)} = \mathcal{E}(t) \ket{\tilde \psi({\bf x}, t)},
\end{align}
where $\mathcal{E}(t)$ is now a linear time-dependent --not necessarily Hermitian-- operator. The dynamical evolution of $\ket{\tilde \psi({\bf x}, t)}$ can be simulated by introducing an ansatz $\ket{\tilde{v}(\boldsymbol\theta(t))} = \alpha(t) \ket{v(\boldsymbol\theta(t))}$ of the form
$
\ket{{v}(\boldsymbol \theta (t))} =  \mathbf{G}(\boldsymbol \theta(t)) \ket{0}^{\otimes n}
$
where $\alpha(t)$ is a parameter of the ansatz that scales the $\ell_2$ normalized quantum state to the desired scale and
$
\mathbf{G}(\boldsymbol \theta(t)) = \prod_{i=1}^N \mathbf{G}_i(\theta_i(t)) 
$
is the product of $N$ parametric unitaries $\mathbf{G}_i$, each composed of one parametric rotation gates $e^{i \theta_k \mathfrak{G}_k}$ with $\mathfrak{G}_k^\dagger = \mathfrak{G}_k$.

\begin{figure}[h!]
\begin{center}
\begin{quantikz}
\lstick{$\tfrac{\ket{0} + e^{i\theta}\ket{1}}{\sqrt{2}}$} & \qw & \qw & \gate{X} & \ctrl{1} & \gate{X} & \qw & \qw & \ctrl{1} & \gate{H} & \meter{}\\
\lstick{$\ket{0}$}  & \gate{\mathbf{G}_N} & \gate{\cdots} & \gate{\mathbf{G}_k} & \gate{\mathfrak{G}_k} & \gate{\mathbf{G}_{k-1}} & \gate{\cdots} & \gate{\Omega} & \gate{\Upsilon} & \qw & \qw
\end{quantikz}
\caption{This quantum circuit evaluates the matrix elements $M_{k,j}$ and the vector elements $V_k$ depending on $\Omega$ and $\Upsilon$.}
\label{fig:expValueCircuit}
\end{center}
\end{figure}
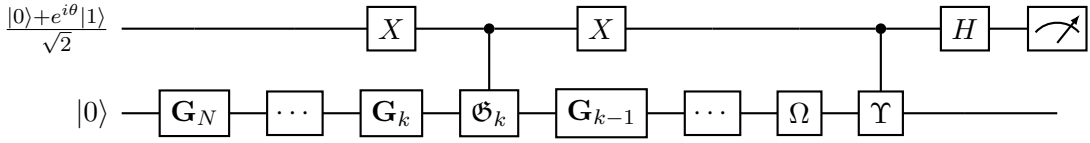

To produce the VarQITE evolution we need to replace $\ket{\tilde \psi({\bf x}, t)}$ by $\ket{{\tilde v}(\boldsymbol \theta (t))}$ in \eqref{masterequation} by first mapping the dynamics of the quantum state to the dynamics of the ansatz. McLachlan's variational principle
\begin{align} \label{mclachlan}
\delta \; \bigg| \bigg| \frac{\partial}{\partial t}\ket{{\tilde v}(\boldsymbol \theta (t))}  - \mathcal{E}(t)\ket{{\tilde v}(\boldsymbol \theta (t))} \bigg| \bigg| = 0
\end{align}
yields the Euler-Lagrange-type of equations \cite{theoryvariational}
\begin{align} \label{preeuler}
\sum_{j=0}^N M_{k,j} \dot \theta_k = V_k, \quad \textnormal{for each} \quad k=0, 1, \cdots N.
\end{align}
Here the elements of the matrix $M_{k,j}$ are given by
\[
M_{k,j} = \real \bigg(\alpha^2(t) \frac{\partial \bra{{v}(\boldsymbol \theta (t))}}{\partial \theta_k} \frac{\partial \ket{{v}(\boldsymbol \theta (t))}}{\partial \theta_j}\bigg), \quad \textnormal{for} \quad k,j \ne 0,
\]
as well as
\[
M_{0,j} = M_{j,0} = \alpha(t) \real \bigg( \bra{{v}(\boldsymbol \theta (t))} \frac{\partial \ket{{v}(\boldsymbol \theta (t))}}{\partial \theta_j}\bigg), \quad \textnormal{for} \quad j>0,
\]
and $M_{0,0}=1$. This matrix is fully dependent on the ansatz circuit. The vector $V_k$, however, depends on the operator $\mathcal{E}$ as follows
\[
V_k = \alpha(t) \real \bigg( \frac{\partial \bra{{ v}(\boldsymbol \theta (t))}}{\partial \theta_k} \mathcal{E} \ket{{ v}(\boldsymbol \theta (t))} \bigg), \quad \textnormal{for} \quad k>0,
\]
and the very first term gets simplified to
$
V_0 = \real (\bra{ v(\boldsymbol \theta (t))} \mathcal{E} \ket{{ v}(\boldsymbol \theta (t))}).
$
Notably, one may have to include so-called phase-fix terms if the global phase of the state evolution underlying $\mathcal{E}$ can change. These terms are explained in \cite{theoryvariational, zoufal_error_bounds}.
Since $\ket{{\tilde v}(\boldsymbol \theta (t))}$ is implemented with parameterized unitaries, the terms $M_{k,j}$ and $V_k$ can be computed parametrically by using the quantum circuit shown in Figure~\ref{fig:expValueCircuit}. 

The extraction of the matrix elements $M_{k,j}$ and the vector elements ${V_k}$ is as follows.  
\begin{itemize}
    \item One takes the sequence of gates $\mathbf{G}_{k-1}$ to $\Omega$ to be $\mathbf{G}_{k-1}$ to $\mathbf{G}_j$ and $\Upsilon = \mathfrak{G}_j$, if we are interested in $M_{k,j}$,
    \item or one sets the sequence of $\mathbf{G}_{k-1}$ to $\Omega$ to be $\mathbf{G}_{k-1}$ to $\mathbf{G}_1$ and then $\Upsilon = U_j$, where $U_n$ are easily-implementable unitary operators such as multi-qubit Pauli operators as represented in \eqref{decompositionE} below, if we are interested in $V_k$.
\end{itemize}
Further details on the respective implementation are given in  \cite{theoryvariational, variationalansatz, zoufal_error_bounds}.

From \eqref{preeuler} and using the above formulas for $M_{k,j}$ and $V_k$, we can now obtain the evolution of the parameters $\dot \theta_k$ using the forward Euler method
\begin{align}
\boldsymbol{\theta}(\tau + \delta \tau) \sim {\boldsymbol \theta}(\tau) + {\dot {\boldsymbol \theta}} \delta \tau = {\boldsymbol \theta}(\tau) + {\bf M}^{-1}(\tau) \cdot {\bf V} \delta \tau. \nonumber
\end{align}
That is, McLachlan's principle defines an ODE which can be numerically solved with an arbitrary ODE solver with $N_T=\tau/\delta \tau$ time steps. 
We would like to point out that the choice of the time step size $\delta \tau$ is crucial to achieve reasonable results. Although, it is generally preferable to choose rather smaller time steps one has to consider the trade off between accuracy and computational cost.
The use of an ODE solver that implements an adaptive step size scheme such as Runge-Kutta can help to find reasonable time steps.
In certain cases the matrix ${\bf M}$ will not be well-conditioned and thus we will need to invoke the Moore–Penrose inverse or use least squares, see e.g. \cite{fontanela2021quantum, variationalansatz}. 
Notably, the construction cost of the terms ${\bf M}$ and ${\bf V}$ scales quadratically in the number of circuit parameters and linearly in the number of Hamiltonian terms.
Furthermore, the computational complexity of solving the underlying system of linear equations scales generally as $\mathcal{O}(N^3)$. This complexity comes from standard Moore-Penrose inverse calculations and solving systems of linear equations where the matrix has dimensions $N \times N$ and the vector on the right hand side has dimension $N$.

One has to decompose the operator $\mathcal{E}(t)$ as a sum of tensor products of Pauli matrices
\begin{align} \label{decompositionE}
\mathcal{E}(t) = \sum_{h=1}^{H} \lambda_h U_h,
\end{align}
where $\lambda_h \in \C$ and $U_h$ is an easily implementable unitary operator made of Pauli operators. This requires an evaluation of $\mathcal{O}(N^2)+\mathcal{O}(N H)$ different quantum circuits. The circuits themselves will each require an ancilla qubit in addition to the qubits needed for the dynamics. This implies that in order to avoid undesirable growth of resources, the VarQITE evolution must be feasible with depths of the unitaries no more than polynomial in $n$, i.e. the depth of the tensor products of Pauli matrices must be $\mathcal{O}(\operatorname{poly}(n))$. This will be explained in Section \ref{sec:decomposition}.

Different manifestations of the Feynman-Kac equation will result in situations where the solution to the PDE might correspond to a probability distribution. To enable this we need to map the probabilities to amplitudes. Since quantum states have to be normalized with respect to the $\ell_2$ norm, while probabilities are normalized with respect to the $\ell_1$ norm, we then need to re-scale the state to handle this mismatch. In other words, the considered states have the form
\begin{align} \label{alphanormalized}
\ket{\tilde \psi(t)} = \alpha(t) \ket{\psi(t)}, \quad \ket{\tilde \psi(t)} = \sum_{i=0}^{2^n-1} p_i(t) \ket{i} \quad \textnormal{with}\quad p_i(t) = u(x_i,t)=\mathbb{P}[X(t)=x_i],
\end{align}
where $X$ is the random variable whose probability distribution function is given by $\ell_1$-normalizable function $u$, and $\alpha(t)$ is given by the $\ell_2$ norm
\[
\alpha(t) = \bigg(\sum_{i=0}^{2^n-1} p_i^2(t) \bigg)^{1/2}.
\]
This means that we embed the probability density into a state-vector which is re-scaled with an additional scalar parameter. This will result in a bijection. Moreover, the scalar parameter $\alpha(t)$ will be provided by the parameterized quantum circuit. It must be noted that indeed this differs from the traditional $\ell_2$ embedding which is of the form
\[
\ket{\psi(t)} = \sum_{i=0}^{2^n-1} \sqrt{p_i(t)} \ket{i}.
\]
Because of this difference of embedding, extraction of quantities involving $p_i$, such as the moments, need to be re-derived in this new context, see Section \ref{readout} and Appendix \ref{sec:moments} for further details.

Suppose we are confronted with solving \eqref{generalizedOp}. Obtaining the full distribution $u(x_i,t)=p_i(t)$ as its solution is the most complete information, but in general quantum algorithms cannot provide this type of result unless full-state tomography is employed, which is an expensive process \cite{d2003quantum}. However, there could be circumstances when knowing the first and second moments (say) of the solution, with quadratic speedup, might be advantageous enough. This also raises the question of approximating a probability distribution from (a few of) its moments. Some of the instances where the moments of the solution are useful are listed below:
\begin{itemize}
    \item The type of heat equation in \eqref{generalizedOp} describes the diffusion of point particles. In this case $u=u(x,t)$, for example in one spatial dimension with $u: \Omega \to \R$, describes the density of particles at a certain point $x$ at time $t$. The first moment $\int_\Omega u(x,t)dx$ is proportional to the total number of particles. If we consider a closed system, which is realized by imposing Neumann boundary conditions, then the conservation of energy means conservation of mass. If Neumann boundary conditions are not imposed, then the total mass is not relevant and it is sensible to instead define the moment $
    F[u(\cdot, t)]:= \int_\Omega u^2(x,t)dx$. Here $F$ is known as the \textit{energy functional} for the heat equation and it measures the distance, in $\ell_2$-norm, from the equilibrium solution $u=0$. Moreover, $F$ is related to the \textit{Dirichlet energy} $E[u(\cdot, t)] := \int_\Omega (\partial_x u)^2 dx$, see \cite[$\mathsection$2.2.5 and $\mathsection$2.3.4]{evansPDEs}, as follows.
    For solutions of the heat equation $u_t = k u_{xx}$ integration by parts yields
\begin{align}
    \frac{d}{dt} F[u(\cdot, t)] &= 2 \int_\Omega u(x,t)u_t(x,t)dx = 2k \int_\Omega u(x,t)u_{xx}(x,t)dx \nonumber \\
    &= -2 k \int_\Omega (u_x(x,t))^2 dx = -2k E[u(\cdot, t)].
\end{align}
When we integrate over $t$, the functional $F[u(\cdot, t)]$ describes the accumulated loss of Dirichlet energy for $t >0$ and equals our scaling factor $\alpha^2(t)$. The time derivative can be approximated by employing finite differences as we have the time evolution of the proxy $\alpha(t)$.
Dirichlet energy plays an important role in soft mapping \cite{solomon2013dirichlet}, machine learning \cite{zhou2021dirichlet} and discrete differential geometry \cite{stein2020simple, stein2020smoothness}.
    \item Suppose that $q$ and $p$ are the coordinate and momenta respectively of a wave function and that $\sigma(x) = \sqrt{v(x)}$ is the standard deviation of the observable $\hat x$, where $v(x) = \mu(x-\mu(x))^2$ is the variance and $\mu(x)$ is the mean of $\hat x$. Reasoning from the Heisenberg uncertainty principle $\Delta p \cdot \Delta x \approx h$ or $\sigma(p) \cdot \sigma(q) \ge \tfrac{1}{2}\hbar$ in \cite{sussmann}, the \textit{uncertainty length} is defined as $\delta[x]:= (\int_{\R} p(x)^2 dx)^{-1}$, where $p(x)$ is the probability distribution of the stochastic variable $\hat x$. This is also known as the \textit{S\"{u}smann measure} and it can be interpreted as the total length of all the intervals that produce rectangles of total areas $h_x \dot \delta[x]=1$. Here the height $h_x$ represents the mean value of the normalized $p(x)$ with $p(x)$ itself as its own weight factor. Its usefulness rests on the fact that other quantities related to width, like variance, might not necessarily exist for certain distributions (e.g. the Cauchy distribution) whereas the S\"{u}ssmann measure will. The inverse of the S\"{u}ssmann measure can be thought of as the average height of the distribution.
    \item The quantum mechanical observables corresponding to Hermitian operators $\hat{O}$ acting on the wave function $\ket{\psi}=\ket{\psi(x,t)}$ obtained via solving Schr\"{o}dinger equation \eqref{schrodingertransformed2} or the Wick-rotated Schr\"{o}dinger equation \eqref{wickrotatednormalized}, are expectation values of the form $\bra{\psi}\hat{O}\ket{\psi}
    $. Thus, all such properties are addressable via the algorithm described in this work and the references mentioned herein.
    \item Let us suppose we have an educated belief that the resulting distribution is normal, or a function thereof, e.g. lognormal. One plausible reason for this belief is that the Central Limit Theorem gives the normal distribution a special status and it is characterised by $\mu$ and $\sigma$. Furthermore, as will be shown in Section \ref{readout} the evolution of the Feynman-Kac system needs to be evaluated only once to extract the final circuit parameters. Once these circuit parameters are obtained, additional higher order moments can be evaluated in parallel using several quantum devices, in those cases where skewness and kurtosis may play significant role in the description of the solution. As the dimension of the problem increases, so will the overhead. Lastly, for time-critical applications one might want to know the first moments using a quantum device while waiting for the full distribution to be computed classically.
\end{itemize}

\subsection{\label{sec:decomposition}Decomposition of the generator}
In order to implement VarQITE, we need to employ the matrix $M$ and vector $V$ from \eqref{preeuler}. We shall start with only one dimension. First, we need to decompose the operator $\mathcal{G}$ into unitaries. The following procedure is borrowed from \cite{stochastic}[$\mathsection$III.C] and then adapted to our operator of interest $\mathcal{G}$ given by \eqref{hamiltonianpaper} below. Start by defining the operators
\begin{align} \label{Vpmn}
V_{+}(n) := \sum_{i=0}^{2^n-2} \ket{i + 1}\bra{i} \quad \textnormal{and} \quad V_{-}(n) := \sum_{i=0}^{2^n-1} \ket{i - 1}\bra{i}.
\end{align}
If we make the boundary identifications $\ket{-1} \leftrightarrow \ket{2^n-1}$ and $\ket{2^n} \leftrightarrow \ket{0}$, then the operators $V_{\pm}$ can be manufactured from the $n$-qubit operators
$ 
\operatorname{Cyc}_{\pm}(n) := \sum_{i=0}^{2^n-1} \ket{i \pm 1}\bra{i}.
$
As per \cite{incrementer}, these operators can be implemented as a product of $\mathcal{O}(n)$ Toffoli, CNOT, and $X$ gates with $\mathcal{O}(n)$ ancilla qubits. The explicit construction of $V_{\pm}$ from $\operatorname{Cyc}_{\pm}$ is achieved by defining an $n$-qubit-control gate $C^{n-1}Z := \sum_{i=0}^{2^n-2} \ket{i}\bra{i}-\ket{2^n-1}\bra{2^n-1}$. According to \cite{nielsenchuang}, then implementation of $C^{n-1} Z$ can be achieved as a product of $\mathcal{O}(n^2)$ Toffolli, CNOT and other single-qubit gates. Using the identity $\tfrac{1}{2}(C^{n-1} Z + I^{\otimes n}) = \sum_{i=0}^{2^n-2} \ket{i}\bra{i}$, one can prove that the $V$'s from \eqref{Vpmn} and the $\operatorname{Cyc}$'s are related by 
\begin{align}
V_+(n) = \operatorname{Cyc}_+(n) \cdot \frac{1}{2}(C^{n-1} Z + I^{\otimes n}) \quad \textnormal{and} \quad
V_-(n) = \frac{1}{2}(C^{n-1} Z + I^{\otimes n}) \cdot \operatorname{Cyc}_-(n). \nonumber
\end{align}
Thus we can decompose $V_{\pm}$ into a sum of two unitaries made of $\mathcal{O}(n^2)$ few-qubit gates. Lastly, we bring in an operator composed of weighted projectors
\begin{align} \label{D(n)}
D(n) := \sum_{i=0}^{2^n-1} i \ket{i}\bra{i} = \frac{2^n-1}{2} I^{\otimes n} - \sum_{i=1}^n 2^{n-i-1}Z_i,
\end{align}
where $Z_i$ is a $Z$ gate acting on the $i$th qubit. This implies that $D(n)$ is now a sum of $\mathcal{O}(n)$ unitaries composed of a single-qubit gate. Utilizing these operators we end up with
\begin{align}
V_{+}(n)(D(n))^m = \sum_{i=0}^{2^n-2} i^m \ket{i+1}\bra{i} \quad \textnormal{and} \quad V_{-}(n)(D(n))^m = \sum_{i=1}^{2^n-1} i^m \ket{i-1}\bra{i}. \nonumber
\end{align}
For example, let us suppose we have an operator $\mathcal{F}$ whose decomposition is given by
\begin{align}
\mathcal{F}(t) :&= \sumtwo_{i,k=0}^{2^n-1} (\mathcal{F}(t))_{i,k} \ket{i}\bra{k} \nonumber \\
&=  \sum_{i=0}^{2^n-2} f_1(x_i,t) \ket{i+1}\bra{i} + \sum_{i=1}^{2^n-1} f_2(x_i,t) \ket{i-1}\bra{i} + \sum_{i=0}^{2^n-1} f_3(x_i,t) \ket{i}\bra{i}, \nonumber
\end{align}
with all other matrix elements $(\mathcal{F}(t))_{i,k}$ being zero, which is a typical situation from the stencil method we are implementing in our formulation. Then we decompose the functions $f_j$ into Taylor expansions 
\begin{align}
f_j(x_i,t) = \sum_m \frac{f_j^{(m)}(0,t)}{m!}x_i^m, \quad j \in \{1,2,3\}, \nonumber
\end{align}
where $f^{(m)}$ denotes the $m$th derivative. Using $x_i = \Delta x \cdot i$ allows us to write
\begin{align}
\mathcal{F}(t) &= \sum_m \bigg(\sum_{i=0}^{2^n-2} \frac{f_1^{(m)}(0,t)}{m!} (\Delta x i)^m \ket{i+1}\bra{i} + \sum_{i=1}^{2^n-1} \frac{f_2^{(m)}(0,t)}{m!}(\Delta x i)^m \ket{i-1}\bra{i} \nonumber \\
& \quad \quad \quad \quad + \sum_{i=0}^{2^n-1} \frac{f_3^{(m)}(0,t)}{m!}(\Delta x i)^m \ket{i}\bra{i} \bigg) \nonumber \\
&= \sum_m \bigg(\frac{f_1^{(m)}(0,t)}{m!} (\Delta x)^m V_+(n)(D(n))^m + \frac{f_2^{(m)}(0,t)}{m!}  (\Delta x)^m V_-(n)(D(n))^m  \nonumber \\
& \quad \quad \quad \quad + \frac{f_3^{(m)}(0,t)}{m!} (\Delta x)^m (D(n))^m \bigg). \nonumber
\end{align}
The operators $V_+(n)(D(n))^m, V_-(n)(D(n))^m$ and $(D(n))^m$ are all made of the sum of $\mathcal{O}(n^m)$ unitaries. Each of these unitaries is made of $\mathcal{O}(n^2)$ few-qubit gates.

Our next step is to now generalize this to multiple dimensions $(x_1, x_2, \cdots, x_D)^T$. This generalization builds up naturally from the one dimensional situation we just described above. In this case, we use multivariate Taylor expansions and the operators
$ 
V_{\pm}^{(k)}(n) := I^{\otimes k-1} \otimes V_{\pm}(n) \otimes I^{\otimes D-k} \quad \textnormal{and} \quad D^{(k)}(n) := I^{\otimes k-1} \otimes D(n) \otimes I^{\otimes D-k}. 
$
These operators will then satisfy 
\begin{align} \label{vpmdDk}
V_{\pm}(n)(D^{(k)}(n))^m = \sum_{i_k=0}^{2^n-2} \sum_{\substack{i_m \in \{0,1,\cdots,2^n-1\} \\ m \ne k}} i_k^m \ket{{\bf i} \pm {\bf e}_k} \bra{{\bf i}},
\end{align}
as well as
\begin{align} \label{vpmDkDell}
V_{\pm}(n)(D^{(k)}(n))^{m_k} V_{\pm}(n)(D^{(\ell)}(n))^{m_\ell} = \sum_{i_k=0}^{2^n-2} \sum_{i_\ell=0}^{2^n-2} \sum_{\substack{i_m \in \{0,1,\cdots,2^n-1\} \\ m \ne k \\ m \ne \ell}} i_k^{m_k} i_\ell^{m_\ell} \ket{{\bf i} \pm {\bf e}_k \pm {\bf e}_\ell} \bra{{\bf i}}.
\end{align}
Expansions for different combinations of $V$'s and different projectors can be obtained by similar methods. With \eqref{vpmdDk} and \eqref{vpmDkDell} we may now decompose the generator $\hat{\mathcal{G}}$ given by \eqref{discreteG1} and \eqref{discreteG2} into easily implementable unitaries composed of few-qubit gates and thereby generate a feasible VarQITE. Specifically, the operators $V_{\pm}(n) (D^{k}(n))^m$ are the sums of $\mathcal{O}(n^m)$ unitaries, like in the one dimensional case, each composed of $\mathcal{O}(n^2)$ few-qubits gates. This procedures scales polynomially in the number of qubits $n$ due to the truncation of the Taylor approximation at a fixed order $m$.

\subsection{\label{readout}Readout and efficiency}

In this section we discuss how to readout results of the solution of a PDE associated to an SDE as well as the resulting total algorithmic complexity. 
Further, we put this into context and identify the situations where this can lead to a potential quantum advantage.

Before we go into the details of the presented algorithm, we discuss the complexity of classical methods.
Suppose we want to solve an SDE in $D$ dimensions and time $t \in [0, T]$ or the corresponding PDE.
Starting with the SDE, we can run a Monte Carlo simulation to generate different scenarios and estimate the solution to the SDE. 
To achieve a target accuracy $\varepsilon > 0$ this will scale as $\mathcal{O}(T / \varepsilon^2)$.
For small $\varepsilon$ and complicated SDEs this can become prohibitive.
An alternative to Monte Carlo simulation is considering the corresponding PDE, discretizing the state space and using a classical PDE solver.
We assume here that a discretization of the state space using $2^{n}$ grid points, i.e., with $2^{n/D}$ grid points per dimension, is sufficient to reach the target accuracy. Then we assume that the total complexity would scale as $\mathcal{O}(T 2^{n})$ to access the final state and evaluate properties of interest of that state, where we assumed the generator $\mathcal{G}$ corresponds to a sparse matrix such that matrix-vector products scale linearly in the dimension and that the number of iterations needed by the ODE solver scales linearly in $T$.

The quantum algorithm introduced in this paper is closely related to the latter approach and in the following we compare a potential advantage between those two.
The comparison between the classical and quantum PDE solvers is not completely fair: the classical solver yields the complete state, while the quantum solver only gives access to some properties of the solution. The classical algorithm allows us to control and evaluate the accuracy, while the quantum algorithm is a heuristic unless the ansatz is universal, which would defeat the purpose as this would scale exponentially in the number of qubits.
We ignore these differences in the remainder of this section, and thus, the following can be seen as necessary -- but not necessarily sufficient -- conditions for a quantum advantage.

While obtaining the full solution to the PDE generated by the presented algorithm is not feasible, we may try to extract certain properties of interest.
Here we need to distinguish between two cases: when we solve our problem forward in time, as, e.g., for the heat equation, or backward in time, as for pricing financial options.
The two situations imply differences in how the initial states are prepared and how the results are read out. 
In the following, we focus on the readout and its efficiency when solving the problem forward in time. 
The case of solving a PDE backward in time using the proposed method is discussed in Appendix \ref{sec:boundaryfinance}.

Suppose the solution to the PDE is given by a scaled $n$-qubit quantum state $\ket{\tilde \psi} = \alpha \ket{\psi} = \sum_{i=0}^{2^n-1} p_i \ket{i}$, i.e., the scaled amplitudes correspond to the elements of the discretized solution to the considered PDE.
We ignore in the following that we only approximate this state.
In \cite{stochastic} the authors discuss how to estimate properties of the form 
\begin{eqnarray}
\left| \sum_{i=0}^{2^n-1} p_i f(x_i) \right|, 
\label{observable_abs}
\end{eqnarray}
where $f$ is some function defined on a grid $x_i$.
If the $p_i$ correspond to a probability distribution of a random variable $X$ (and for simplicity assuming $f(x) \geq 0$), then this results in an expected value 
$\mathbb{E}[f(X)]$.

This can be achieved if we can efficiently construct an operator $S_f$ such that
\begin{eqnarray}
S_f \ket{0} = \sum_{i=0}^{2^n-1} f(x_i) \ket{i}, \nonumber
\end{eqnarray}
or by a weighted sum of such operators \cite{stochastic}.
Note that this imposes an $\ell_2$ normalization constraint on the function $f$, i.e. 
$
\sum_{i=0}^{2^n-1} |f(x_i)|^2 = 1.
$
If $f$ is not $\ell_2$ normalized, the normalization factor needs to be multiplied on the result later on, amplifying the estimation error, as we discuss later in this section.
The operator $S_f$ can then be used to construct
\begin{eqnarray}
\bra{\tilde \psi} S_f \ket{0}\bra{0} S_f^{\dagger} \ket{\tilde \psi} = \alpha^2 \bra{\psi} S_f \ket{0}\bra{0} S_f^{\dagger} \ket{ \psi} = \left| \sum_{i=0}^{2^n-1} p_i f(x_i) \right|^2.
\label{eq:exp_val_sf}
\end{eqnarray}
Since 
$$
\ket{0}\bra{0} = (\mathbb{I} - X^{\otimes n} \cdot C^{n-1}Z \cdot X^{\otimes n})/2,
$$ we can estimate \eqref{eq:exp_val_sf} as
\begin{eqnarray}
\alpha^2 (1 - \bra{\psi} S_f X^{\otimes n} \cdot C^{n-1}Z \cdot X^{\otimes n} S_f^{\dagger} \ket{\psi}) / 2, \nonumber
\end{eqnarray}
i.e., by estimating the expected value of $C^{n-1}Z$ for the state $X^{\otimes n} S_f^{\dagger} \ket{\psi}$.
As the authors of \cite{stochastic} point out, this can be achieved, for instance, up to accuracy $\varepsilon > 0$ with a Hadamard test, using $\mathcal{O}(1/\varepsilon^2)$ samples or with Quantum Phase Estimation (QPE), using $\mathcal{O}(\log(1/\varepsilon))$ samples but with a $\mathcal{O}(1/\varepsilon)$-times longer circuit.
Note that we evaluate the square of the expected value using the quantum computer and then multiply the result by $\alpha^2$. 
Thus, to achieve an overall accuracy $\varepsilon$ for $\mathbb{E}[f(X)]$, we need to estimate the expected value with an accuracy $\mathcal{O}(\varepsilon / \alpha)$ \cite{stochastic}.
In the case of the $p_i$ being probabilities, it follows from the relation between $\ell_1$ and $\ell_2$ norms that $\alpha \in [1/\sqrt{2^n}, 1]$, i.e., $\alpha$ may have a positive impact on the estimation error.

Although this looks promising, there are a couple of potential pitfalls that can diminish or at least significantly reduce the potential quantum advantage:
\begin{enumerate}
    \item The operator $S_f$ is often difficult to construct and in general requires a number of terms exponential in the represented dimension that usually scales linearly in the number of qubits.
    
    \item The function $f$ needs to be $\ell_2$-normalized, that is we estimate $f/ \|f\|_2$ for a non-normalized function. When going back to the original scale, the resulting estimation error will be $\|f\|_2 \varepsilon$. Thus, the expected value needs to be estimated more accurately to compensate that scaling. This factor is maximal for constant functions, where it scales as  $\|f\|_{\infty}\sqrt{2^n}$. With the Hadamard test, this implies $\mathcal{O}(2^n)$ samples which would diminish the quantum advantage, while QPE would at least achieve a quadratic advantage, depending on $\|f\|_{\infty}$.
    
    \item Since $\alpha \in [1/\sqrt{2^n}, 1]$ (for probability distributions) it may help to partially compensate the error amplification introduced by the normalization of $f$.  
\end{enumerate}

These points are critical to be taken into account and determine the potential advantage of the presented technique, and are often ignored in the related literature. 
For solving PDEs backwards in time, similar phenomena appear and we discuss them in more depth in Appendix \ref{sec:boundaryfinance}.

For illustration, we assume the solution represents the probability distribution $p$ of a $D$-dimensional random variable $X$ and we will focus on how to estimate its expected value $\mathbb{E}[X]$, for more details, see Appendix \ref{sec:moments}.
To this extent we assume that the scaled quantum state $\alpha \ket{\psi}$ on $n$ qubits represents the final solution, i.e., we represent each dimension with $n/D$ qubits.

Further, we assume an $n/D$-qubit operator 
$$
S_x: \ket{0} \mapsto \frac{1}{C} \sum_{x=0}^{2^{n/D}-1} x \ket{x},
$$
i.e., we set $f(x) = x$, and 
$$
C = \sqrt{\sum_{x=0}^{2^{n/D}-1} x^2} = \theta(\sqrt{(2^{3n/D})}) = \theta(\|f\|_{\infty}\sqrt{(2^{n/D})})
$$ 
is the required normalization constant. 
There are different ways to construct $S_x$, e.g., via a piece-wise constant approximation as suggested in \cite{stochastic} or following the approach suggested in \cite{groverrudolph}, which may be simplified due to the simple structure of the function considered.
In the following, we focus on how to estimate a single component 
$$
\mathbb{E}[X_i] = \sum_{X_1=0}^{2^{n/D}-1} \cdots \sum_{X_D=0}^{2^{n/D}-1} p(X_1, \ldots, X_D) X_i,
$$
with $i \in \{1, \ldots, D\}$, of the expected value.
This needs to be repeated $D$ times to estimate the full $D$-dimensional result.
Without loss of generality we now assume $i = 1$ and define 
$$
S_f = S_x \otimes \bigotimes_{i=2}^D H^{\otimes n/D}.
$$
It is easy to see that this corresponds to the objective function
\begin{eqnarray} 
f(X_1, \ldots, X_D) &=& \frac{X_1}{C \sqrt{2^{n-n/D}}}, \nonumber
\end{eqnarray}
and we can apply the above approach to estimate $\mathbb{E}[X_1] / (C \sqrt{2^{n-n/D}})$.
To achieve a target accuracy $\varepsilon > 0$ for $\mathbb{E}[X_1]$ using QPE requires an estimation accuracy of $\varepsilon / (C\sqrt{2^{n-n/D}})$ which results in a $\theta(2^{n/D}\sqrt{2^{n}})$-times longer circuits (ignoring $\alpha$).

To summarize, the proposed quantum algorithm can achieve an exponential speed-up for the (approximate) state evolution, but only up to a quadratic speed-up for the extraction of solution properties, which results in an overall complexity of $\mathcal{O}(T \operatorname{poly}(n) + \sqrt{2^n})$ since the readout can be done using the final parameters without rerunning the evolution.
Thus, the proposed algorithm might be particularly suitable for problems with a large time horizon.
However, it could also accumulate a higher approximation error for those, and understanding where these two effects balance is subject to further research.

\subsection{{\label{sec:enforce}}Enforcing $\ell_1$ normalization}

Many PDEs describe the evolution of probability distributions, i.e., the components of a discretized solution should be in $[0, 1]$ and sum up to one. Although we allow the scaling factor $\alpha_t$ to overcome the $\ell_2$-normalization of the quantum state $\ket{\psi_t}$, the variational principle in general does neither preserve the $\ell_1$-norm for $\alpha \ket{\psi_t}$ nor that the elements should be within $[0, 1]$. 
Since it minimizes a geometric distance, the closest representable state might not be $\ell_1$-normalized.
In the following we describe how the procedure could be adjusted to (approximately) enforce $\ell_1$-normalization while sacrificing a bit of accuracy following the approach introduced in \cite{vazquez2021enhancing}.

As shown in Section \ref{readout}, we can estimate properties of the form
$\left| \sum_{i=0}^{2^n-1} p_i f(x_i) \right|$ for an $\ell_2$-normalized function $f$ and amplitudes $p_i$ of the current state $\ket{\psi_t}$ up to an accuracy $\varepsilon > 0$. If we now set $f(x_i) = \sqrt{1/2^n}$ for all $i$, then we can easily construct $S_f = H^{\otimes n}$, i.e., just a single layer of Hadamard gates, which will result in an estimation of
$\left| \sum_{i=0}^{2^n-1} p_i \right| / \sqrt{2^n}$.
If we estimate this to an accuracy $\varepsilon / \sqrt{2^n}$ using QPE, this can be used as an approximation to the $\ell_1$-norm of the state and allows to adjust the $\alpha_t$ such that the scaled quantum state would be (approximately) $\ell_1$-normalized.
Note that the accuracy requirements on the estimation of the $\ell_1$-norm likely limit the potential advantage to a quadratic speed-up and increase the overall runtime to $\mathcal{O}(T\sqrt{2^n})$.
We will illustrate the result of this approximation numerically in Section ~\ref{sec:plots}. Appendix ~\ref{sec:appendixnorms} contains additional comparisons with and without the $\ell_1$-enforcement procedure described in this section.

\newpage
\subsection{{\label{sec:algosummary}Summarizing the algorithm}}

The algorithm we have described is now summarized in Algorithm \ref{alg:FKalg}. For clarity, we use one spatial dimension as well as the forward Euler ODE solver. However, in practice, more sophisticated solvers such as for instance Runge-Kutta methods are usually more suitable.

\begin{algorithm}
\caption{Variational quantum algorithm for the Feynman-Kac formula}
\label{alg:FKalg}
\DontPrintSemicolon
  \KwInput{$n_q \in \N$, $D = 1$, $N_T>0$, $\delta \tau > 0$, $N_T=\tau_{\textnormal{total}}/\delta \tau$, $\delta x>0$ and initial condition $\psi$. SDE of the form $dX_t = \mathfrak{b}(X_t,t)dt + \sqrt{2 \mathfrak{a}(X_t,t)} dW_t$.}
  \KwOutput{Normalized state kets for solution of 
    \begin{align} \label{algopde}
  \frac{\partial u}{\partial t} + \mathfrak{a}(x,t) \frac{\partial^2 u}{\partial x^2} + \mathfrak{b}(x,t) \frac{\partial u}{\partial x} - \mathfrak{c}(x,t)u = 0.
  \end{align}
  }
  Apply appropriate Feynman-Kac formulae to switch SDE to PDE where 
  \[u(x,t)=\mathbb{E}[e^{-\int_0^t \mathfrak{c}(X_\tau,\tau) d\tau} \psi(X_T) \; | \; X_t =x] \textnormal{ for $t> 0$ and $u(x,0)=\psi(x)$.}\]
  
  Determine the appropriate normalization ($\ell_1, \ell_2, \cdots$) of \eqref{algopde}.
  
  Extract generator/Hamiltonian $\mathcal{G}$ for \eqref{algopde}.
  
  Discretize $\mathcal{G}$ with periodic boundary conditions.
  
  Decompose $\mathcal{G}=\sum_{h=1}^H \lambda_h U_h$, where $\lambda_h \in \C$ and $U_h$ are implementable unitary operators composed of tensor products of Pauli operators. 
  
  Choose ansatz circuit with $N+1$ parameterized gates.
  
  Define initial condition $\ket{\tilde v({\boldsymbol \theta}(0))}$.
  
  Determine initial parameters $\boldsymbol\theta_0 = (\theta_1(0), \cdots, \theta_N(0), \theta_{N+1}(0))$ where $\theta_{N+1}(\tau)=\alpha(\tau)$, see \eqref{minimization}.
  
  Initialize $\tau = 0$.
  
  \While{$\tau \le \tau_{\operatorname{total}}$}
    {
        Compute ansatz-dependent matrix elements $M_{k,j}=\real(\alpha^2(\tau)\frac{\partial \bra{v(\boldsymbol \theta (\tau))}}{\partial \theta_k} \frac{\partial \ket{v(\boldsymbol \theta (\tau))}}{\partial \theta_j})$ for $k,j \ne 0$ and $M_{0,j}=M_{j,0}=\alpha(\tau)\real(\bra{v(\boldsymbol \theta (\tau))} \frac{\partial \ket{v(\boldsymbol \theta (\tau))}}{\partial \theta_j})$ for $j>0$ and $M_{0,0}=1$.    
        
        \If{${\bf M}^{-1}$ is singular}
        {Use least squares or Moore-Penrose inversion.}
        \If{$\mathcal{G}$ is time-dependent}
        {Update generator $\mathcal{G}(\tau)$ and its associated Pauli decomposition $\sum_{h=1}^H \lambda_h(\tau) U_h(\tau)$.}
        Compute generator-dependent vector elements $V_{k}=\alpha(\tau)\real(\frac{\partial \bra{v(\boldsymbol \theta (\tau))}}{\partial \theta_k} \mathcal{G} \ket{v(\boldsymbol \theta (\tau))}$ for $k>0$ and $V_0 = \real (\bra{ v(\boldsymbol \theta (\tau))} \mathcal{G} \ket{{ v}(\boldsymbol \theta (\tau))})$.   
        
        Update values of $\boldsymbol \theta$ with $\boldsymbol{\theta}(\tau + \delta \tau) \approx {\boldsymbol \theta}(\tau) + {\bf M}^{-1}(\tau) \cdot {\bf V} \delta \tau$.
        
        Update $\tau \to \tau + \delta \tau$.
    }
\end{algorithm}

\section{\label{sec:numerical}Numerical experiments} 
We now provide numerical evidence of the algorithm we have proposed. To do so, we first need to establish the exact set of SDEs that we wish to study, in our case it will be a simple set of two correlated Brownian motions. From there we will write down the PDE for the expectation of these Brownian motions. This PDE is an anistropic heat equation that mixes the space variables $x$ and $y$ due to the correlation between the Brownian motions. The full solution to this PDE is provided in Appendix \ref{sec:analytical} by means of techniques from Mellin and Laplace transforms. The ansatz circuit will be explained and we will also illustrate the behavior of the $\ell_1$ and $\ell_2$ norms during the evolution. The VarQITE algorithm outputs the basis kets in one dimension, and since we shall use six and eight qubits, we will obtain $2^6$ and $2^8$ basis kets. Although the problem is two dimensional, it is instructive to take amplitude vector in the computational basis and the comparison against the classical methods of forward Euler as well as Monte Carlo with periodic boundary conditions.

\subsection{\label{sec:anisotropic}Two-dimensional anistropic heat equation}

We shall now specialize the number of dimensions, the number of Brownian motions and the coefficients in the SDE to illustrate the Feynman-Kac formula on a quantum computer. Taking $D=N=2$ in \eqref{systemSDEs} yields
\begin{align} \label{oursystemSDEs}
\begin{cases}
dX_t^1 = \mu_1(t, X_t^1, X_t^2) dt + \sigma_{11}(t, X_t^1, X_t^2) dW_t^1 + \sigma_{12}(t, X_t^1, X_t^2) dW_t^2, \\
dX_t^2 = \mu_2(t, X_t^1, X_t^2) dt + \sigma_{21}(t, X_t^1, X_t^2) dW_t^1 + \sigma_{22}(t, X_t^1, X_t^2) dW_t^2. 
\end{cases}
\end{align}
Next, we specialize to $\mu_1 = \mu_2 = 0$ as well as 
\begin{align}
\sigma_{11} = \sqrt{{\tilde \sigma}_1}, \quad \sigma_{12} = 0, \quad \sigma_{21} = \sqrt{{\tilde \sigma}_2} \rho, \quad\sigma_{22} = \sqrt{{\tilde \sigma}_2} \sqrt{1-\rho^2}, \nonumber
\end{align}
where $\rho$ is a real in the interval $[-1,1]$ which now is interpreted as the correlation between the two Brownian motions $W_t^1$ and $W_t^2$. The resulting matrix becomes
\begin{align}
{\bf \Sigma} =
\begin{pmatrix}
\sqrt{{\tilde \sigma}_1} & 0\\
\sqrt{{\tilde \sigma}_2} \rho  & \sqrt{{\tilde \sigma}_2}\sqrt{1-\rho^2}
\end{pmatrix}. \nonumber
\end{align}
The infinitesimal generator $\hat{\mathcal{G}}$ reduces to
\begin{align}
\hat{\mathcal{G}} = \frac{{\tilde \sigma}_1}{2} \frac{\partial^2}{\partial x_1^2} + \rho \sqrt{{\tilde \sigma}_1 {\tilde \sigma}_2} \frac{\partial^2}{\partial x_1 \partial x_2} + \frac{{\tilde \sigma}_2}{2} \frac{\partial^2}{\partial x_2^2} -r(x_1,x_2,t). \nonumber
\end{align}
in this particular case. We shall also relabel $x_1=x$ and $x_2=y$. Moreover for simplicity our discount function will be $r(x,y,t)=0$ for all $t$, and we shall choose $\sigma_1$ and $\sigma_2$ to be $1$.

\begin{figure}[h!] 
	\centering
	\includegraphics[scale=0.36]{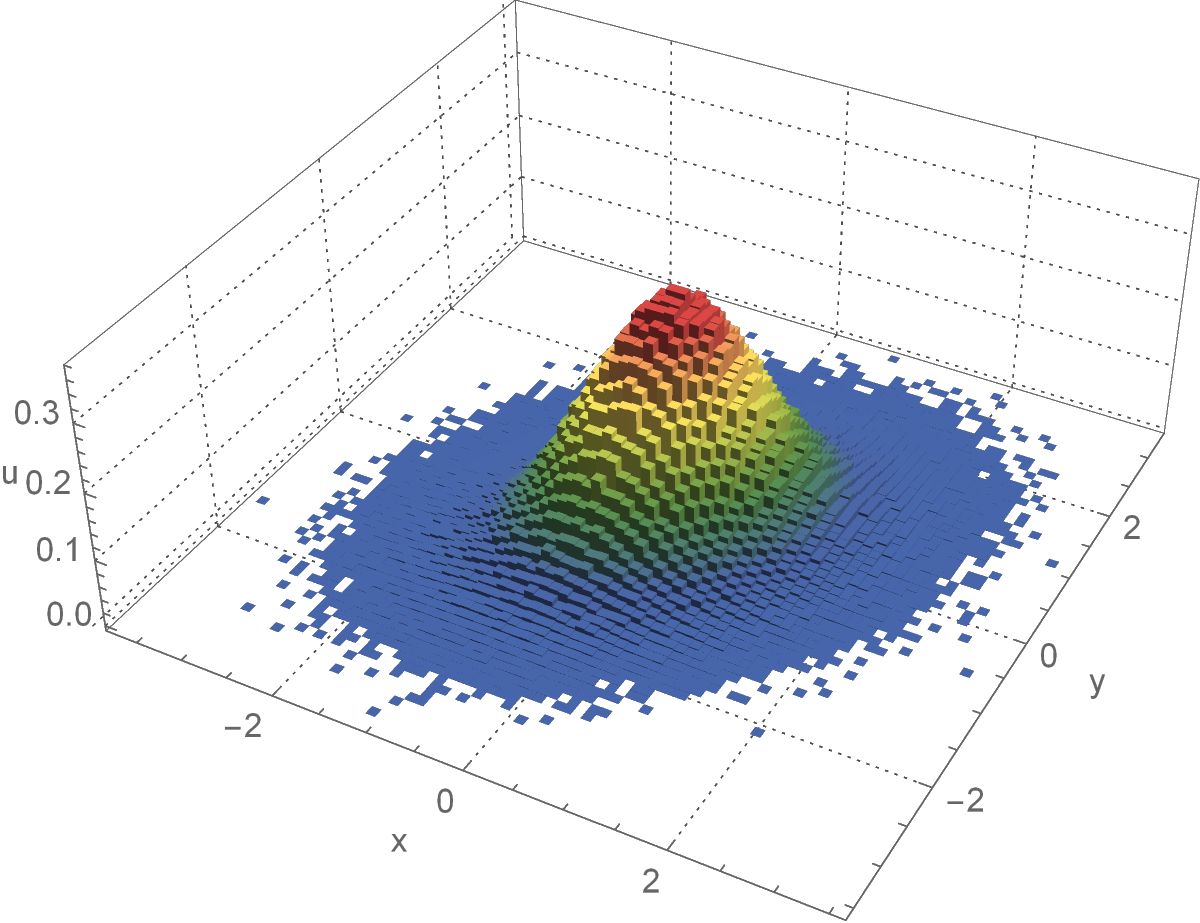}
	\includegraphics[scale=0.36]{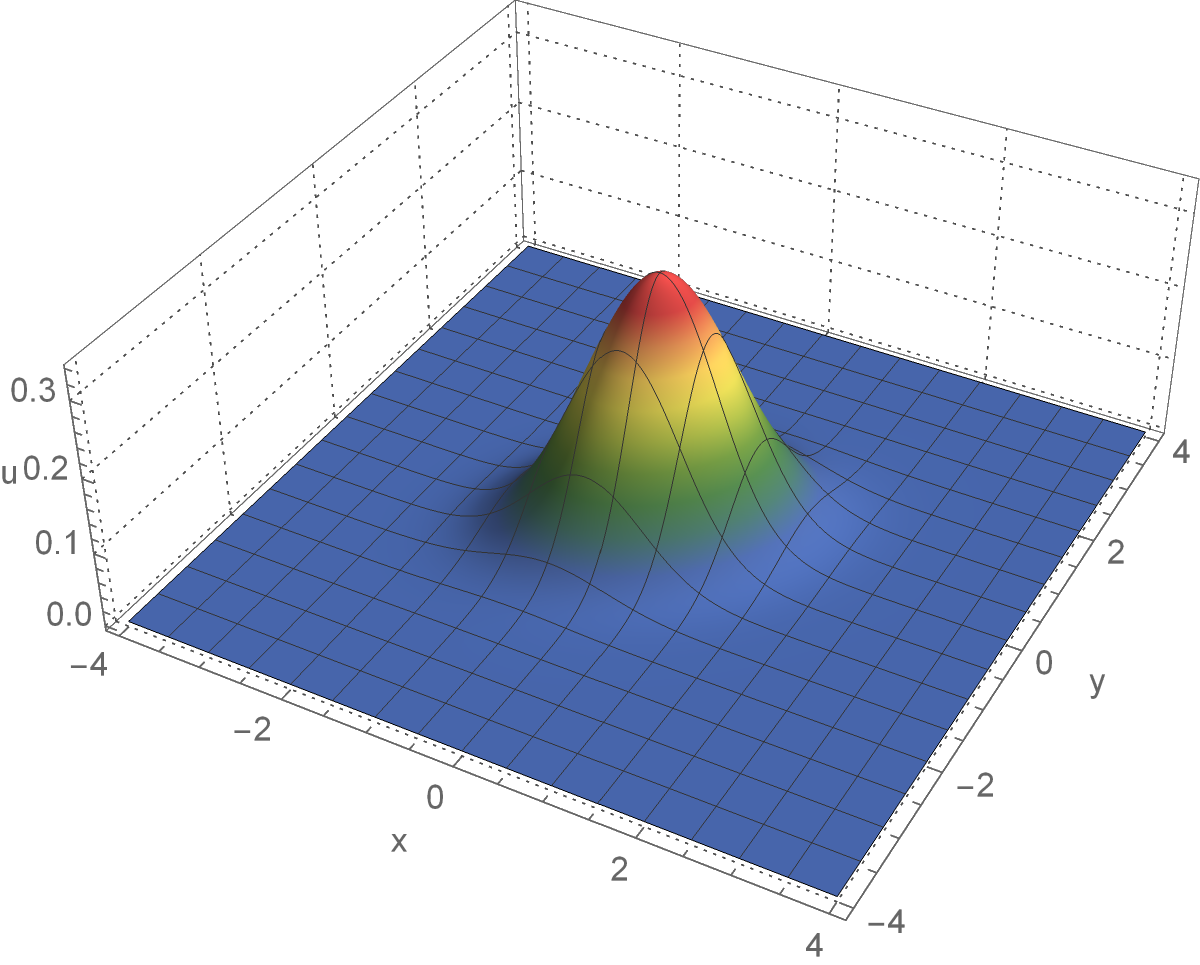}
	\caption{\underline{Left}: probability distribution of 1 million paths governed by the system of SDEs from \eqref{oursystemSDEs}. \underline{Right}: Corresponding analytical solution to \eqref{ourPDE} given by \eqref{oursolution} with $\rho=\tfrac{1}{3}$ at $t=\tfrac{1}{10}$.}
	\label{fig:realizationsanalytical} 
\end{figure}

In summary, we are working with the differential operator and conditional expectation 
\begin{align} \label{hamiltonianpaper}
\mathcal{G} = \rho \frac{\partial^2 }{\partial x \partial y} + \frac{1}{2} \frac{\partial^2 }{\partial x^2} + \frac{1}{2} \frac{\partial^2 }{\partial y^2} = \mathcal{G}^{\dagger} \quad \quad \textnormal{and} \quad u(x,y, t) = \mathbb{E} [\psi(X_t, Y_t) \; | \; X_t = x, Y_t=y].
\end{align}
The associated Feynman-Kac PDE for the expectation of the system of SDEs $dX_t = dW_t^1$ and $dY_t = dW_t^2$ is
\begin{align} \label{ourPDE}
\frac{\partial u}{\partial t} = \mathcal{G} u \quad t>0,
\end{align}
with initial value condition $\psi$ given by
\begin{align} \label{diracdeltas}
u(x,y,0) = \delta(x-x_0)\delta(y-y_0), 
\end{align}
where $\delta$ is the Dirac delta function. The operator $\mathcal{G}$ is an anisotropic Hamiltonian that includes the Laplacian operator
$
\nabla^2 = \frac{\partial^2}{\partial x^2} + \frac{\partial^2}{\partial y^2}, \nonumber
$
as well as the term mixing both first order derivatives
$
\frac{\partial^2}{\partial x \partial y}.
$
Equation \eqref{ourPDE} admits an exact heat kernel solution, which we derive -- in a more general setting involving arbitrary constant coefficients -- in Appendix \ref{sec:analytical}, 
\begin{align} \label{oursolution}
u(x,y,t) = \frac{1}{2 \pi t \sqrt{1-\rho^2}} \exp\bigg[-\frac{1}{t (1-\rho^2)} \bigg(\rho (x_0-x)(y-y_0) + \frac{1}{2}(x-x_0)^2 + \frac{1}{2}(y-y_0)^2 \bigg)\bigg].
\end{align}
Incidentally, while the solution to \eqref{ourPDE} without the mixed term $\frac{\partial^2}{\partial x \partial y}$ and with initial condition \eqref{diracdeltas} is well-known, the general solution \eqref{oursolution} was not immediately available in the literature \cite{heatkernels, LorincziHiroshimaBetz+2011, schaumpdes}, see Mehler's formula. The results of a large number of Monte Carlo simulations and the plot of exact solution are shown in Figure~\ref{fig:realizationsanalytical}.

\subsection{\label{sec:ansatz}Choice of ansatz, expressivity, barren plateaus and initialization of parameters}
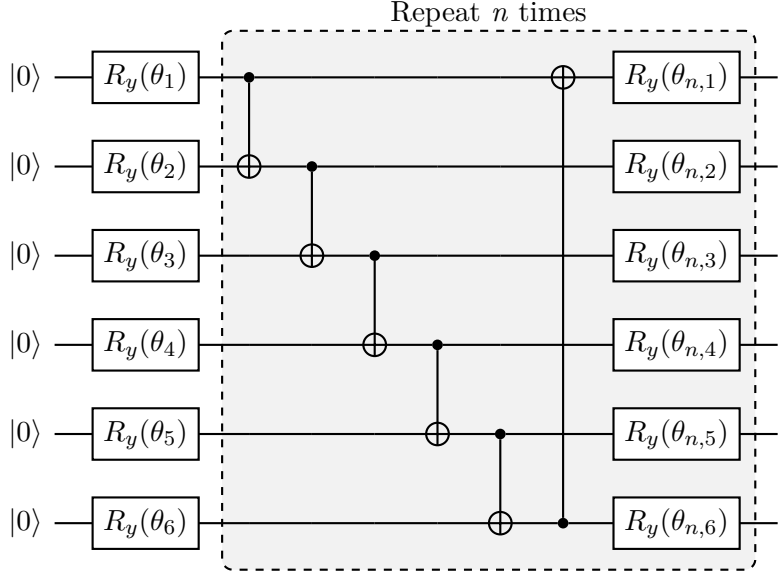
\begin{figure}[h!]
\begin{center} 
\begin{quantikz}
\lstick{$\ket{0}$} & \gate{R_y(\theta_1)} & \ctrl{1}\gategroup[wires=6, steps=7, style={dashed, rounded corners,fill=gray!10, inner xsep=2pt}, background]{Repeat \textit{n} times} & \qw & \qw & \qw &  \qw & \targ{} & \gate{R_y(\theta_{n,1})}& \qw \\ 
\lstick{$\ket{0}$} & \gate{R_y(\theta_2)} & \targ{} & \ctrl{1} & \qw & \qw & \qw & \qw & \gate{R_y(\theta_{n,2})}& \qw \\ 
\lstick{$\ket{0}$} & \gate{R_y(\theta_3)}  & \qw & \targ{} & \ctrl{1} & \qw  & \qw & \qw & \gate{R_y(\theta_{n,3})}& \qw \\ 
\lstick{$\ket{0}$} & \gate{R_y(\theta_4)} & \qw & \qw & \targ{} & \ctrl{1}  & \qw & \qw & \gate{R_y(\theta_{n,4})}& \qw \\ 
\lstick{$\ket{0}$} & \gate{R_y(\theta_5)} & \qw & \qw & \qw & \targ{}   & \ctrl{1} & \qw & \gate{R_y(\theta_{n,5})} & \qw\\ 
\lstick{$\ket{0}$} & \gate{R_y(\theta_6)} & \qw & \qw & \qw & \qw       & \targ{} & \ctrl{-5} & \gate{R_y(\theta_{n,6})} & \qw 
\end{quantikz}
\caption{Parametric quantum circuit used in the numerical experiments in this work. It is composed of a layer of $R_y$ rotation gates on every qubit, a block of nonparametric CNOT gates connecting successive qubits, and another layer of $R_y$ rotation gates on every qubit. This parametric quantum circuit is also referred to as \texttt{RealAmplitudes} ansatz due to the real-valued nature of the amplitudes in the resulting state vector.}
\label{ansatzcircuit}
\end{center}
\end{figure}

The choice of ansatz will be what is called \texttt{RealAmplitudes} from \texttt{Qiskit}
\cite{qiskitrealamplitudes} and it can be found in Figure \ref{ansatzcircuit}. 

It is a circuit that consists of six qubits with several layers of single qubit rotations as well as CNOT gates with circular entanglement. The circuit always returns real-valued amplitudes. Other ansatze are indeed possible for example using $R_x$ and $R_z$ gates. However, this comes at the cost of taking the real part of the resulting ket at the end of the evolution, see e.g. \cite{theoryvariational}. Moreover, the non-parametric controlled gates could be replaced by parameter gates (such as controlled rotations) for more generality, see for instance the ansatz proposed in \cite{fontanela2021quantum}.

In order to get reliable simulation results, we need to ensure that the model ansatz is able to represent the problem in state space.
However, we also need to consider that an ansatz which is sufficiently expressive, such that it corresponds to an (approximate) t-design  \cite{Hayashi_2005tdesigns, Ambainis07t-designs}, is prone to result in barren plateaus \cite{Clean_2018_BarrenPlateaus}. More explicitly, it is known that for an ansatz that corresponds to an (approximate) t-design the gradient operator $\bf{V}$ becomes exponentially small in the system size which in turn may cause problems in the model simulation.
It is thus vital to trade off the expressivity of a chosen ansatz with potential propagation issues.
One particularly promising approach relies on the use of additional information to design problem-specific ans\"{a}tze.

The choice of an appropriate ansatz for a given problem is still an active area of research. In our case, we are solving a PDE, \eqref{ourPDE}, that admits only real solutions, which are in fact probability distributions. Therefore, it is convenient to keep the ansatz from producing complex amplitudes effectively avoiding dealing with the imaginary part of these amplitudes. Consequently, for the problem at hand it is reasonable to use the \texttt{RealAmplitudes} ansatz put forward in this section.

The parameters $\boldsymbol\theta_0 = (\theta_1, \cdots, \theta_{12}, \theta_{13})$, where $\theta_{13}=\alpha(t)$ are initialized classically using
\begin{align} \label{minimization}
\boldsymbol\theta_0 = \operatorname{arg min}\limits_{\boldsymbol\theta \in \R^n} \{|| \ket{\psi({\boldsymbol \theta})} - \ket{\psi(0)} ||\}.
\end{align}
In \cite[$\mathsection 4$]{fontanela2021quantum}, the authors make the point that care is needed when solving \eqref{minimization} due to the fact that the functions involved might not be convex and certain optimization techniques could be trapped in local minima. Moreover, the design of the ansatz should also go hand in hand with the optimization \eqref{minimization}. Note that in general this scales poorly with the number of qubits. As with many quantum algorithms, one needs to be careful to ensure that the general state preparation techniques do not add costly overhead. Note that this initialisation step might also be performed with a quantum computer using SWAP test when the dimension of the problem starts to be too large for classical computers. 

\subsection{\label{sec:plots}VarQITE vs forward Euler vs Monte Carlo in computational basis}
Although we are dealing with the three-dimensional problem in \eqref{ourPDE}, it is instructive to plot the amplitudes in the computational basis. The parameters are set to $dx=dy=1$ as well as $dt=0.001$. The number of evolutions is $N_T=1000$. We first use six qubits in the ansatz with twelve parameterized gates, i.e. $n=1$, and $\alpha(t) = \theta_{13}(t)$ as the norm encoder variable. With a circuit of six qubits and $n=3$ layers, we have $\alpha(t)= \theta_{25}(t)$ as norm encoder. Then we use eight qubits with $n=5$ layers and $\alpha(t)= \theta_{41}(t)$ as the norm enconder variable.

\begin{figure}[h!]
	\centering
	\includegraphics[scale=0.34]{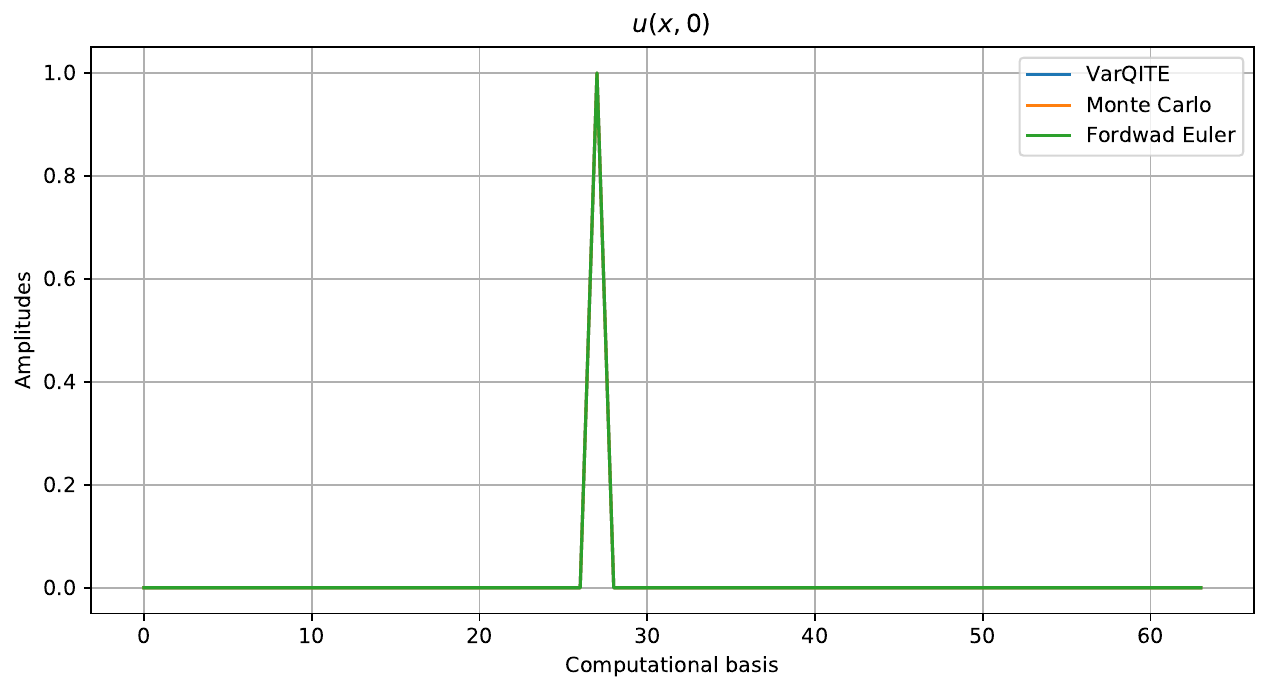}
    \includegraphics[scale=0.34]{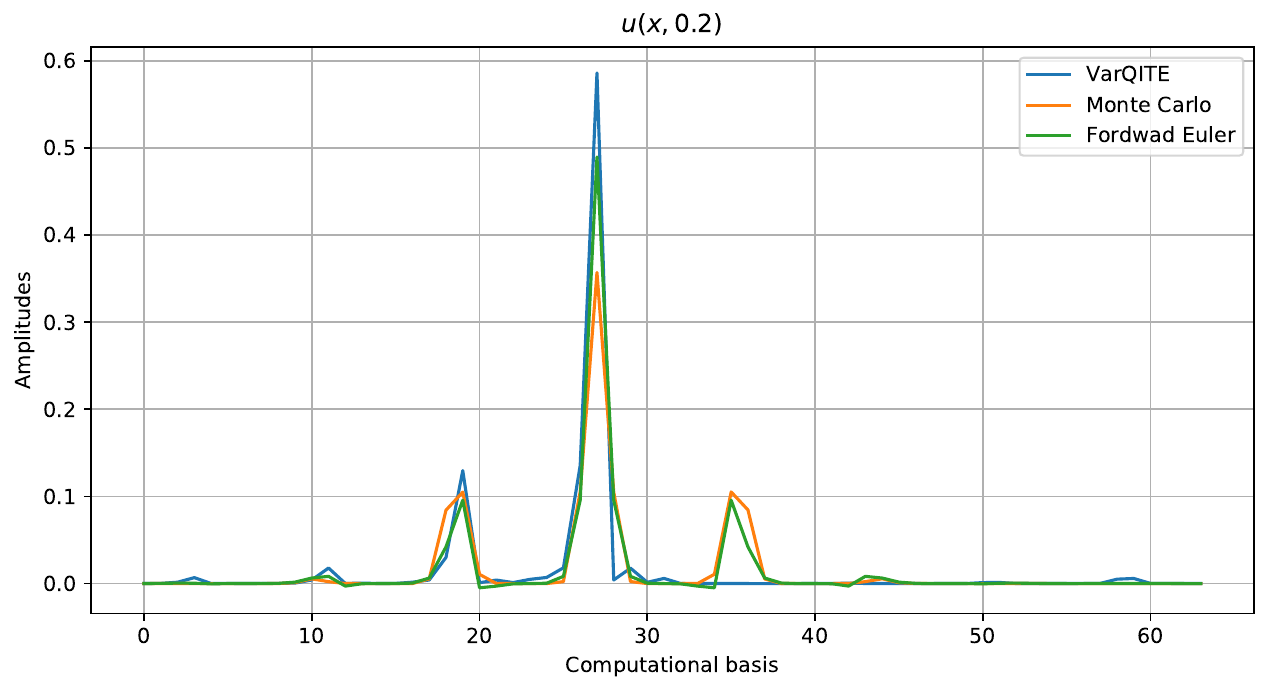}
	\centering
	\includegraphics[scale=0.34]{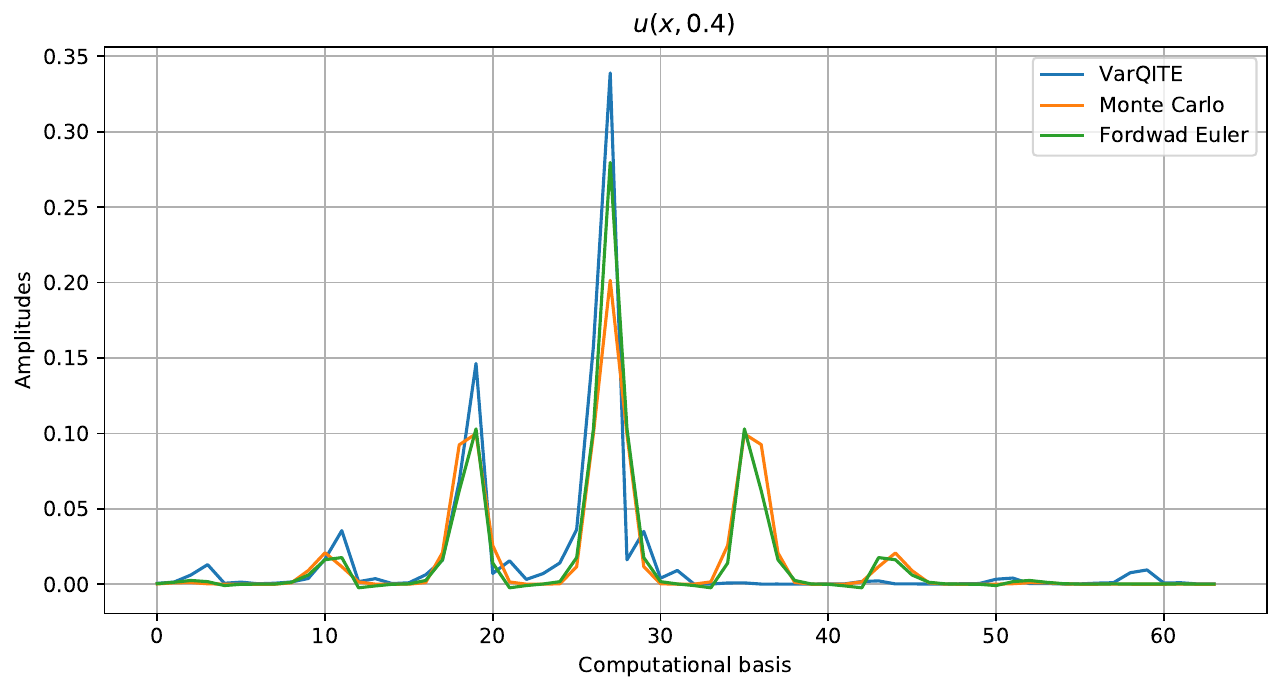}
    \includegraphics[scale=0.34]{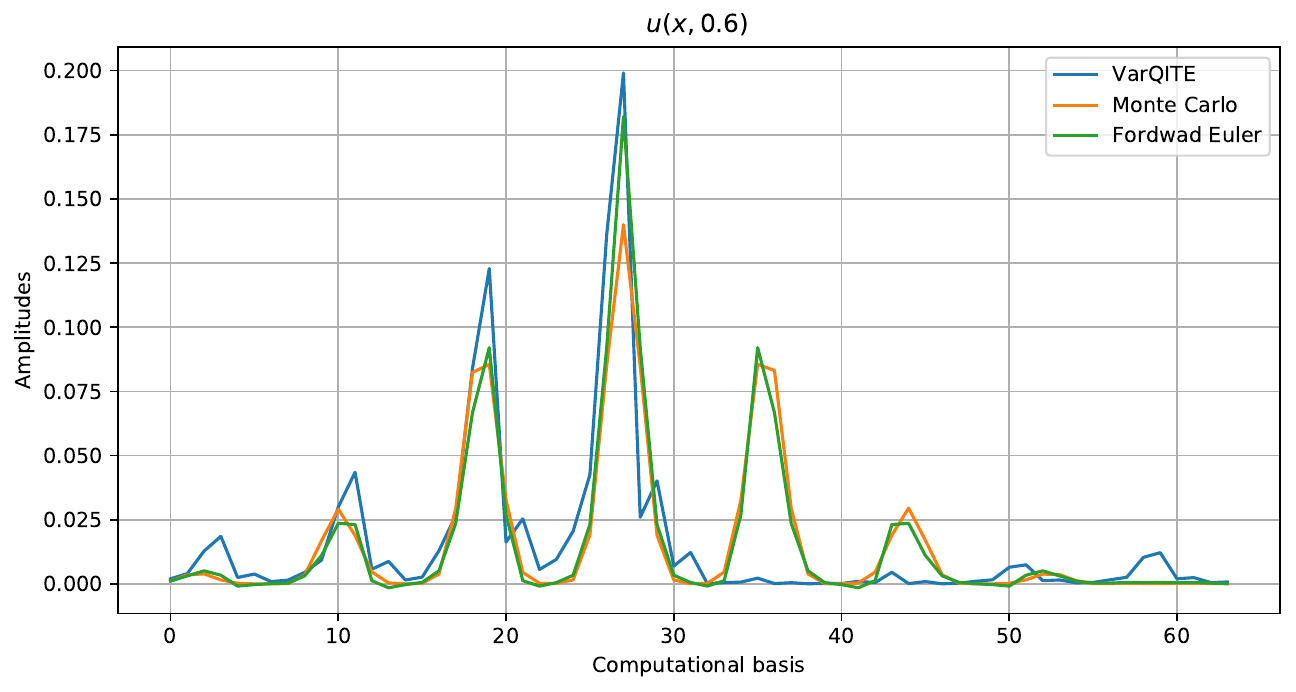}
	\centering
	\includegraphics[scale=0.34]{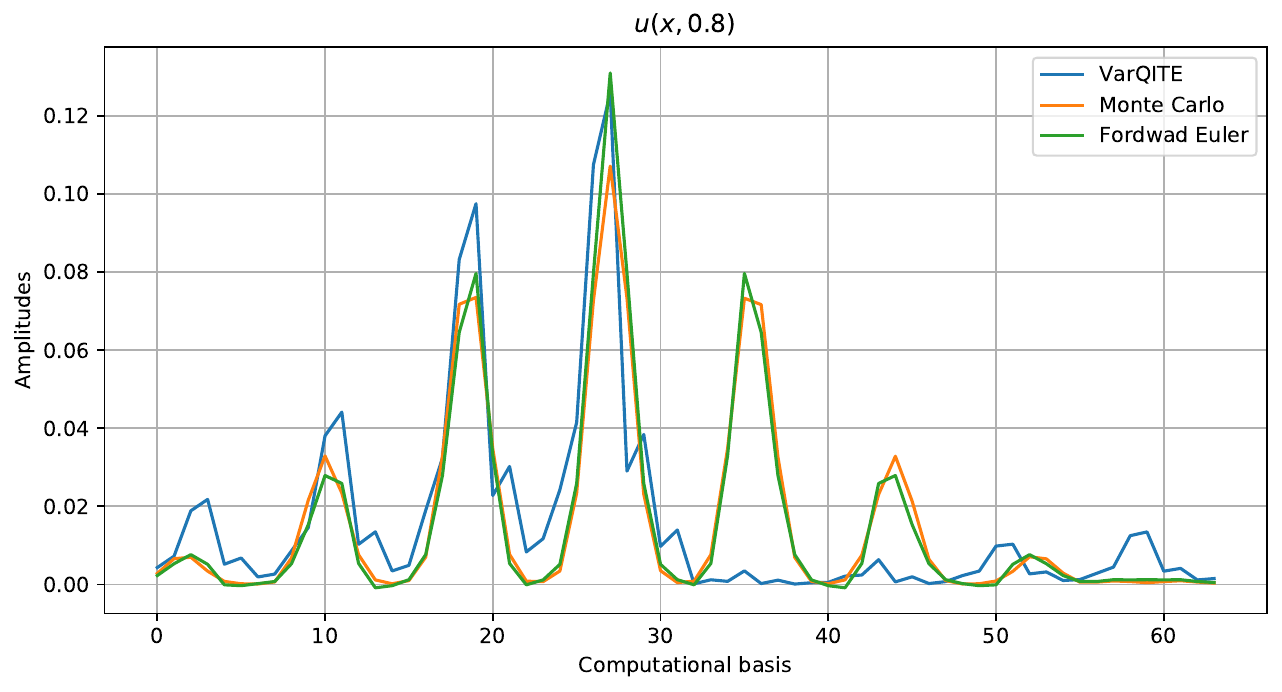}
    \includegraphics[scale=0.34]{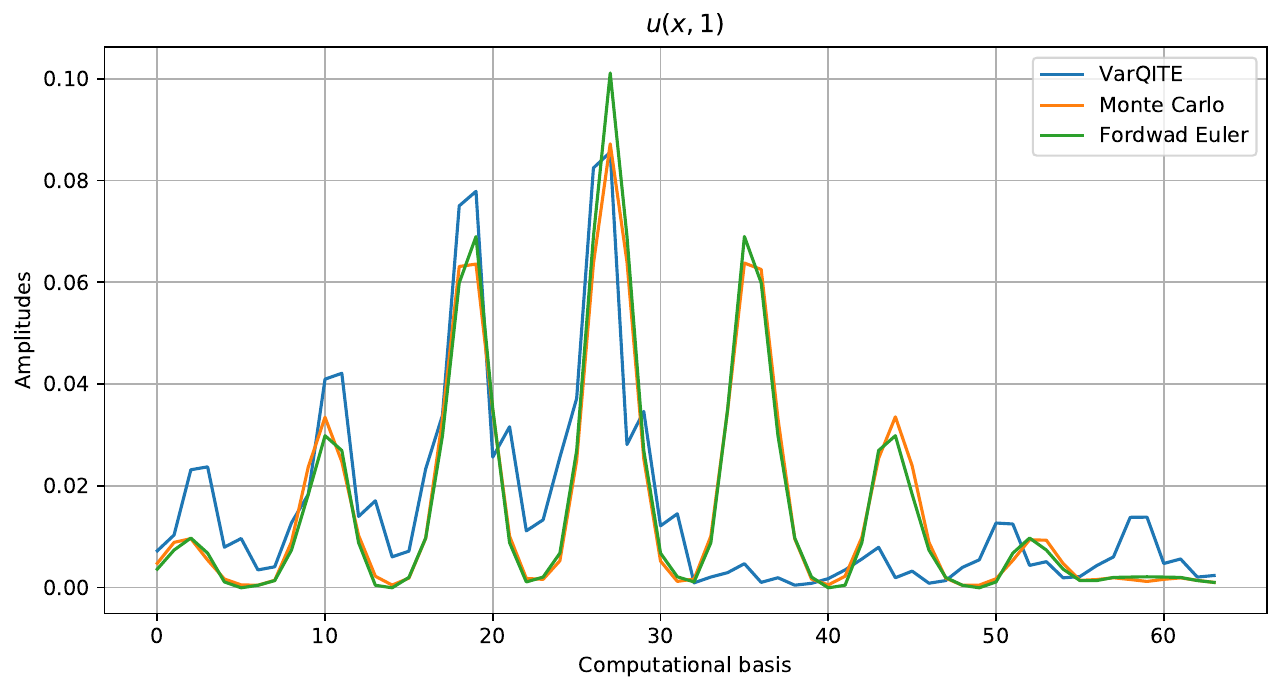}
	\caption{Plots of $u_{\operatorname{VarQITE}}(x,t), u_{\operatorname{MC}}(x,t)$ and $u_{\operatorname{FE}}(x,t)$ for $t \in \{0, 0.2, 0.4, 0.6, 0.8, 1\}$ in computational basis states with an ansatz consisting of six qubits and $n=1$ layer. The horizontal axis represents $2^6=64$ computational basis and vertical axis represents $u(x,t)$ with $dt=0.001$ and 1000 evolutions.}
	\label{ansatz6qubits1layer}
	\includegraphics[scale=0.34]{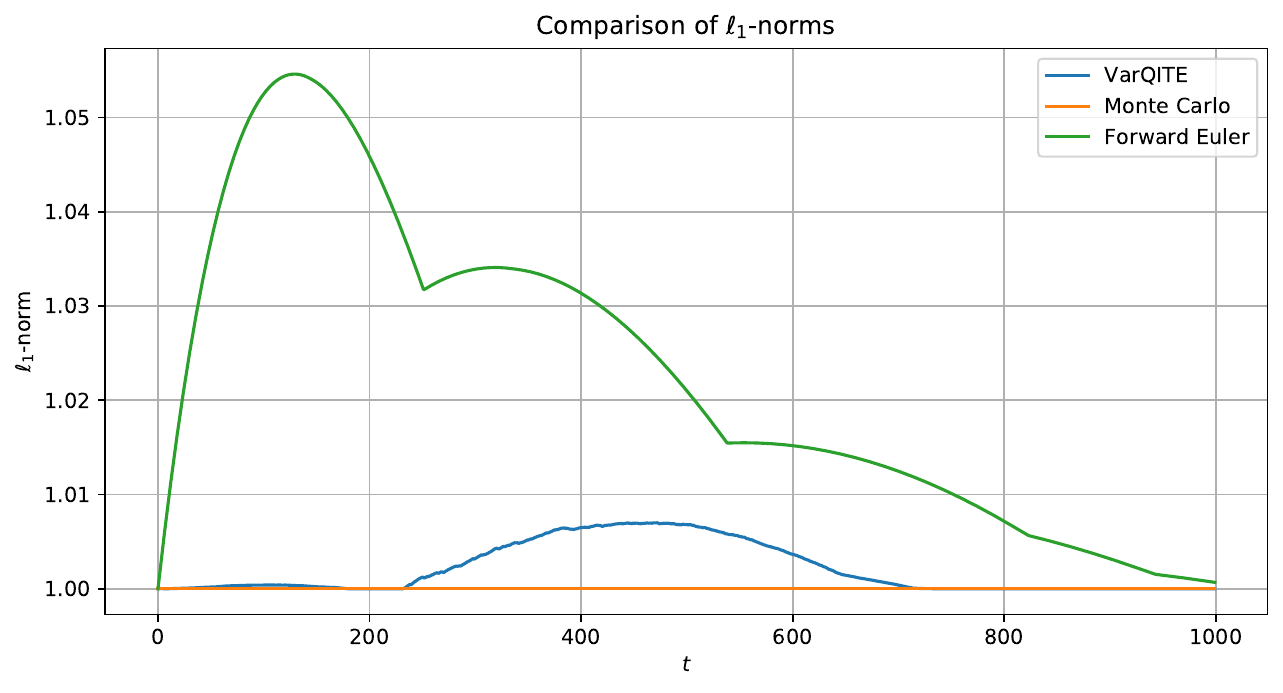}
    \includegraphics[scale=0.34]{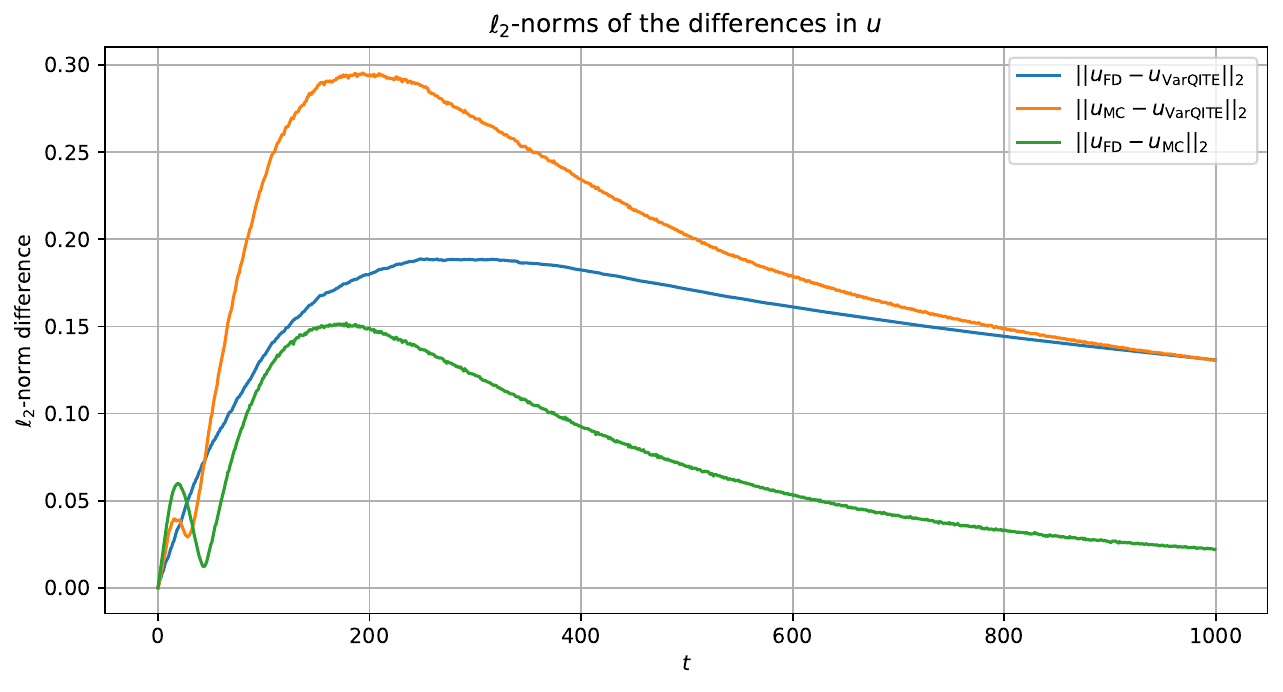}
    \caption{\underline{Left}: $\ell_1$ norms for the first 1000 evolutions are plotted using three different methods: VarQITE, Monte Carlo, and forward Euler method. Note that the scale on the $y$-axis has been narrowed down to demonstrate the approximate preservation of the $\ell_1$ norms for all three methods. The $x$-axis represents time-steps and $y$ axis represents $\ell_1$-norm.
\underline{Right}: This error plot shows the $\ell_2$ norms of the differences between the solutions $u$ obtained via VarQITE and forward Euler (blue) and between VarQITE and Monte Carlo method (yellow). The $x$-axis represents time and $y$ axis represents $\ell_2$-norm of the differences in $u$.
}
\label{ansatz6qubits1layernorms}
\end{figure}

\begin{figure}[h!]
	\centering
	\includegraphics[scale=0.34]{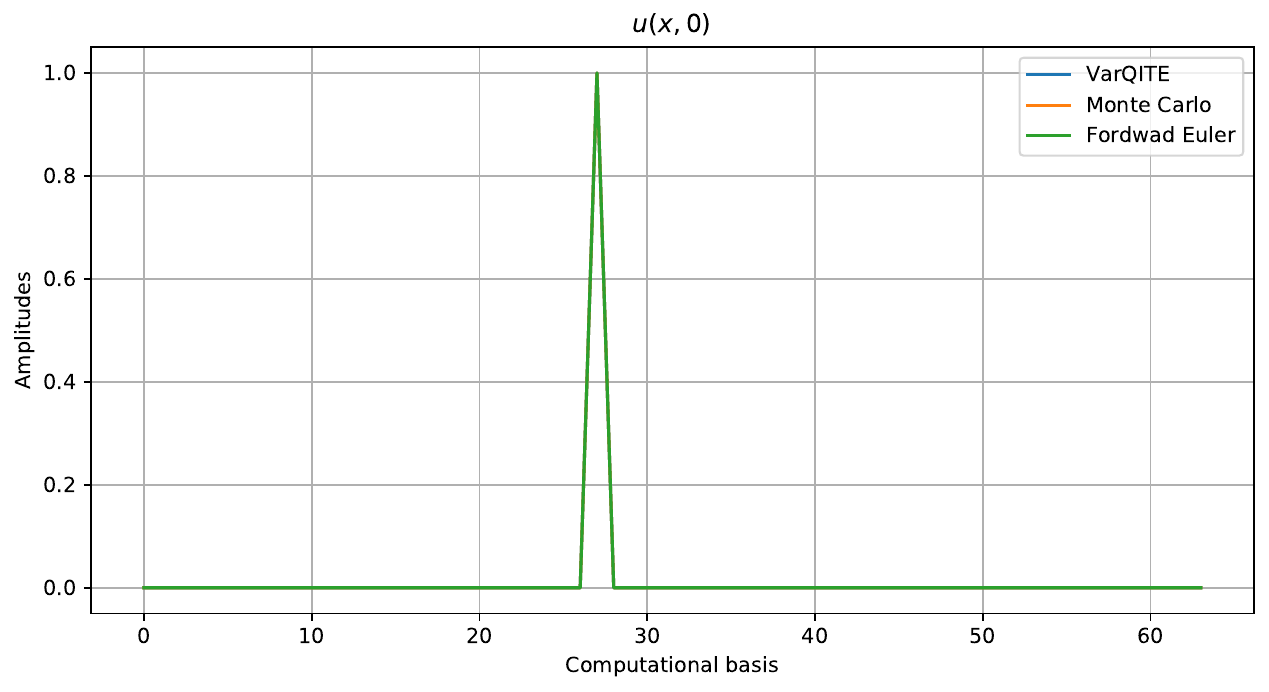}
    \includegraphics[scale=0.34]{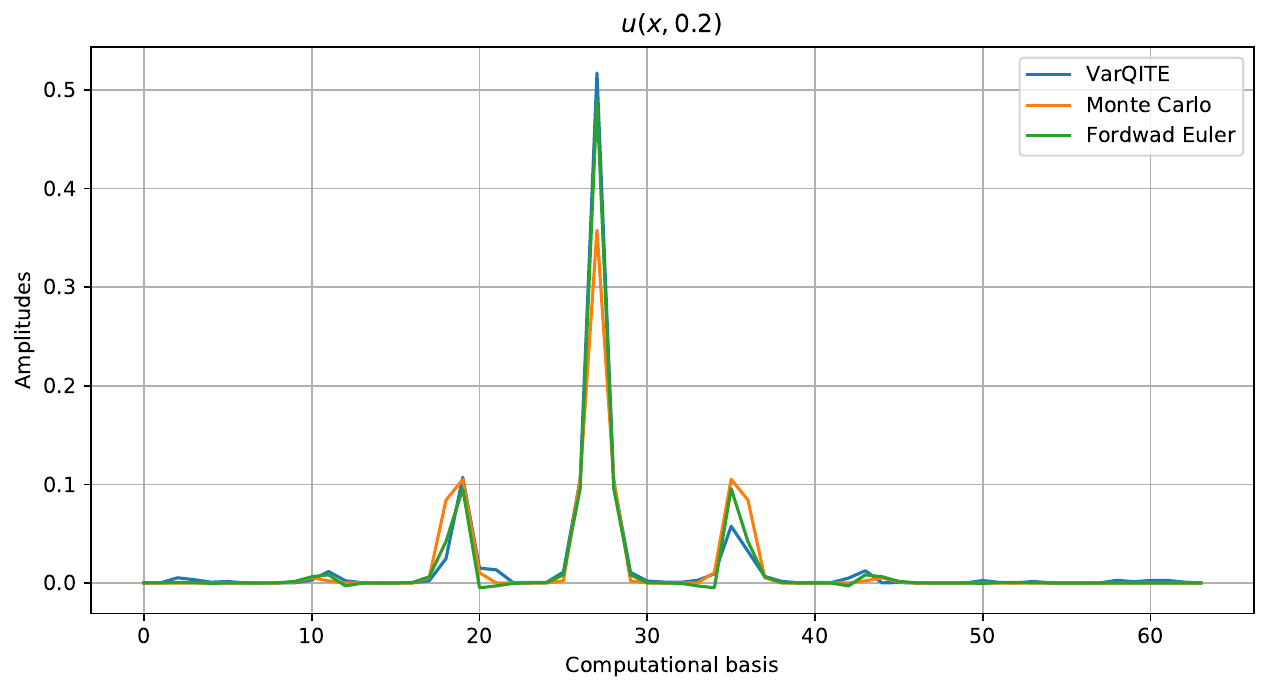}
	\centering
	\includegraphics[scale=0.34]{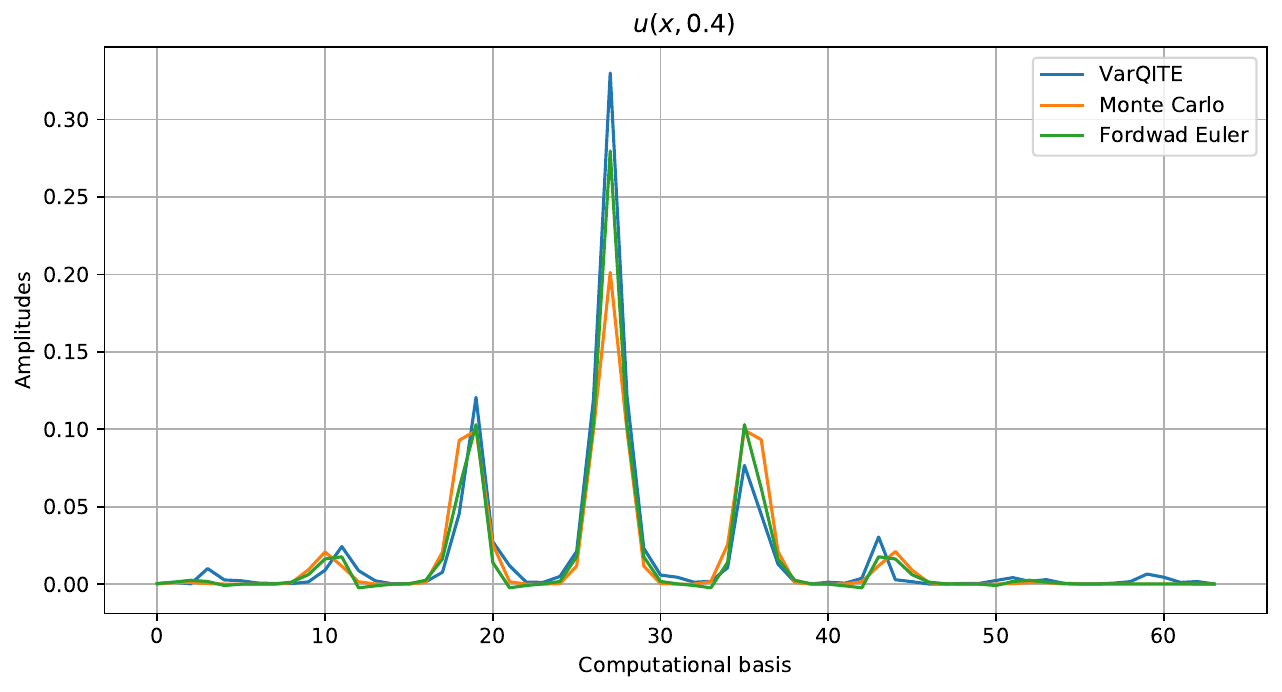}
    \includegraphics[scale=0.34]{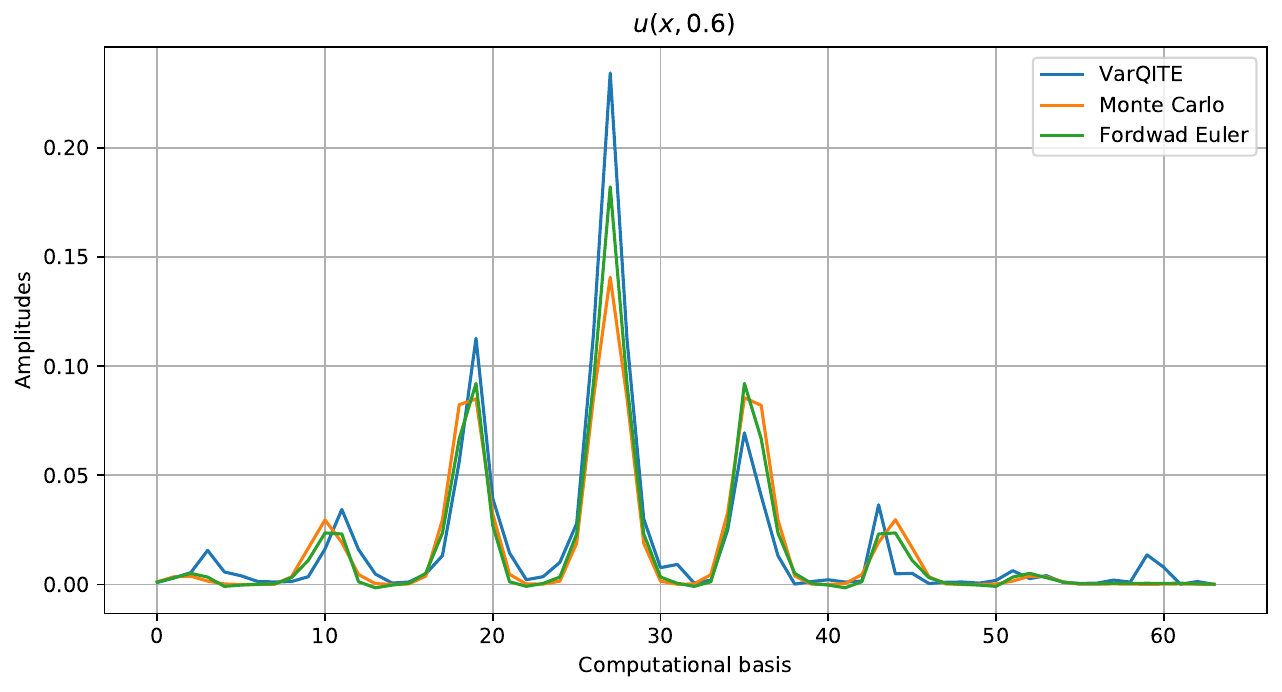}
	\centering
	\includegraphics[scale=0.34]{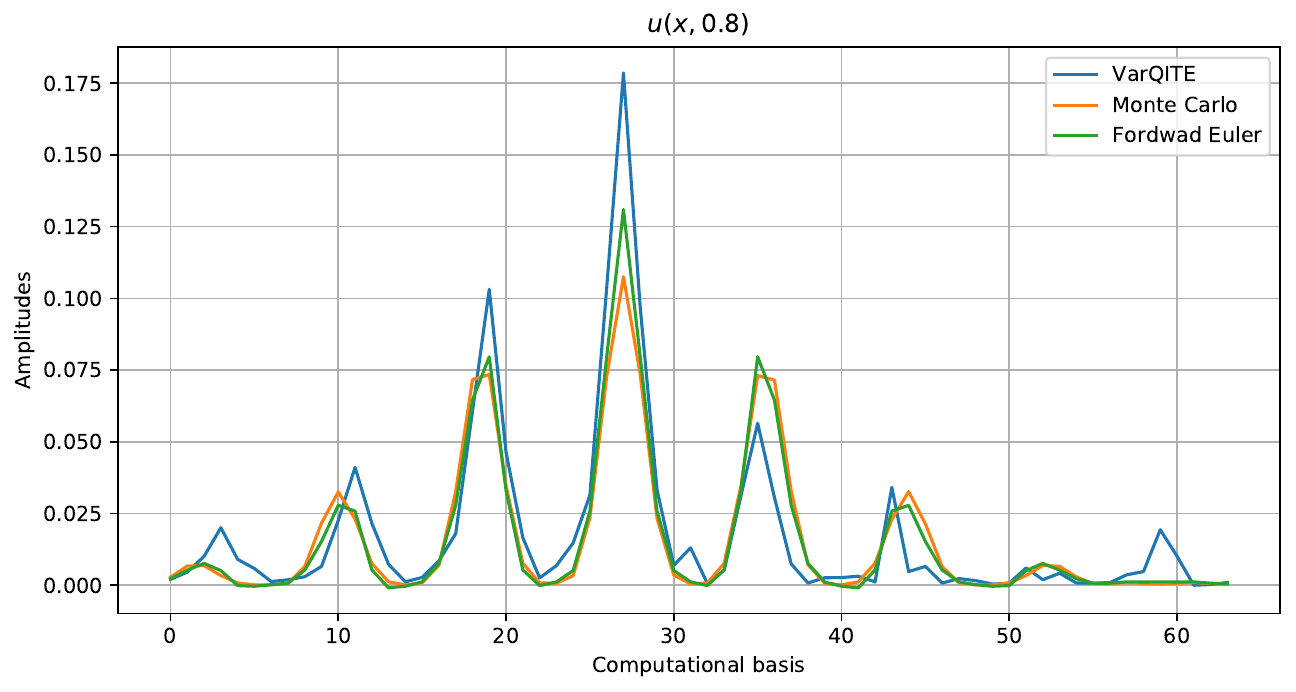}
    \includegraphics[scale=0.34]{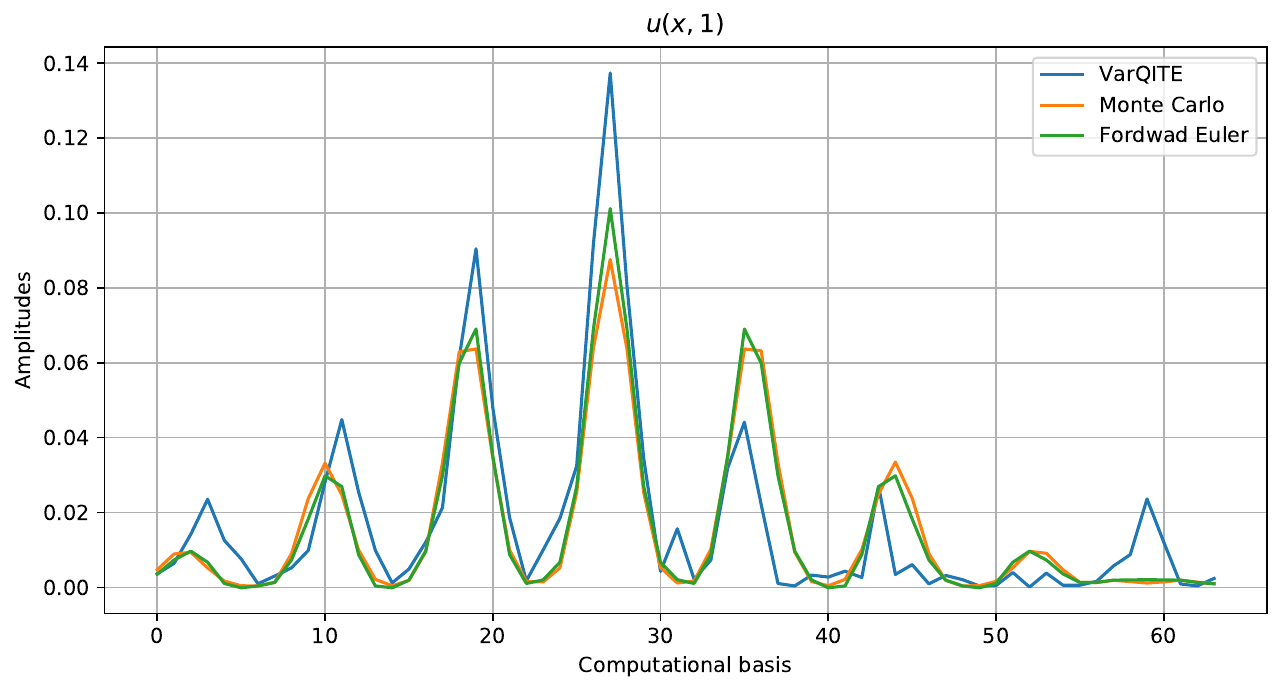}
	\caption{Plots of $u_{\operatorname{VarQITE}}(x,t), u_{\operatorname{MC}}(x,t)$ and $u_{\operatorname{FE}}(x,t)$ for $t \in \{0, 0.2, 0.4, 0.6, 0.8, 1\}$ in computational basis states with an ansatz consisting of six qubits and $n=3$ layers. The horizontal axis represents $2^6=64$ computational basis and vertical axis represents $u(x,t)$ with $dt=0.001$ and 1000 evolutions.}
	\label{ansatz6qubits3layers}
	\includegraphics[scale=0.34]{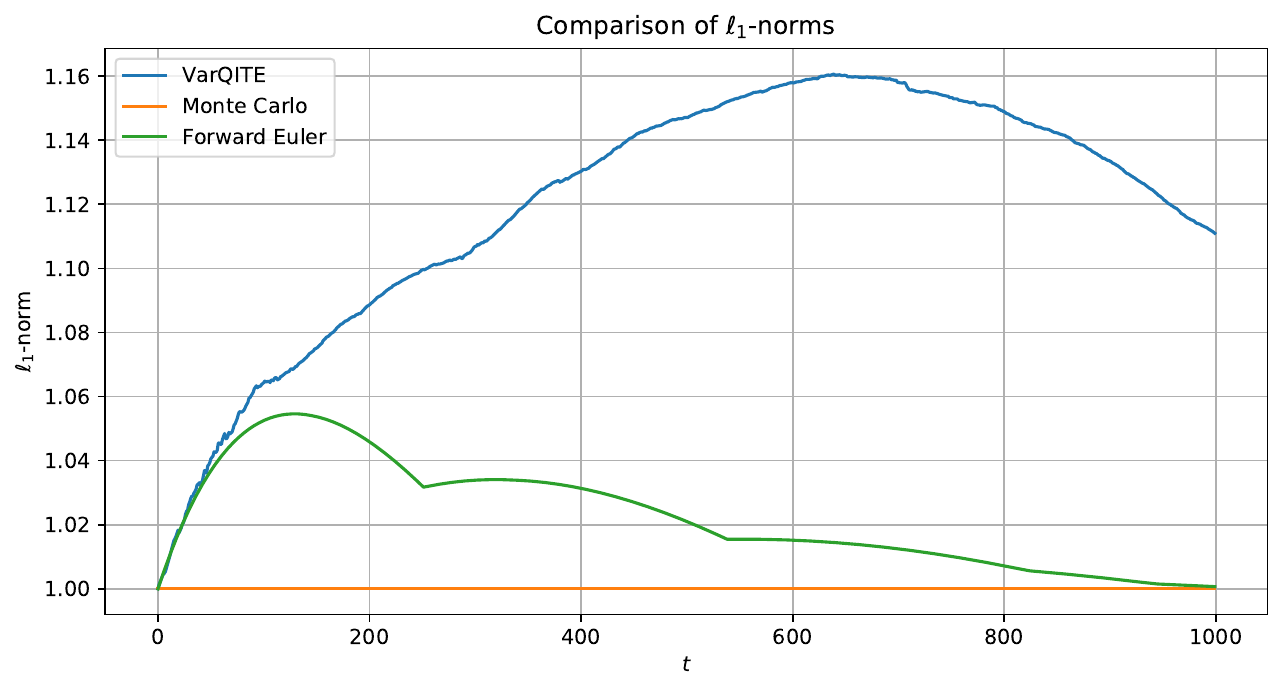}
    \includegraphics[scale=0.34]{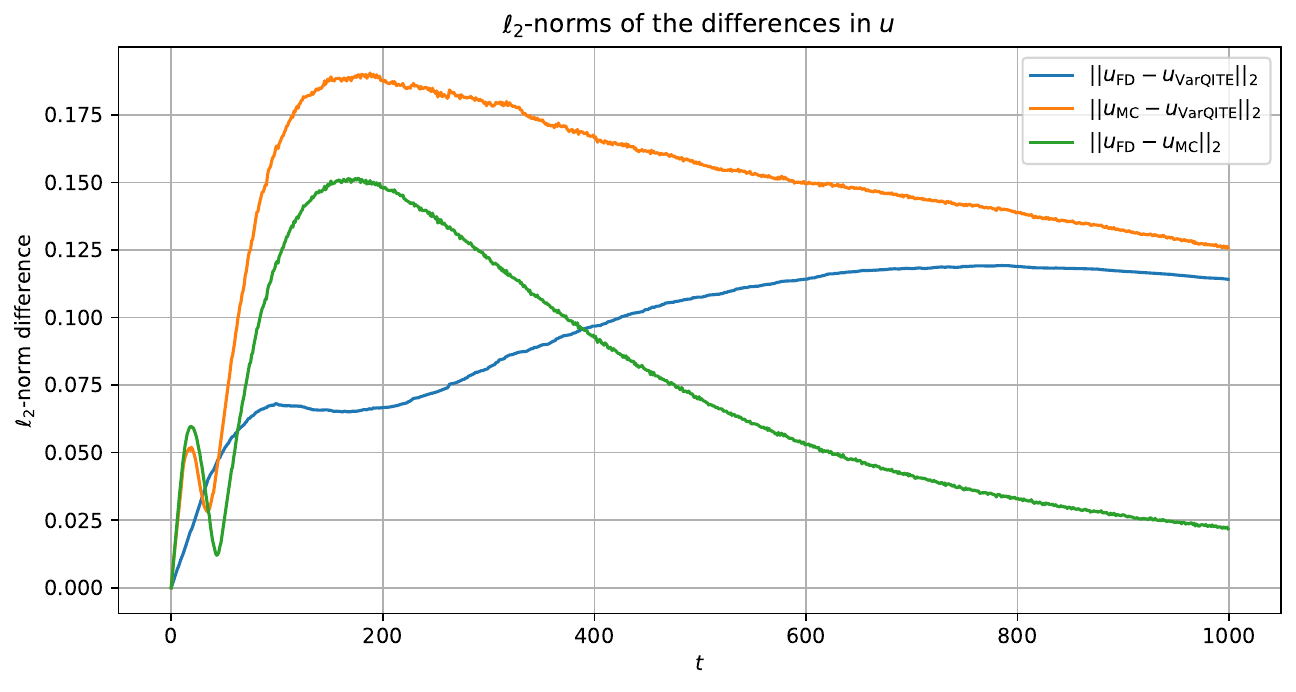}
    \caption{\underline{Left}: $\ell_1$ norms for the first 1000 evolutions are plotted using three different methods: VarQITE, Monte Carlo, and forward Euler method. Note that the scale on the $y$-axis has been narrowed down to demonstrate the approximate preservation of the $\ell_1$ norms for all three methods. The $x$-axis represents time-steps and $y$ axis represents $\ell_1$-norm.
\underline{Right}: This error plot shows the $\ell_2$ norms of the differences between the solutions $u$ obtained via VarQITE and forward Euler (blue) and between VarQITE and Monte Carlo method (yellow). The $x$-axis represents time and $y$ axis represents $\ell_2$-norm of the differences in $u$.
}
\label{ansatz6qubits3layersnorms}
\end{figure}

\begin{figure}[h!]
	\centering
	\includegraphics[scale=0.34]{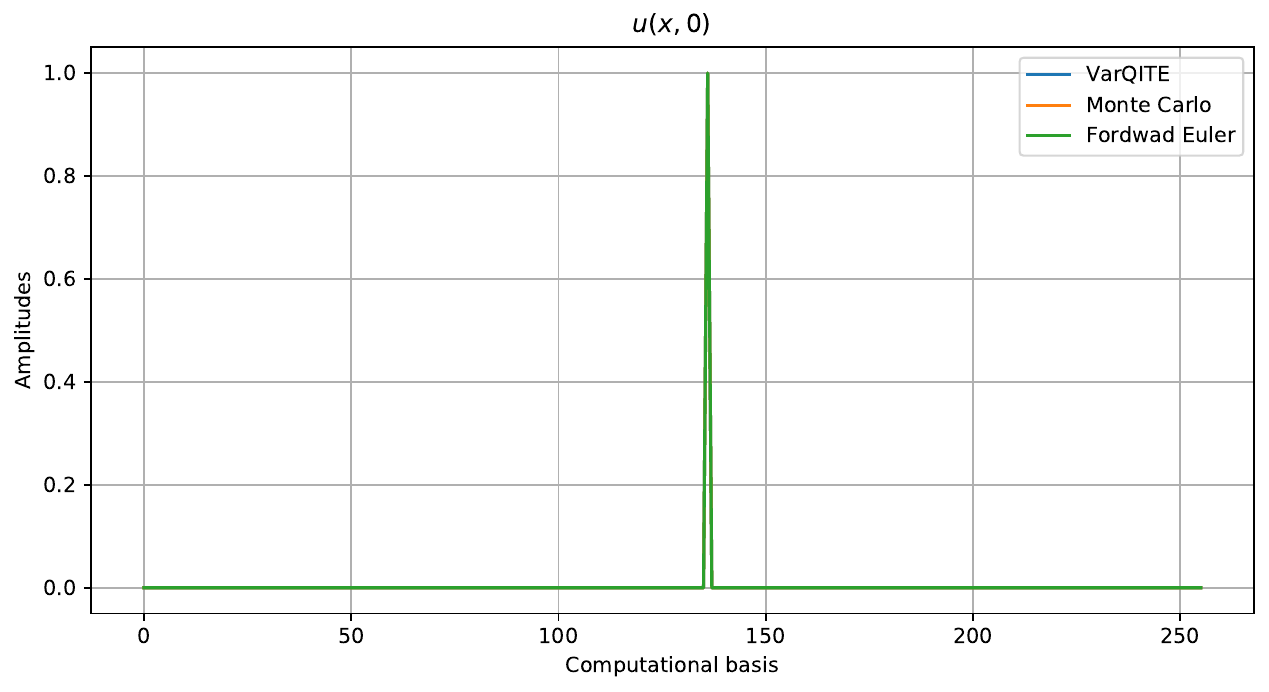}
    \includegraphics[scale=0.34]{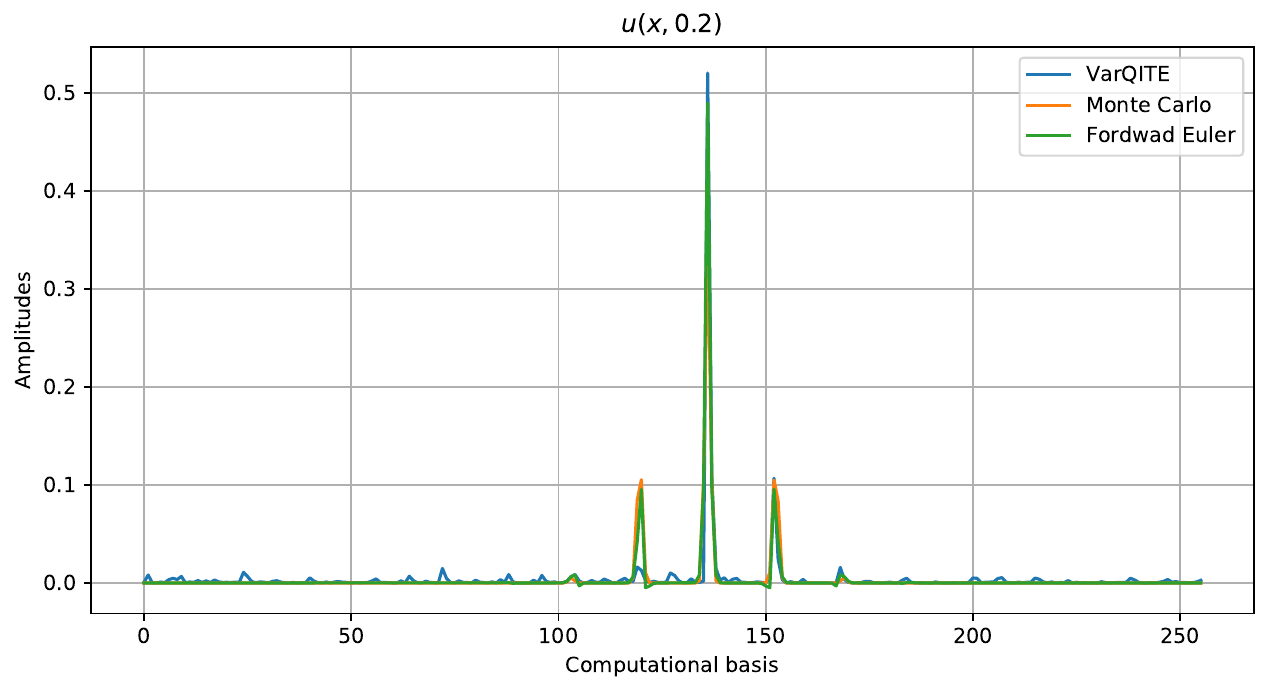}
	\centering
	\includegraphics[scale=0.34]{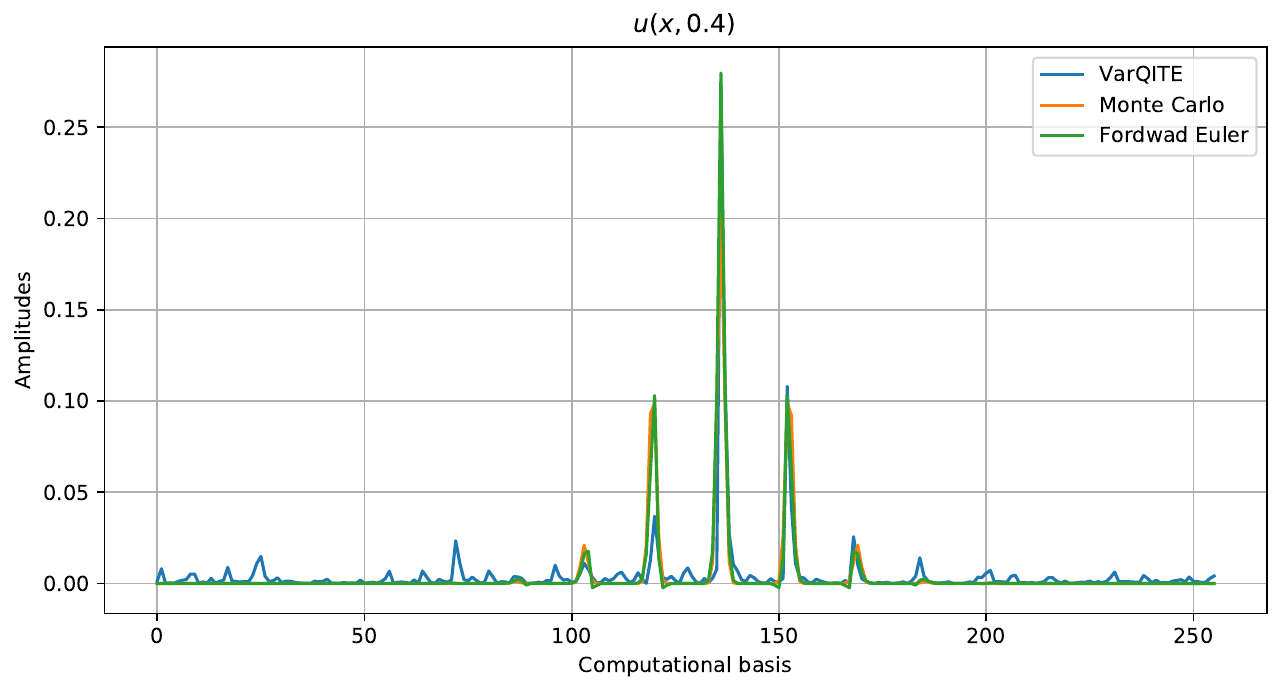}
    \includegraphics[scale=0.34]{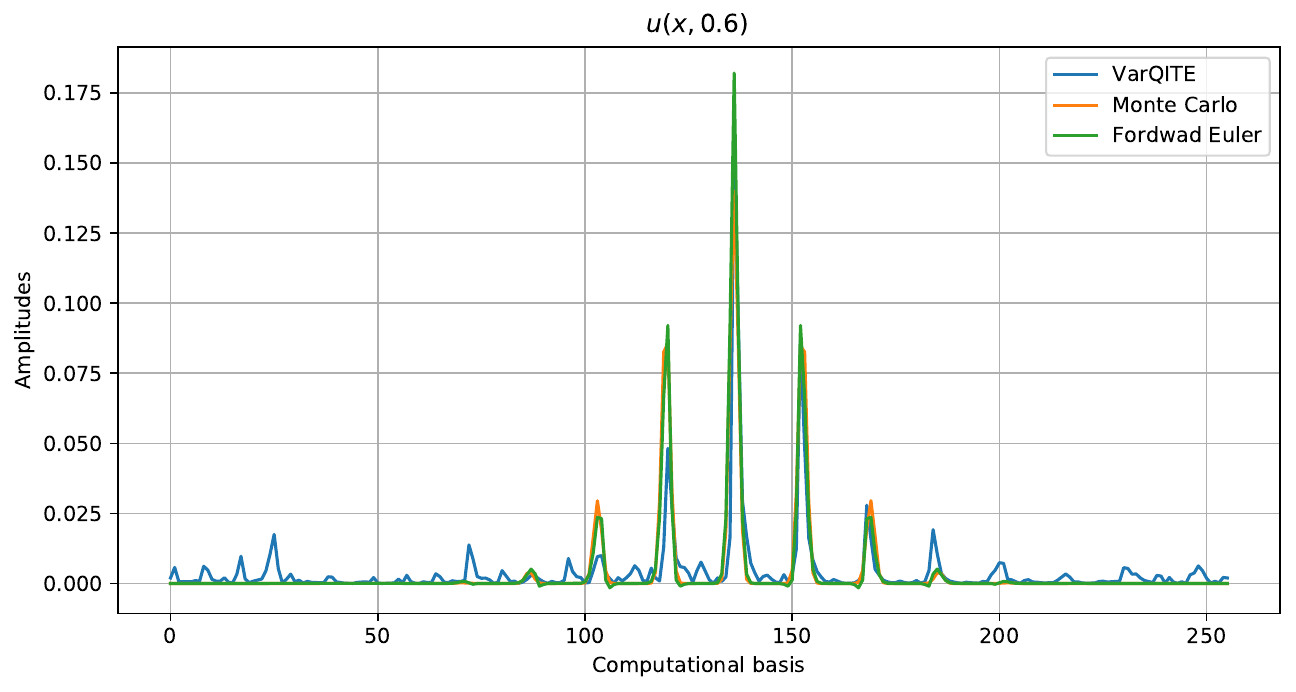}
	\centering
	\includegraphics[scale=0.34]{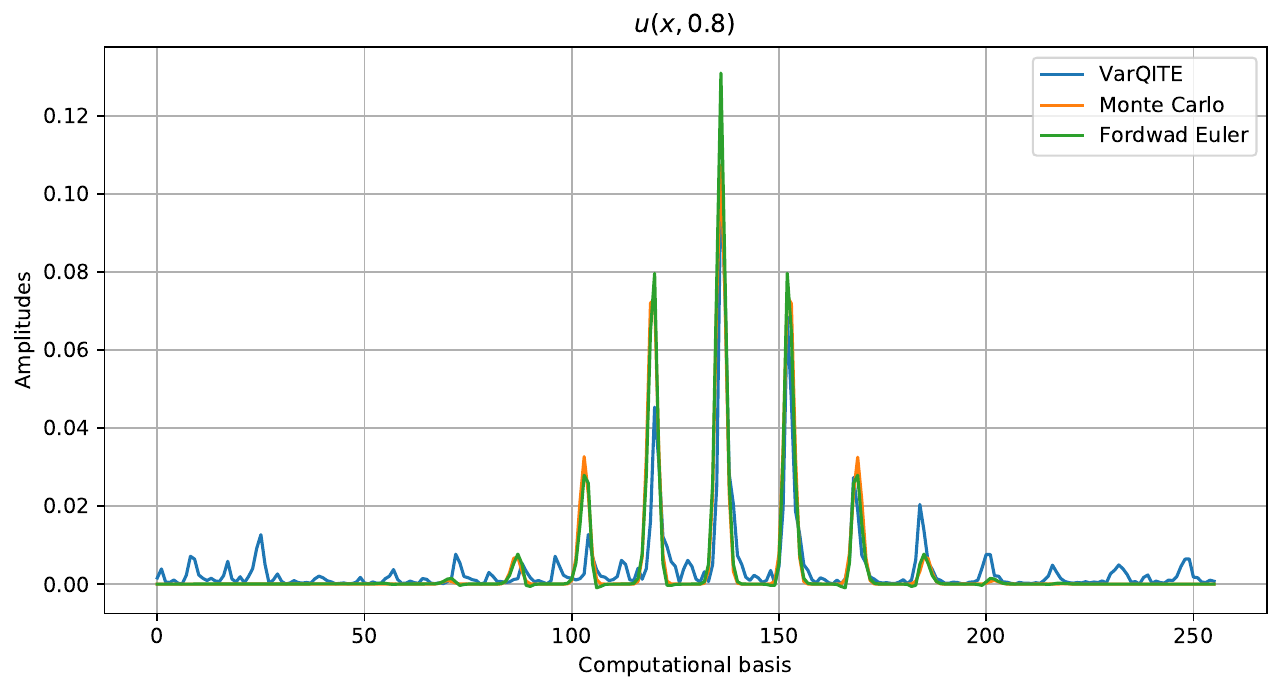}
    \includegraphics[scale=0.34]{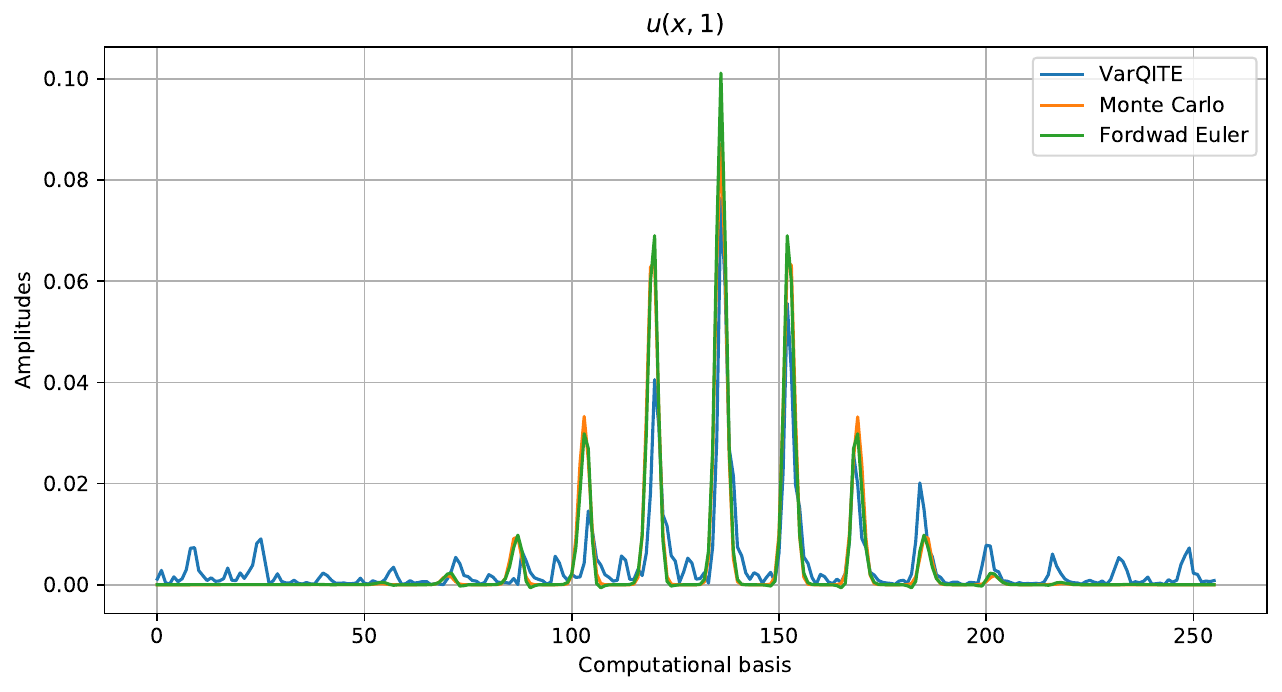}
	\caption{Plots of $u_{\operatorname{VarQITE}}(x,t), u_{\operatorname{MC}}(x,t)$ and $u_{\operatorname{FE}}(x,t)$ for $t \in \{0, 0.2, 0.4, 0.6, 0.8, 1\}$ in computational basis states with an ansatz consisting of eight qubits and $n=5$ layers. The horizontal axis represents $2^8=256$ computational basis and vertical axis represents $u(x,t)$ with $dt=0.001$ and 1000 evolutions.}
	\label{ansatz8qubits5layers}
    \includegraphics[scale=0.34]{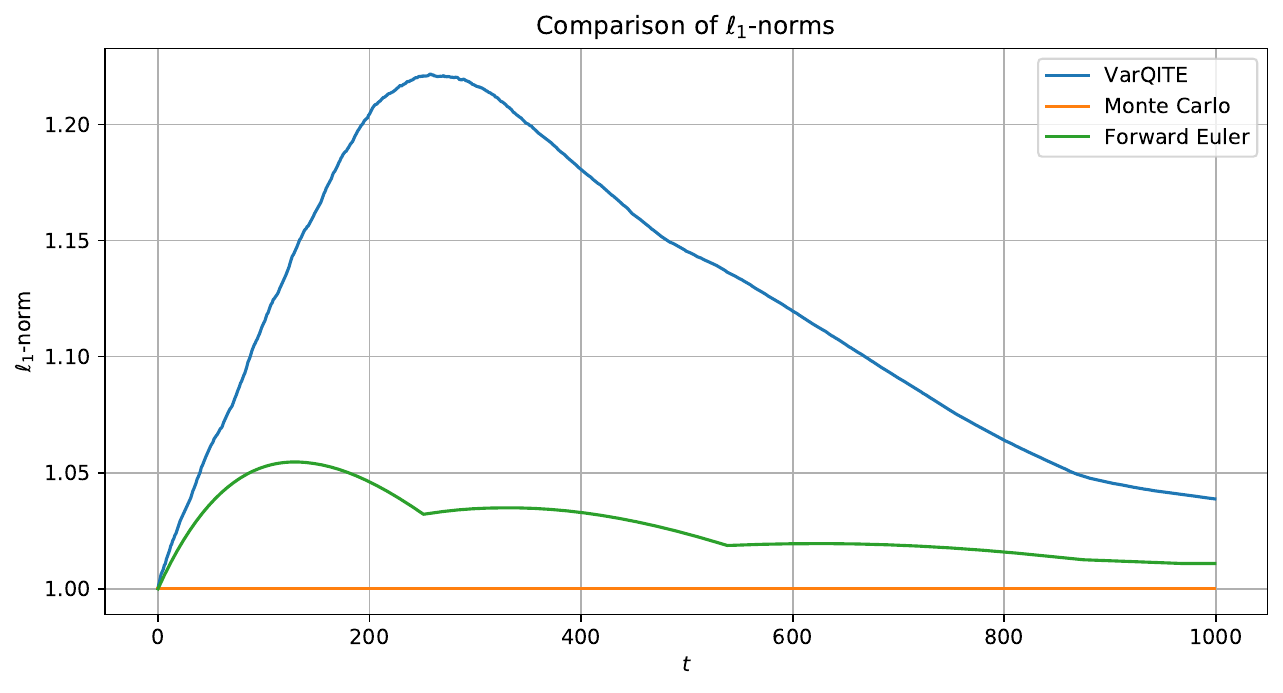}
    \includegraphics[scale=0.34]{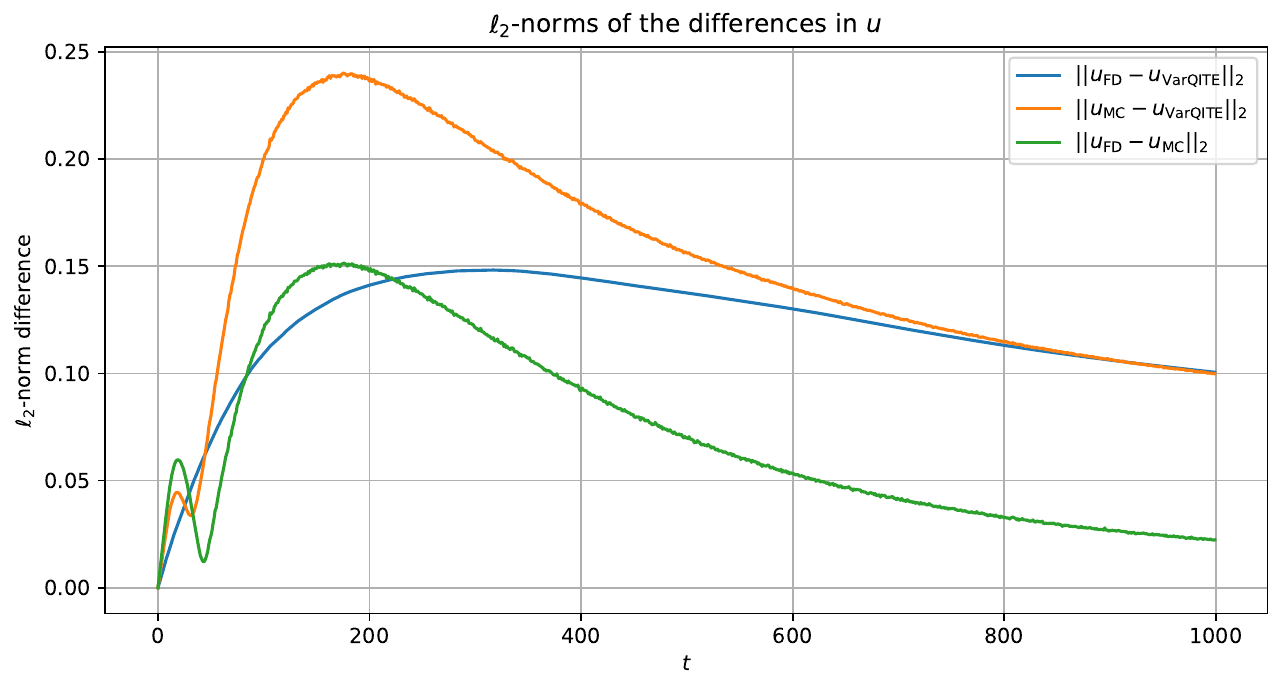}
\caption{\underline{Left}: $\ell_1$ norms for the first 1000 evolutions are plotted using three different methods: VarQITE, Monte Carlo, and forward Euler method. Note that the scale on the $y$-axis has been narrowed down to demonstrate the approximate preservation of the $\ell_1$ norms for all three methods. The $x$-axis represents time-steps and $y$ axis represents $\ell_1$-norm.
\underline{Right}: This error plot shows the $\ell_2$ norms of the differences between the solutions $u$ obtained via VarQITE and forward Euler (blue) and between VarQITE and Monte Carlo method (yellow). The $x$-axis represents time and $y$ axis represents $\ell_2$-norm of the differences in $u$.
}
\label{ansatz8qubits5layersnorms}
\end{figure}

The $\ell_1$ norm enforcement of the evolution discussed in \ref{sec:enforce} can be seen in these plots of 1000 evolutions on the bottom left of Figures \ref{ansatz6qubits1layernorms}, \ref{ansatz6qubits3layersnorms} and \ref{ansatz8qubits5layersnorms}. Earlier in the evolution period, the first kets came out with a flipped sign due to a phase. Moreover, periodic boundary conditions and the nature of the forward Euler method also yielded a norm slightly larger than one for the classical methods. Small deviations in the $\ell_1$-norm from VarQITE are coming from the approximate nature of the $\ell_1$-norm enforcement.
Another useful metric to keep in mind is the difference between the $\ell_2$ norms of forward Euler and VarQITE as well as Monte Carlo and VarQITE as shown on the bottom right of Figures \ref{ansatz6qubits1layernorms}, \ref{ansatz6qubits3layersnorms} and \ref{ansatz8qubits5layersnorms}. 

We note the following analysis of the experiments:
\begin{itemize}
    \item[(1)] One can see in Figure \ref{ansatz6qubits1layer} that VarQITE was significantly off around the computational basis states 32 to 38 with six qubits and only $n=1$ layer. This situation is due to the choice of ansatz and its associated expressivity.
    \item[(2)] To see the effect of increasing the expressivity of the ansatz, we added more layers by increasing $n$ from 1 to 3 while keeping the number of qubits constant at six. The results are plotted on Figure \ref{ansatz6qubits3layers}. 
    Lastly, we then chose an eight qubit ansatz and added two additional layers of CNOTs and $R_y$ rotations, i.e. $n=5$. The results are in Figure \ref{ansatz8qubits5layers}.
    \item[(3)] In both instances (2) and (3), the match of the plots of VarQITE and forward Euler no longer show the discrepancy in the computational bits 32 to 38 for six qubits (respectively, 170 to 180 for eight qubits) while still maintaining a tight $\ell_1$ normalization.
    \item[(4)] To perform the simulations with additional layers and additional qubits, numerical differentiation needed to be performed which added some instability. We expect these results to match more tightly with symbolic or circuit-based differentiation and this will be studied in future research.
    \item[(5)] We do not include finite-sampling noise or hardware noise. All numerical simulations have been tested and produced using \texttt{Python} as well as the \texttt{NumPy} library. The effect of noise on the results is an interesting topic and will be left for future research.
    \item[(6)] Although this methodology works with a high degree of generality, for high dimensional cases, the expressivity of the ansatz employed in VarQITE will limit the accuracy of the results.
\end{itemize}

\section{\label{sec:conclusion}Conclusion}
We have proposed a quantum algorithm to transition from stochastic differential equations involving Wiener processes $X_1, X_2, \cdots, X_D$ to partial differential equations $\tfrac{\partial u}{\partial t} = \mathcal{G} u$ where $u$ is the expectation of the Wiener processes. The resulting PDEs have been simulated through variational quantum imaginary time evolution (VarQITE). We have seen that well-known PDEs such as Schr\"{o}dinger, heat and Fokker-Planck equations can be recast through this type of methodology and thereby can be unified in the single framework of the Feynman-Kac formula. Moreover, we have maintained a high degree of generality by allowing arbitrarily many Brownian motions per stochastic process and high dimensionality. However when dealing with dimensional cases, the expressivity of the ansatz will limit the accuracy of the results. Therefore, as the dimension of the problem increases one has to improve or enhance the ansatz accordingly. Some of the resulting PDEs do not preserve $\ell_2$ norm, such as the heat equation which preserves probability distributions. This represents a non-trivial obstacle and we have utilized novel ideas to maintain the solution approximately normalized throughout the quantum evolution by introducing a proxy to the $\ell_1$ norm.

The algorithm has shown substantial agreement in the solution of the PDE through VarQITE, classical Monte Carlo and forward Euler in a sufficiently complex and non-trivial problem involving an anisotropic Fokker-Planck equation in $(2+1)$ dimensions. Because obtaining the full solution to the PDE is an expensive problem, we need to instead address obtaining the multidimensional moments of the solution through quantum techniques. In order to do so, we have decomposed the second order differential operator $\mathcal{G}$ from the PDE into a feasible set of unitaries composed of few-qubit gates. For a suitable function $f$, the moments $\mathbb{E}[f(X_1, X_2, \cdots, X_D)]$ can be computed by approximating $f$ with its multidimensional Taylor series to a fixed order and this approximation can be improved by increasing the number of qubits to allow for finer resolution. These moments can be computed through QPE or Hadamard tests with potential advantages depending on the concrete situation.

The methodology provided here supplements the existing quantum-enhanced Monte Carlo algorithms and both techniques can coexist and complement each other. The scope of the topics we have addressed and their mapping to quantum computers allows us to open the possibilities for other interesting items for future work such as option pricing with stochastic volatility, see \cite{wilmott}, as well as stochastic control problems such as American and decision-embedded options which require the Hamilton-Jacobi-Bellman equation, see \cite{vecer}. These topics might have to be explored in the context of real, as opposed to imaginary, time evolution. Lastly, the heat equation is not the only candidate for this algorithm. It\^{o}'s lemma can be used \cite[$\mathsection$4.1.1]{nolen} to transform a conditional expectation arising from a vectorized SDE into the solution of the PDE. Another venue would be to consider the wave or Klein-Gordon equations (or in general, other type of hyperbolic differential equations). In this case, the time derivative is now of second order, unlike in the heat equation, which was of first order. For these situations, other types of probability distributions and Brownian motions need to be added, specially by considering jumps and Poisson processes, see \cite{dalang}.

\section{Acknowledgments}
The authors would like to thank Peter Tysowski and Vaibhaw Kumar for constructive discussions regarding these topics. Antoine Jacquier was kind enough to read the manuscript and provide some corrections and suggestions. Oleksandr Kyriienko also provided insightful remarks.
CZ acknowledges support from the National Centre of Competence in Research \emph{Quantum Science and Technology} (QSIT). The authors would like to acknowledge the referees for the very helpful comments that have greatly increased the accuracy, precision, and readability of the manuscript.\\

IBM, the IBM logo, and ibm.com are trademarks of International Business Machines Corp., registered in many jurisdictions worldwide. Other product and service names might be trademarks of IBM or other companies. The current list of IBM trademarks is available at \url{https://www.ibm.com/legal/copytrade}.

\section{Code availability}
The code that supports the findings of this study is available from the corresponding author NR upon reasonable request.

\bibliographystyle{arxiv_no_month}
\bibliography{May30QuantumTemplate}

\appendix

\section{\label{sec:details}Details on Feynman-Kac, It\^{o}'s lemma and the heat equation}
Suppose that $u=u(x,t)$ with $x \in \R$ and $0 \le t \le T$ is a solution to the `backward' heat equation:
\begin{align} \label{heateqA}
\begin{dcases}
\frac{\partial u}{\partial t} + \frac{1}{2}\frac{\partial^2 u}{\partial x^2} =0, \quad \textnormal{for} \quad t<T  \\
~  \
u(x,T) = \psi(x).
\end{dcases}
\end{align}
Now let $(W_t)_{t \ge 0}$ be Brownian motion and consider $u(W_t,t)$. Performing a small variation yields
\begin{align} \label{variationu}
u(W_t+dW_t,t+d t) = u(W_t,t) + \frac{\partial u}{\partial t}dt + \frac{\partial u}{\partial x}dW_t + \frac{1}{2}\frac{\partial^2 u}{\partial x^2}(dW_t)^2 + O(dt^2, dW_t^3).
\end{align}
From It\^{o}'s lemma \eqref{ito} we have the rule of thumb
\begin{align}
(dW_t)^k = 
\begin{cases}
dW_t, \quad &\mbox{if $k=1$}, \nonumber \\
dt, \quad &\mbox{if $k=2$}, \nonumber \\
0, \quad &\mbox{if $k>2$}.
\end{cases}
\end{align}
Moreover, $(dt)^k=0$ for $k \ge 2$. Thus, we will ignore the last terms in \eqref{variationu}. Passing the term $u(W_t,t)$ to the left-hand side we see that \eqref{variationu} becomes
\begin{align} 
u(W_t+dW_t,t+dt) - u(W_t,t) &= du(W_t,t) \nonumber \\
&= \frac{\partial u}{\partial t}dt + \frac{\partial u}{\partial x}dW_t + \frac{1}{2}\frac{\partial^2 u}{\partial x^2}(dW_t)^2 \nonumber \\
&= \bigg(\frac{\partial u}{\partial t} + \frac{1}{2}\frac{\partial^2 u}{\partial x^2}\bigg)dt  + \frac{\partial u}{\partial x}dW_t. \nonumber
\end{align}
By \eqref{heateqA} we see that the term involving $dt$ is zero, and therefore we are left with $du(W_t,t) = \frac{\partial u}{\partial w}dW_t$.
Integrate this from $t$ to $T$ so that
\begin{align}
\int_t^T du(W_s,s) = u(W_T,T)-u(W_t,t) = \int_t^T \frac{\partial u}{\partial x}(W_s,s)dW_s, \nonumber
\end{align}
which, after re-arrangement, leads to
\begin{align} \label{a6}
u(W_t,t) = \psi(W_T)-\int_t^T \frac{\partial u}{\partial x}(W_s,s)dW_s,
\end{align}
where we used the final boundary condition of \eqref{heateqA}. The key part is that we must now take expectations, conditioned on $W_t=x$, i.e. $u(x,t) = \mathbb{E}[\psi(W_T) \; | \; W_t=x]$,
so that picking up from \eqref{a6} we get 
\begin{align} \label{aux11}
u(x,t) = \mathbb{E}[\psi(W_T) \; | \; W_t=x] + \mathbb{E} \bigg[\int_t^T \frac{\partial u}{\partial x}(W_s,s) dW_s \; | \; W_t=x \bigg].
\end{align}
Let us deal with the second term first. Using tools from analysis, such as a generalization of Fubini's theorem, we can swap the expectation and the integral and therefore
\begin{align}
 \mathbb{E} \bigg[\int_t^T \frac{\partial u}{\partial x} (W_s,s)dW_s \; | \; W_t=x \bigg] =  \int_t^T \mathbb{E} \bigg[ \frac{\partial u}{\partial x} (W_s,s)dW_s \; | \; W_t=x \bigg]. \nonumber
\end{align}
From the properties of Brownian motion we know that $dW_s$ is independent of $W_s$ or, for that matter, any other function of $W_s$, hence by independence we can separate the expectation of the product into the product of the expectations as
\begin{align}
\mathbb{E} \bigg[ \frac{\partial u}{\partial x}(W_s,s) dW_s \; | \; W_t=x \bigg] = \mathbb{E} \bigg[ \frac{\partial u}{\partial x}(W_s,s) \; | \; W_t=x \bigg] \times \mathbb{E} [ dW_s \; | \; W_t=x ]. \nonumber
\end{align}
Invoking again the independence of Brownian increments we get $\mathbb{E} [dW_s \; | \; W_t=x] = \mathbb{E} [dW_s] = 0$,
since the Brownian motion is associated with a normal distribution with mean zero. Thus the expectation of the integral in \eqref{aux11} is simply zero and we are left with $u(x,t) = \mathbb{E}[\psi(W_T) \; | \; W_t=x]$.
The task at hand now is to compute this expectation and to this end, we use $W_T=W_t+(W_T-W_t)$ inside the expectation so that
\begin{align}
\mathbb{E}[\psi(W_T) \; | \; W_t=x] &= \mathbb{E}[\psi(W_t+(W_T-W_t)) \; | \; W_t = x] \nonumber \\
&= \mathbb{E}[\psi(x+(W_T-W_t)) \; | \; W_t = x] \nonumber \\
&= \mathbb{E}[\psi(x+(W_T-W_t))]. \nonumber
\end{align}
This is because the expressions $W_T-W_t$ and $W_t$ are independent, therefore conditioning on $W_t=x$ is irrelevant. The conclusion is that if $u(x,t)$ is a solution to \eqref{heateqA}, then $u(x,t) = \mathbb{E}[\psi(w+W_T-W_t)]$.
Now we need to remove the expectation. For that, recall that $W_T-W_t$ is normally distributed and therefore $W_T-W_t = \sqrt{T-t} Z$ with $Z \sim \mathcal{N}(0,1)$.
Therefore, using the probability distribution function of the Gaussian distribution, we see that
\begin{align}
u(x,t) &= \mathbb{E}[\psi(x+\sqrt{T-t} Z)] = \frac{1}{\sqrt{2\pi}}\int_{-\infty}^\infty \psi(x+\sqrt{T-t}z)e^{-z^2/2} dz, \nonumber
\end{align}
this is to be compared to \eqref{babyintegral}. This is the solution to the heat equation that would have been obtained through Fourier transform methods or through similarity reductions.
\section{\label{sec:fullfk}The full Feynman-Kac formula in one dimension}
This idea can be generalized with very little work to the following situation. Suppose we have
\begin{align} 
\frac{\partial u}{\partial t} + a(x)\frac{\partial^2 u}{\partial x^2} +b(x) \frac{\partial u}{\partial x} = c(x)u, \quad \textnormal{for} \quad t<T,
 \nonumber
\end{align}
and $u(x,T) = \psi(x)$. We can associate to this PDE a stochastic process $X_t$ manufactured from these coefficients as the SDE $dX_t = b(X_t)dt + \sqrt{2 a (X_t)} dW_t$.
Then Feynman-Kac is the result that $u(x,t) = \mathbb{E} [\exp(-\int_t^T c(X_u)du) \psi(X_T) \; | \; X_t =x]$.
All issues of existence and uniqueness can be found in Shreve and Karatzas \cite[$\mathsection$4]{karatzas}. The sketch of the argument leading to Feynman-Kac formula is to first set
\begin{align}
Y_t := \exp \bigg[ -\int_0^t c(X_s)ds \; u(X_t,t)\bigg], \nonumber
\end{align}
and to then employ It\^{o} to show that
\begin{align} 
\frac{dY_t}{\exp[ -\int_0^t c(X_s)ds ]} = \bigg(\frac{\partial u}{\partial t} + a(X_t) \frac{\partial^2 u}{\partial x^2} + b(X_t) \frac{\partial u}{\partial x} - c(X_t) u \bigg)dt + \frac{\partial u}{\partial x}dW_t, \nonumber
\end{align}
with the derivatives of $u$ evaluated at $(X_t,t)$. This allows us to establish that if $u$ does solve the PDE, then $Y_t$ above has zero drift (i.e. it is a martingale) and then write $\mathbb{E}[Y_t \;|\; X_t = x] = u(x,t)$. The details can be found in \cite{brummelhuis}.

\section{\label{sec:fkschro}Linking the Feynman-Kac PDE to the Schr\"{o}dinger equation}
Let us make the substitution $u(x,t) = e^{g(x,t)} v(x,t)$,
for some analytic functions $g$ and $v$. 
Substituting the newly transformed derivatives appearing into \eqref{shortPDE} and re-grouping terms, we obtain
\begin{align} \label{aux1}
- \frac{\partial v}{\partial t} &= \mathfrak{a}(x,t)\frac{\partial^2 v}{\partial x^2} + \bigg(2\mathfrak{a}(x,t) \frac{\partial g}{\partial x} + \mathfrak{b}(x,t)\bigg) \frac{\partial v}{\partial x} \nonumber \\
&\quad +  \bigg(\frac{\partial g}{\partial t} + \mathfrak{a}(x,t) \bigg(\frac{\partial g}{\partial x}\bigg)^2  +\mathfrak{a}(x,t) \frac{\partial^2 g}{\partial x^2} +\mathfrak{b}(x,t) \frac{\partial g}{\partial x} - \mathfrak{c}(x,t)\bigg)v.
\end{align}
Setting the coefficient of the unwanted term to be zero yields
\begin{align} \label{aux2}
2\mathfrak{a}(x,t) \frac{\partial g}{\partial x} + \mathfrak{b}(x,t) = 0  \Leftrightarrow \frac{\partial g}{\partial x} =  - \frac{\mathfrak{b}(x,t)}{2\mathfrak{a}(x,t)}.
\end{align}
We may integrate this to
\begin{align} \label{integratingfactor}
g(x,t) = -\int \frac{\mathfrak{b}(x,t)}{2\mathfrak{a}(x,t)} dx + h(t),
\end{align}
where $h$ is an arbitrary function of $t$ only. Using \eqref{aux2} on the coefficient of the linear term $v$ appearing in \eqref{aux1} we get the expression
\begin{align} 
\mathfrak{w}(x,t) :&= \frac{\partial g}{\partial t} + \mathfrak{a}(x,t) \bigg(\frac{\partial g}{\partial x}\bigg)^2  +\mathfrak{a}(x,t) \frac{\partial^2 g}{\partial x^2} +\mathfrak{b}(x,t) \frac{\partial g}{\partial x} - \mathfrak{c}(x,t) \nonumber \\
&= \frac{\partial h}{\partial t} - \frac{\partial}{\partial t} \int \frac{\mathfrak{b}(x,t)}{2 \mathfrak{a}(x,t)}dx - \frac{\mathfrak{b}^2(x,t)}{4 \mathfrak{a}(x,t)} - \frac{1}{2} \mathfrak{a}(x,t) \bigg(\frac{\partial}{\partial x} \frac{\mathfrak{b}(x,t)}{\mathfrak
{a}(x,t)}\bigg) - \mathfrak{c}(x,t). \nonumber
\end{align}
Since $h$ was an arbitrary function we may take it to be $h(t)=0$ so that $\tfrac{dh}{dt}=0$ as well. Effectively, this implies that
\begin{align}
g(x,t) = -\int^x \frac{\mathfrak{b}(y,t)}{2\mathfrak{a}(y,t)} dy, \nonumber
\end{align}
where the integration sign $\int^x$ indicates that the result of the integration with respect to $y$ is a function of $x$ without a constant of integration. This implies that
\begin{align} \label{effectivew}
\mathfrak{w}(x,t) = - \frac{\partial}{\partial t} \int \frac{\mathfrak{b}(x,t)}{2 \mathfrak{a}(x,t)}dx - \frac{\mathfrak{b}^2(x,t)}{4 \mathfrak{a}(x,t)} - \frac{1}{2} \mathfrak{a}(x,t) \bigg(\frac{\partial}{\partial x} \frac{\mathfrak{b}(x,t)}{\mathfrak
{a}(x,t)}\bigg) - \mathfrak{c}(x,t).
\end{align}
In a substantial number of interesting cases the coefficients $\mathfrak{a}, \mathfrak{b}$ and $\mathfrak{c}$ are functions of $x$ only, which will further simplify \eqref{effectivew}. Therefore, using these auxiliary results, \eqref{shortPDE} becomes
\begin{align} 
0 = \frac{\partial v}{\partial t} + \mathfrak{a}(x,t) \frac{\partial^2 v}{\partial x^2}  +\mathfrak{w}(x,t) v. \nonumber
\end{align}
There are two additional substitutions we need to make. First, we need to turn this into an initial value problem by setting $\tau = T-t$, in which case we end up with
\begin{align} \label{negativetime}
\frac{\partial v}{\partial \tau} = \mathfrak{a}(x,\tau)\frac{\partial^2 v}{\partial x^2}  +\mathfrak{w}(x,\tau) v. 
\end{align}
Lastly, the Wick rotation $\xi = - i \tau$ transforms \eqref{negativetime} into
\begin{align}
-i \frac{\partial v}{\partial \xi} = \mathfrak{a}(x,\xi) \frac{\partial^2 v}{\partial x^2}  +\mathfrak{w}(x,\xi) v. \nonumber
\end{align}
Note that some straightforward modifications to the time variable need to be applied in these last two transformations to the functions $u$ and $v$ to keep consistency.

\section{\label{sec:appendixnorms}Additional norm plots}

\begin{figure}[h!]
	\centering
    \includegraphics[scale=0.34]{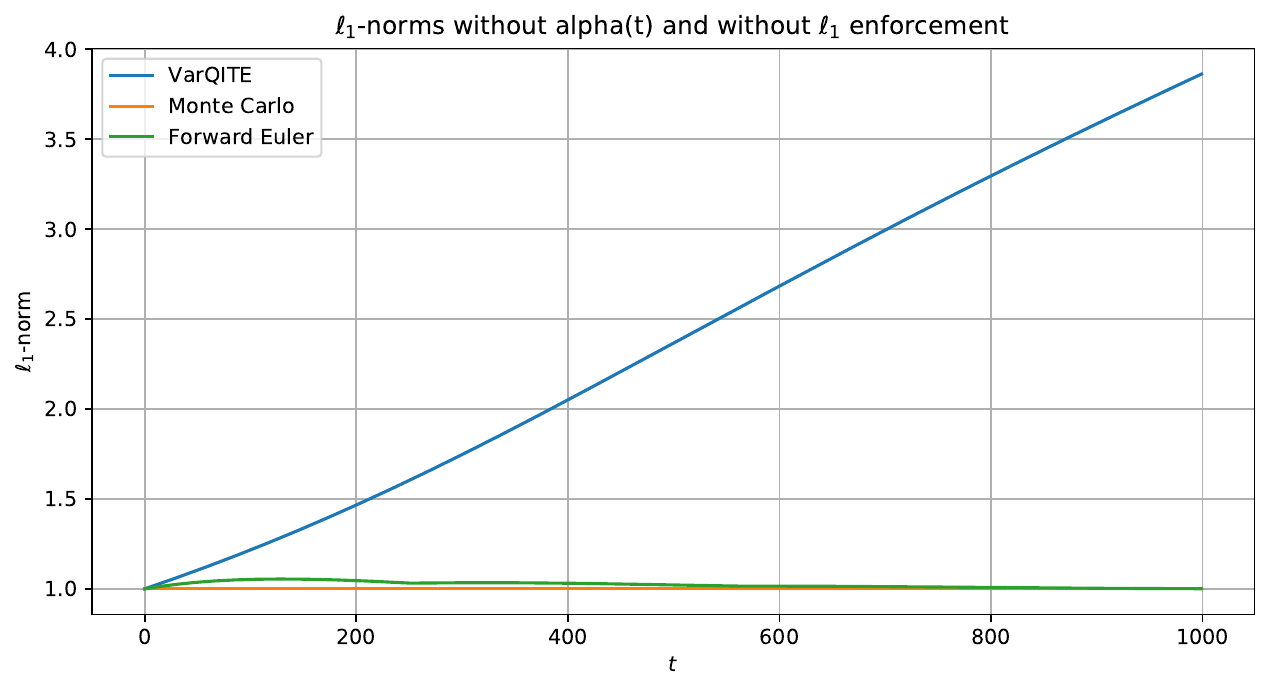}
    \includegraphics[scale=0.34]{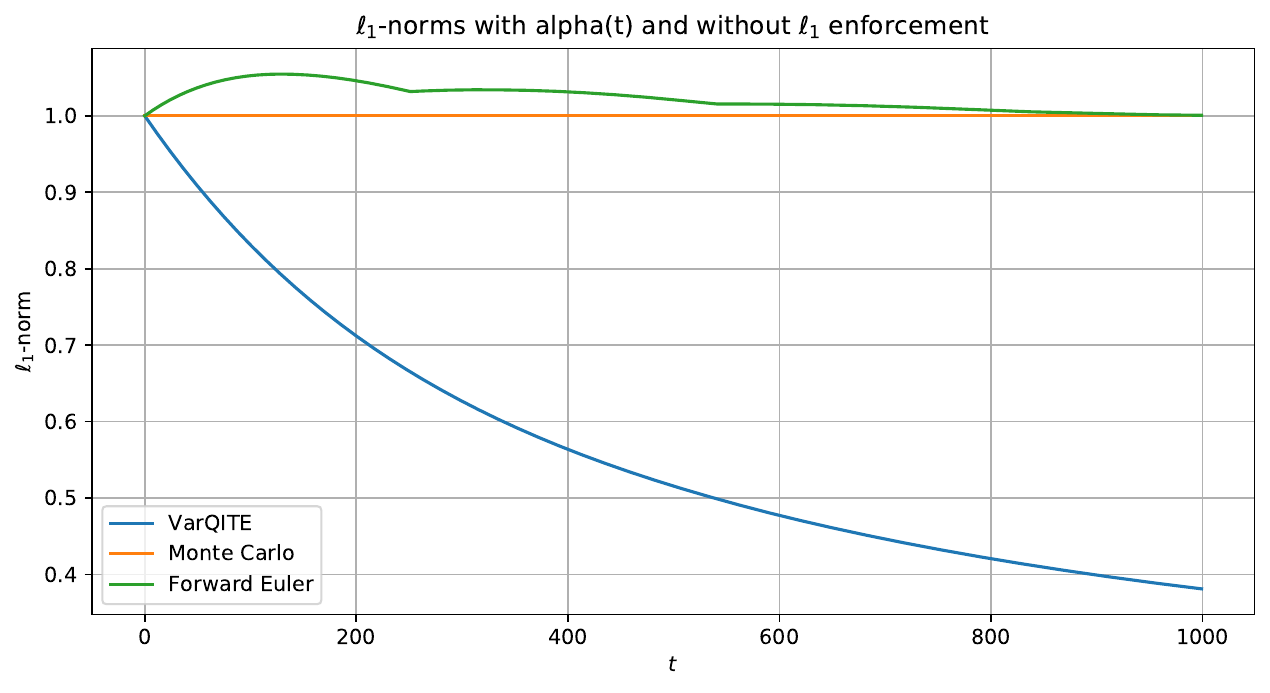}
\caption{Plots of $\ell_1$ norms for the first 100 evolutions with six qubits and $n=1$ layers of ansatz,
\underline{left}: without $\alpha(t)$ and without the $\ell_1$-norm enforcement procedure and 
\underline{right}: with $\alpha(t)$ but without the $\ell_1$-norm enforcement procedure.
}
    \label{fig:comparison_without_enforcement} 
\end{figure}

To demonstrate the efficacy of the $\ell_1$-norm enforcement procedure as explained in Sec \ref{sec:enforce}, we provide additional plots in Figure \ref{fig:comparison_without_enforcement} comparing the VarQITE method against Monte Carlo and forward Euler methods. 

\section{\label{sec:boundaryfinance}Backward propagation and option pricing}

To apply the introduced technique to option pricing, a different application of the Feynman-Kac formula is required: one considers propagation backwards in time instead of forward, i.e., the final payoff is used as initial condition and we propagate back in time to the starting point of the process.
Every element of the result then corresponds to the fair option price for the corresponding initial state.

For instance, one  may wish to solve the heat equation by specializing $\mathfrak{a}(x,\xi)=\tfrac{1}{2}$ and $\mathfrak{w}(x,\xi)=0$ as well as the payoff function
$ 
\psi(X_T) = \max(X_T-K,0), \nonumber
$
where $K$ now represents the strike of a European option with maturity $T$. The rationale behind is that the Black-Scholes equation can be written as the heat equation using suitable transformations. Stepping into the domain of path-dependent options, another popular financial derivative is the Asian option, see \cite{vecer}. To price it, we specialize the coefficients of the Hamiltonian to $\mathfrak{a}(x,\xi)=\tfrac{(q(\xi)-x)^2}{2}$ and $\mathfrak{w}(x,\xi)=0$, where $q(t)=\tfrac{1-e^{-r(T-t)}}{rT}$ if $r \ne 0$ and $q(t)=1-\tfrac{t}{T}$ if $r = 0$. The payoff is given by the average of process $X_t$, i.e.
\begin{align} \label{asian}
\psi(X_T,T) = \max \bigg(\frac{1}{T} \int_0^T X_u du - K, 0\bigg).
\end{align}
In this case, \eqref{asian} represents an Asian option with strike $K$ and maturity $T$. We remark that the payoff function $\psi$ now depends on $T$. These cases are studied in \cite{fontanela2021quantum, santosh, montecarloxanadu, ibmoption}, not always under the optic of variational quantum imaginary time evolution. The European case is showcased in \cite{gonzalezconde2021pricing, unary} by amplitude estimation and without applying variational principles.

Much like in the case of forward propagation, cf. Section \ref{readout}, the performance of the algorithm when going backward in time needs to be carefully analyzed.
First, the payoff function, which is now considered as the initial condition, needs to be loaded into a quantum state, i.e., it needs to be normalized and we need to prepare a corresponding quantum circuit.
Let us consider a European call option for illustration with $X_T \in [X_l, X_u]$, $X_u - X_l = B$, and $K = (X_l + X_u)/2$.
If we discretize $X_T$ using $n$ qubits, i.e. with $2^n$ grid points and grid size $h = B/2^n$, then the normalization factor of the quantum state scales as $\theta(B \sqrt{2^n})$.
As in forward propagation, this limits the potential advantage already to a quadratic speed-up, since we need to re-scale any result --including the estimation error-- by a factor of size $\theta(B \sqrt{2^n})$ to get back to the original scale.
In addition to the normalization, constructing a circuit to prepare arbitrary quantum states can be prohibitive \cite{shende2006}. 
For the simple payoffs considered efficient circuits should exist, and for more generic functions variational approach as suggested in \cite{fontanela2021quantum} or \cite{zoufal_qgan} may be used.

Suppose the initial state has been loaded, the algorithm has been applied to estimate a prepared quantum state that corresponds to the solution and one is interested in the result. The impact of the $\alpha$-factor might be negligible here, since we already start in a quantum state with rather uniform amplitudes. However, the option price for a particular target initial price corresponds now to the amplitude of a single computational basis state.
The readout of this can be achieved via QAE up to accuracy $\varepsilon > 0$ at cost $\mathcal{O}(1/\varepsilon)$. Since the amplitude of interest likely will be of order $\mathcal{O}(1/\sqrt{2^n})$, the required $\varepsilon$ will be of equal scale. 
Combining the requirements coming from the normalization likely renders this approach impractical for option pricing.

\section{\label{sec:analytical}Analytical solution}

The equation we need to solve is
\begin{align} \label{original}
\frac{\partial u}{\partial t} = A \frac{\partial^2 u}{\partial x \partial y} + B \frac{\partial^2 u}{\partial x^2} + C \frac{\partial^2 u}{\partial y^2} - ru,
\end{align}
with $u=u(x,y,t)$, and initial condition $u(x,y,0) = \delta(x-x_0) \delta(y-y_0),$ with $x_0, y_0 > 0$.
Here $A,B, C$ and $r$ are all real constants such that $4BC - {A^2} > 0$. The idea is to apply a Laplace transform on the time variable and a Mellin transform on each of the space variables. This will allow us to change the PDE into an algebraic equation which we can solve trivially. Then we proceed to invert the Laplace transforms as well as the Mellin transforms to recover the solution of \eqref{original}.

Let us start with
\begin{align} \label{extendedxy}
\frac{\partial u}{\partial t} = A xy \frac{\partial^2 u}{\partial x \partial y} + B x^2 \frac{\partial^2 u}{\partial x^2} + C y^2\frac{\partial^2 u}{\partial y^2} - ru + D x\frac{\partial u}{\partial x}  + E y\frac{\partial u}{\partial y},
\end{align}
with $D$ and $E$ real constants. Make the changes
\begin{align} \label{transformations}
X = \log x, \quad Y = \log y.
\end{align}
Then from \eqref{transformations} we see that
\begin{align} \label{transformations2}
  \frac{\partial}{{\partial x}} = \frac{1}{x}\frac{\partial}{{\partial X}},\quad \frac{{{\partial^2}}}{{d{x^2}}} = \frac{1}{{{x^2}}}\left( {\frac{{{\partial^2}}}{{\partial{X^2}}} - \frac{\partial}{{\partial X}}} \right) \quad \textnormal{and} \quad
  \frac{\partial}{{\partial y}} = \frac{1}{y}\frac{\partial}{{\partial Y}},\quad \frac{{{\partial^2}}}{{\partial{y^2}}} = \frac{1}{{{y^2}}}\left( {\frac{{{\partial^2}}}{{\partial{Y^2}}} - \frac{\partial}{{\partial Y}}} \right) . 
\end{align} 
Using \eqref{transformations2} on \eqref{extendedxy} yields
\begin{align} \label{aux001}
  \frac{{\partial u}}{{\partial t}} &= Axy\frac{1}{x}\frac{\partial }{{\partial X}}\frac{1}{y}\frac{\partial }{{\partial Y}}u + B{x^2}\frac{1}{{{x^2}}}\left( {\frac{{{\partial ^2}}}{{\partial {X^2}}} - \frac{\partial }{{\partial X}}} \right)u + C{y^2}\frac{1}{{{y^2}}}\left( {\frac{{{\partial ^2}}}{{\partial {Y^2}}} - \frac{\partial }{{\partial Y}}} \right)u - ru  \nonumber \\
  &\quad + Dx\frac{1}{x}\frac{\partial }{{\partial X}}u + Ey\frac{1}{y}\frac{\partial }{{\partial Y}}u \nonumber \\
   &= A\frac{{{\partial ^2}u}}{{\partial X\partial Y}} + B\frac{{{\partial ^2}u}}{{\partial {X^2}}} + C\frac{{{\partial ^2}u}}{{\partial {Y^2}}} + \left( {D - B} \right)\frac{{\partial u}}{{\partial X}} + \left( {E - C} \right)\frac{{\partial u}}{{\partial Y}} - ru .
\end{align} 
Now, \eqref{aux001} will have the same structure as \eqref{original} if
\begin{align} \label{conditionsconstants}
D = B \quad \textnormal{and} \quad E=C,
\end{align}
minding that $x,y$ are related to $X,Y$ by \eqref{transformations}. Thus, we need to solve \eqref{extendedxy} with $D$ and $E$ satisfying the conditions given by \eqref{conditionsconstants}. The solution to \eqref{extendedxy} when \eqref{conditionsconstants} holds is given by
\[
u(x,y,t) = \frac{1}{{2\pi \sqrt \Delta  }}\frac{{{e^{ - rt}}}}{{{x_0}{y_0}t}}\exp \left[ { - \frac{B}{{\Delta t}}{{\left( {\frac{A}{{2B}}\log \frac{x}{{{x_0}}} - \log \frac{y}{{{y_0}}}} \right)}^2} - \frac{1}{{4Bt}}{{\log }^2}\frac{x}{{{x_0}}}} \right]
\]
where $\Delta  = 4BC - {A^2} > 0$. Using ${X_0} = \log {x_0}$ and ${Y_0} = \log {y_0}$ in addition to \eqref{transformations} (that is, reversing the change of variables), then the solution to \eqref{original} when \eqref{conditionsconstants} holds is given by
\[
u(x,y,t) = \frac{{{e^{ - rt - {x_0} - {y_0}}}}}{{2\pi t\sqrt \Delta  }}\exp \left[ { - \frac{B}{{\Delta t}}{{\left( {\frac{A}{{2B}}\left( {x - {x_0}} \right) - \left( {y - {y_0}} \right)} \right)}^2} - \frac{1}{{4Bt}}{{\left( {x - {x_0}} \right)}^2}} \right].
\]
We start by taking the Laplace transform of $u$ with respect to $t$ so that
\begin{align}
  {\hat u}(x,y,s) &= \mathcal{L}\{ u(x,y,t)\} (s) = \int_0^\infty  {{e^{ - st}}u(x,y,t)dt}  \nonumber \\
   &=  - \frac{1}{s}{e^{ - st}}u(x,y,t)|_{t = 0}^{t = \infty } + \frac{1}{s}\int_0^\infty  {{e^{ - st}}\left( {\frac{\partial }{{\partial t}}u(x,y,t)} \right)dt}  \nonumber \\
   &\quad + \frac{1}{s}\left( {Axy\frac{{{d^2}}}{{dxdy}} + B{x^2}\frac{{{d^2}}}{{d{x^2}}} + C{y^2}\frac{{{d^2}}}{{d{y^2}}} - r + Dx\frac{d}{{dx}} + Ey\frac{d}{{dy}}} \right) \nonumber \\
   & \quad \quad \quad \quad \times \int_0^\infty  {{e^{ - st}}u(x,y,t)dt}  \nonumber \\
   &= \frac{1}{s}\delta (x - {x_0})\delta (y - {y_0}) \nonumber \\
   &\quad + \frac{1}{s}\left( {Axy\frac{{{d^2}}}{{dxdy}} + B{x^2}\frac{{{d^2}}}{{d{x^2}}} + C{y^2}\frac{{{d^2}}}{{d{y^2}}} - r + Dx\frac{d}{{dx}} + Ey\frac{d}{{dy}}} \right)\hat u(x,y,s) . \nonumber
\end{align}
Next take a Mellin transform of ${\hat u}$ with respect to $x$. This gives us
\begin{align} \label{aux002}
  {\tilde {\hat u}}({w_1},y,s) &= \mathcal{M}\{ \hat u(x,y,s)\} ({w_1}) = \int_0^\infty  {{x^{{w_1} - 1}}\hat u(x,y,s)dx}  \nonumber \\
   &= \frac{1}{s}\delta (y - {y_0})\mathcal{M}\{ \delta (x - {x_0})\} ({w_1}) + \frac{1}{s}Ay\frac{d}{{dy}}\mathcal{M}\left\{ {x\frac{d}{{dx}}\hat u(x,y,s)} \right\}({w_1}) \nonumber \\
   &\quad + \frac{1}{s}B\mathcal{M}\left\{ {{x^2}\frac{{{d^2}}}{{d{x^2}}}\hat u(x,y,s)} \right\}({w_1}) + \frac{1}{s}C{y^2}\frac{{{d^2}}}{{d{y^2}}}\mathcal{M}\left\{ {\hat u(x,y,s)} \right\}({w_1}) \nonumber \\
   & \quad - \frac{r}{s}\mathcal{M}\left\{ {\hat u(x,y,s)} \right\}({w_1}) + \frac{1}{s}D\mathcal{M}\left\{ {x\frac{d}{{dx}}\hat u(x,y,s)} \right\}({w_1}) \nonumber \\
   & \quad + \frac{1}{s}Ey\frac{d}{{dy}}\mathcal{M}\left\{ {\hat u(x,y,s)} \right\}({w_1}).   
\end{align}
We shall need to employ \cite{obermellin}
\begin{align} 
\mathcal{M}\{ xf'(x)\} (w) =  - w\mathcal{M}\{ f(x)\} (w)\quad \textnormal{and }\quad \mathcal{M}\{ {x^2}f''(x)\} (w) = w(1 + w)\mathcal{M}\{ f(x)\} (w)  \nonumber
\end{align}
to evaluate the terms appearing in \eqref{aux002}. Using these Mellin transforms leads us to
\begin{align}
  {\tilde {\hat u}}({w_1},y,s) &= \frac{{x_0^{{w_1} - 1}}}{s}\delta (y - {y_0}) - \frac{{Ay}}{s}{w_1}\frac{d}{{dy}}{\tilde {\hat u}}({w_1},y,s) + \frac{B}{s}{w_1}(1 + {w_1}){\tilde {\hat u}}({w_1},y,s) \nonumber \\
   &\quad + \frac{C}{s}{y^2}\frac{{{d^2}}}{{d{y^2}}}{\tilde {\hat u}}({w_1},y,s) - \frac{r}{s}{\tilde {\hat u}}({w_1},y,s) - \frac{D}{s}{w_1}{\tilde {\hat u}}({w_1},y,s) + \frac{E}{s}y\frac{d}{{dy}}{\tilde {\hat u}}({w_1},y,s). \nonumber
\end{align}
Another Mellin transform of ${\tilde {\hat u}}({w_1},y,s)$, this time with respect to $y$, yields
\begin{align}
  {\tilde {\tilde {\hat u}}}({w_1},{w_2},s) &= \mathcal{M}\{ {\tilde {\hat u}}({w_1},y,s)\} ({w_2}) = \int_0^\infty  {{y^{{w_2} - 1}}{\tilde {\hat u}}({w_1},y,s)dy}  \nonumber \\
   &= \frac{{x_0^{{w_1} - 1}y_0^{{w_2} - 1}}}{s} + \frac{{A}}{s}{w_1}{w_2}{\tilde {\tilde {\hat u}}}({w_1},{w_2},s) + \frac{B}{s}{w_1}(1 + {w_1}){\tilde {\tilde {\hat u}}}({w_1},{w_2},s) \nonumber \\
   &\quad + \frac{C}{s}{w_2}(1 + {w_2}){\tilde {\tilde {\hat u}}}({w_1},{w_2},s) - \frac{r}{s}{\tilde {\tilde {\hat u}}}({w_1},{w_2},s)  - \frac{D}{s}{w_1}{\tilde {\tilde {\hat u}}}({w_1},{w_2},s) \nonumber \\
   & \quad - \frac{E}{s}{w_2}{\tilde {\tilde {\hat u}}}({w_1},{w_2},s). \nonumber  
\end{align} 
Now we solve for ${\tilde {\tilde {\hat u}}}({w_1},{w_2},s)$ in the above equation
\begin{align} \label{generalw1w2s}
  {\tilde {\tilde {\hat u}}}({w_1},{w_2},s)
   = \frac{{x_0^{{w_1} - 1}y_0^{{w_2} - 1}}}{{s - (A{w_1}{w_2} + {w_1}(B - D + B{w_1}) + {w_2}(C - E + C{w_2}) - r)}}.
\end{align} 
We shall now bring in \eqref{conditionsconstants}. Strictly speaking it is not necessary to use them and there is some loss of generality. However, it makes the problem much easier and using more generality in the coefficients is straightforward. Thus using \eqref{conditionsconstants} we see that \eqref{generalw1w2s} reduces to
\begin{align} \label{preinverse}
{\tilde {\tilde {\hat u}}}({w_1},{w_2},s) = \frac{{x_0^{{w_1} - 1}y_0^{{w_2} - 1}}}{{s - (A{w_1}{w_2} + Bw_1^2 + Cw_2^2 - r)}}.
\end{align}
The task, now that we have found ${\tilde {\tilde {\hat u}}}({w_1},{w_2},s)$ in terms of $w_1, w_2$ and $s$, is to revert back all the transforms and translate the answer to the space of solutions of \eqref{original}. Apply the inverse Laplace transform 
$
\mathcal{L}^{-1} \{\frac{1}{s-a}\}(t) = e^{at}
$
to \eqref{preinverse} so that we end up with
\begin{align} 
{\tilde {\tilde u}}({w_1},{w_2},t) &= x_0^{{w_1} - 1}y_0^{{w_2} - 1}\exp \left[ {\left( {A{w_1}{w_2} + Bw_1^2 + Cw_2^2 - r} \right)t} \right] \nonumber \\ 
   &= {e^{ - rt}}x_0^{{w_1} - 1}y_0^{{w_2} - 1}\exp \left[ {Ctw_2^2} \right]\exp \left[ {\alpha {w_1} + \beta w_1^2} \right] , \nonumber
\end{align}
where $\alpha  = At{w_2},$ and $\beta  = Bt$. To invert the Mellin transform with respect to $x$ we use the following
$
G(s) = {e^{\alpha s + \beta {s^2}}} = {e^{\alpha s}}{e^{\beta {s^2}}} = {G_1}(s){G_2}(s).
$
Then the inverses of $G_1$ and $G_2$ are given by \cite{obermellin}
\[
{g_1}(x) = {\mathcal{M}^{ - 1}}\{ {e^{\alpha s}}\} (x) = \delta (\alpha  - \log x)
\]
as well as
\[
{g_2}(x) = {\mathcal{M}^{ - 1}}\{ {e^{\beta {s^2}}}\} (x) = \frac{1}{{2\sqrt {\pi \beta } }}\exp \left( { - \frac{{{{\log }^2}x}}{{4\beta }}} \right).
\]
Thus the inverse of $G$ can be found by computing the convolution
\begin{align}
  g(x) = \int_0^\infty  {{g_1}(z){g_2}\left( {\frac{x}{z}} \right)\frac{{dz}}{z}} &= \int_0^\infty  {\delta (\alpha  - \log z)\frac{1}{{2\sqrt {\pi \beta } }}\exp \left( { - \frac{{{{\log }^2}\tfrac{x}{z}}}{{4\beta }}} \right)\frac{{dz}}{z}}  \nonumber \\
  &= \frac{1}{{2\sqrt {\pi \beta } }}\exp \left[ { - \frac{{{{(\alpha  - \log x)}^2}}}{{4\beta }}} \right], \nonumber
\end{align} 
since $\alpha \in \R$. Moving on from ${\tilde {\tilde u}}(w_1, w_2, t)$ we invert $w_1$ back into $x$ and see that
\begin{align}
  \tilde u(x,{w_2},t) &= {e^{ - rt}}y_0^{{w_2} - 1}\exp \left[ {Ctw_2^2} \right]{\mathcal{M}^{ - 1}}\left\{ {x_0^{{w_1} - 1}\exp \left[ {\alpha {w_1} + \beta w_1^2} \right]} \right\}(x) \nonumber \\
   &= {e^{ - rt}}y_0^{{w_2} - 1}\exp \left[ {Ctw_2^2} \right]\int_0^\infty  {\delta (z - {x_0})\frac{1}{{2\sqrt {\pi \beta } }}\exp \left[ { - \frac{{{{(\alpha  - \log \tfrac{x}{z})}^2}}}{{4\beta }}} \right]\frac{{dz}}{z}}  \nonumber \\
   &= \frac{{{e^{ - rt}}}}{{2\sqrt {\pi \beta } }}y_0^{{w_2} - 1}\exp \left[ {Ctw_2^2} \right]\frac{1}{{{x_0}}}\exp \left[ { - \frac{{{{(\alpha  - \log \tfrac{x}{{{x_0}}})}^2}}}{{4\beta }}} \right] \nonumber \\
   &= \frac{{{e^{ - rt}}}}{{2{x_0}\sqrt {B\pi t} }}\exp \left[ { - \frac{{{{\log }^2}\tfrac{x}{{{x_0}}}}}{{4Bt}}} \right]y_0^{{w_2} - 1}\exp \left[ {\alpha {w_2} + \beta w_2^2} \right], \nonumber
\end{align}
where we have recycled $\alpha$ and $\beta$ to now be
$
\alpha  = \frac{A}{2B} \log \frac{x}{x_0}
$
and
$\beta  = Ct - \frac{{{A^2}t}}{{4B}}$. Lastly, we Mellin invert $w_2$ back into $y$ and we arrive at
\begin{align}
  u(x,y,t) &= \frac{{{e^{ - rt}}}}{{2{x_0}\sqrt {B\pi t} }}\exp \left[ { - \frac{{{{\log }^2}\tfrac{x}{{{x_0}}}}}{{4Bt}}} \right]{\mathcal{M}^{ - 1}}\left\{ {y_0^{{w_2} - 1}\exp \left[ {\alpha {w_2} + \beta w_2^2} \right]} \right\}(y) \nonumber \\
   &= \frac{{{e^{ - rt}}}}{{2{x_0}\sqrt {B\pi t} }}\exp \left[ { - \frac{{{{\log }^2}\tfrac{x}{{{x_0}}}}}{{4Bt}}} \right]\int_0^\infty  {\delta (z - {y_0})\frac{1}{{2\sqrt {\pi \beta } }}\exp \left[ { - \frac{{{{(\alpha  - \log \tfrac{y}{z})}^2}}}{{4\beta }}} \right]\frac{{dz}}{z}}  \nonumber \\
   &= \frac{{{e^{ - rt}}}}{{2{x_0}{y_0}\sqrt {B\pi t} }}\frac{1}{{2\sqrt {\pi \beta } }}\exp \left[ { - \frac{{{{\log }^2}\tfrac{x}{{{x_0}}}}}{{4Bt}}} \right]\exp \left[ { - \frac{{{{(\alpha  - \log \tfrac{y}{{{y_0}}})}^2}}}{{4\beta }}} \right] . \nonumber
\end{align}
By setting $\Delta  = 4BC - {A^2}$, and doing some mild algebraic re-arrangements we end up with
\begin{align} 
u(x,y,t) = \frac{1}{{2\pi \sqrt \Delta  }}\frac{{{e^{ - rt}}}}{{{x_0}{y_0}t}}\exp \left[ { - \frac{B}{{\Delta t}}{{\left( {\frac{A}{{2B}}\log \frac{x}{{{x_0}}} - \log \frac{y}{{{y_0}}}} \right)}^2} - \frac{1}{{4Bt}}{{\log }^2}\frac{x}{{{x_0}}}} \right], \nonumber
\end{align}
which is the solution we were seeking.

\section{\label{sec:moments}Moments}

While obtaining the full solution $u$ to the PDF \eqref{generalizedOp} generated by the Feynman-Kac algorithm is an expensive problem, familiar within the field of quantum metrology, we may nevertheless extract information about the moments of $u$ with a quadratic advantage using QPE which are deep circuits, see Section \ref{readout}. Alternatively we could use Hadamard tests which are not as efficient but they consist of shallower circuits. This section generalizes to several dimensions some of the concepts introduced in \cite{stochastic}, which then get applied to the system of SDEs introduced in \eqref{systemSDEs}. Along the way, we take this chance to amend some inaccuracies from \cite{stochastic}. To avoid confusion with exponentiation, the labeling $i=1,2,\cdots,D$ of the processes $X_t^i$ will now be subscripted and the time variable $t$ will be suppressed. 

Recall that $n_m=2^{n_q/D}$, where $n_q$ is the total number of qubits and $D$ the number of dimensions, represents the number of grid points allocated to each dimension. Suppose we have a function $f: \R^D \to \R$ and consider the expectation
\begin{align} \label{expectationDimensional}
\mathbb{E}[f(X_1(T), X_2(T), &\cdots, X_D(T))] \nonumber \\
&= \sum_{i_1=0}^{n_m-1}\sum_{i_2=0}^{n_m-1} \cdots \sum_{i_D=0}^{n_m-1} f(x_{i_1}, x_{i_2}, \cdots, x_{i_D}) \nonumber \\
& \quad \times \mathbb{P}_{X_1, X_2, \cdots, X_D}[X_1(T)=x_{i_1}, X_2(T)=x_{i_2}, \cdots, X_D(T)=x_{i_D}].
\end{align}
Here $u(x_{i_1}, x_{i_2}, \cdots, x_{i_D}, t)=\mathbb{P}_{X_1, X_2, \cdots, X_D}$ denotes the joint probability distribution function of the random variables $X_1, X_2, \cdots, X_D$ in situations where we have $\ell_1$ normalization. We have $D$ dimensions of equal length $[0, x_{i,\max}]=[0, x_{\max}]$ and in each one of them we split the interval $[0, x_{i,\max}]$ into $N_{x_i}=N_x=d$ sub-intervals spaced by the same lattice spacing $\Delta x_i = \Delta x = h$ for $i=1,2,\cdots, D$ and the total number of qubits gets evenly allocated to each of the $D$ dimensions. The splitting of $[0, x_{i,\max}]$ is
\begin{align} \label{splittinginterval}
\{ [0, a_{i,1}], [a_{i,1}, a_{i,2}], \cdots, [a_{i,d-1}, x_{i,\max}] \} =: \Upsilon(i),
\end{align}
that is, we have $d$ intervals in each dimension $D$, yielding a total of $d^D$ intervals in all dimensions altogether. For an integer $n$ we have $x_n = n \Delta x$. We further set the notation
\begin{align} \label{alphaik}
\alpha_{i,k} := [a_{i,k}, a_{i,k+1}] \in \Upsilon(i), \quad \textnormal{for} \quad k=0,1,2,\cdots,d-1, \quad \textnormal{and} \quad i=1,2,\cdots,D. 
\end{align}

Let us recall Taylor's expansion in several dimensions. In order to do so, we need to introduce some notation first. If $\alpha = (\alpha_1, \alpha_2, \cdots, \alpha_n)$ denotes an $n$-tuple of nonnegative integers, then ${\bf x}^\alpha = x_1^{\alpha_1} x_2^{\alpha_2} \cdots x_n^{\alpha_n}$ where ${\bf x} = (x_1, x_2, \cdots, x_n) \in \R^n$. The sum $|\alpha| = \alpha_1 + \alpha_2 + \cdots + \alpha_n$ is the order, or the degree, of $\alpha$. Moreover $\partial^{\alpha} = \partial_1^{\alpha_1}\partial_2^{\alpha_2} \cdots \partial_n^{\alpha_n}f = \frac{\partial^{|\alpha|}f}{\partial x_1^{\alpha_1} \partial x_2^{\alpha_2} \cdots \partial x_n^{\alpha_n}}$ and $\alpha! = \alpha_1! \alpha_2! \cdots \alpha_n!$. Thus the order of $\alpha$ is the same as the order of ${\bf x}^{\alpha}$ as a monomial or the order or $\partial^{\alpha}$ as a partial derivative.

In the interval $\alpha_{i,k}$, the function $f({\bf x}) = f(x_1, x_2, \cdots, x_D)$ is assumed to be well approximated by its $L$th order Taylor series ($i$ labels dimension and $k$ labels interval in the dimension)
\begin{align}
f_{i,k}({\bf x}) = \sum_{|m| \le L_{i,k}} \frac{\partial^m f_{i,k}({\bf 0})}{m!} {\bf x}^{m} =: \sum_{|m| \le L_{i,k}} f_{i,k,m} {\bf x}^{m}.
\end{align}
A natural assumption that we shall use towards the end is that $L_{i,k}$ is the same for each sub-interval of each dimension and hence $L_{i,k}=L$. The error in the expectation \eqref{expectationDimensional} due to this approximation will be discussed shortly. Due to the finiteness of ${\bf x}$, we can find an appropriate shift to have the range of $f$ be positive and this will ensure that the expectation \eqref{expectationDimensional} will also be positive.

In $D$ dimensions the unnormalized state that we are working with is given by
\begin{align} \label{unnormalizedstateDimensional}
\ket{\tilde{\psi}(t)} = \sum_{i_1=0}^{n_m-1}\sum_{i_2=0}^{n_m-1} \cdots \sum_{i_D=0}^{n_m-1} \mathbb{P}_{X_1, X_2, \cdots, X_D}[X_1(t)=x_{i_1}, X_2(t)=x_{i_2}, \cdots, X_D(t)=x_{i_D}] \ket{{\bf i}}
\end{align}
where ${\bf i} = \ket{i_1, i_2, \cdots, i_D}$. To compute the expectation \eqref{expectationDimensional} we first consider the non-unitary operator satisfying
\begin{align} \label{Sfon0}
S_f \ket{{\bf 0}} = \sum_{i_1=0}^{n_m-1}\sum_{i_2=0}^{n_m-1} \cdots \sum_{i_D=0}^{n_m-1} f(x_{i_1}, x_{i_2}, \cdots, x_{i_D}) \ket{{\bf i}}.
\end{align}
This operator can be decomposed into easily implementable unitaries $Q$, i.e.
\begin{align} \label{SfDDef}
S_f = \bigotimes_{d=1}^D \sum_{k_d} \xi_{k_d} Q_{k_d},
\end{align}
where $\xi_{k_d} \in \C$. Taking $\ket{{\bf 0}} \bra{{\bf 0}} = \ket{0, 0, \cdots, 0}\bra{0, 0, \cdots, 0}$, a calculation shows that
\begin{align} \label{psiSfSfpsi}
\bra{\tilde{\psi}(T)} (S_f \ket{{\bf 0}} \bra{{\bf 0}} S_f^\dagger) \ket{\tilde{\psi}(T)} &= \prod_{i=1}^D \sum_{a_i=0}^{n_m-1} \mathbb{P}_{X_1, \cdots, X_D}[X_1(T)=x_{a_1}, \cdots, X_D(T)=x_{a_D}] \nonumber \\
& \quad \times \bra{{\bf a}} \prod_{j=1}^D \sum_{b_j=0}^{n_m-1} f(x_{b_1}, \cdots, x_{b_D}) \ket{{\bf b}}\nonumber \\
&\quad \times \prod_{k=1}^D \sum_{c_k=0}^{n_m-1} \mathbb{P}_{X_1, \cdots, X_D}[X_1(T)=x_{c_1}, \cdots, X_D(T)=x_{c_D}] \nonumber \\
& \quad \times \bra{{\bf c}} \prod_{\ell=1}^D \sum_{d_\ell=0}^{n_m-1} f^*(x_{d_1}, \cdots, x_{d_D}) \ket{{\bf d}}\nonumber \\
&=\sum_{a_1=0}^{n_m-1} \sum_{a_2=0}^{n_m-1} \cdots \sum_{a_D=0}^{n_m-1} f(x_{a_1}, x_{a_2}, \cdots, x_{a_D}) \nonumber \\
& \quad \times \mathbb{P}_{X_1, X_2, \cdots, X_D}[X_1(T)=x_{a_1}, X_2(T)=x_{a_2}, \cdots, X_D(T)=x_{a_D}] \nonumber \\
&\quad \times \sum_{c_1=0}^{n_m-1} \sum_{c_2=0}^{n_m-1} \cdots \sum_{c_D=0}^{n_m-1} f(x_{c_1}, x_{c_2}, \cdots, x_{c_D}) \nonumber \\
& \quad \times \mathbb{P}_{X_1, X_2, \cdots, X_D}[X_1(T)=x_{c_1}, X_2(T)=x_{c_2}, \cdots, X_D(T)=x_{c_D}] \nonumber \\
&= (\mathbb{E}[f(X_1(T), X_2(T), \cdots, X_D(T))])^2,
\end{align}
since $f$ was a real function. We may now use some of the results presented in \ref{sec:decomposition}. For instance $\ket{0}\bra{0} = (\mathbb{I}-X^{\otimes n} \cdot C^{n-1} Z \cdot X^{\otimes n})/2$ is a sum of easily implementable unitaries, and therefore $\ket{{\bf 0}} \bra{{\bf 0}}$ will also be implementable as such a decomposition by the use of $\ket{{\bf 0}} \bra{{\bf 0}} = (\mathbb{I}-X^{\otimes nD} \cdot C^{nD-1} Z \cdot X^{\otimes nD})/2$. 

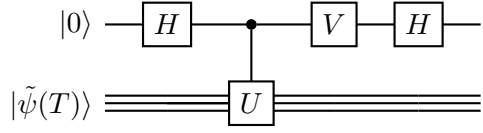
\begin{figure}[h!]
\begin{center}
\begin{quantikz}
 \lstick{$\ket{0}$}  & \gate{H} & \ctrl{1} & \gate{V} & \gate{H} & \qw \\
 \lstick{$\ket{\tilde{\psi}(T)}$} & \qwbundle[alternate]{} & \gate{U}\qwbundle[alternate]{} & \qwbundle[alternate]{}  & \qwbundle[alternate]{} & \qwbundle[alternate]{}
\end{quantikz}
\caption{This circuit evaluates the Hadamard tests for \eqref{expectationDimensional}, with $U$ from \eqref{unitaryQuantities} and $V$ selects the real and imaginary parts.}
\label{fig:momentcircuit}
\end{center}
\end{figure}

Using \eqref{SfDDef} we can write the first and last terms of the left-hand side of \eqref{psiSfSfpsi} as
\begin{align} \label{Sfbf0bf0Sf}
S_f \ket{{\bf 0}} \bra{{\bf 0}} S_f^\dagger = \bigg(\bigotimes_{d=1}^D \sum_{i_d} \xi_{i_d} Q_{i_d} \bigg)\bigg( \frac{\mathbb{I}-X^{\otimes nD} \cdot C^{nD-1} Z \cdot X^{\otimes nD}}{2}\bigg) \bigg(\bigotimes_{d'=1}^D \sum_{i'_{d'}} \xi_{i'_{d'}}^* Q_{i'_{d'}}^\dagger \bigg)
\end{align}
This is the sum of tensor products of unitaries. Thus in \eqref{psiSfSfpsi} we will only have expressions of the type
\begin{align} \label{unitaryQuantities}
\bra{\tilde{\psi}(T)} Q_{i_d} Q_{i'_{d'}}^\dagger \ket{\tilde{\psi}(T)} \quad \textnormal{and} \quad \bra{\tilde{\psi}(T)} Q_{i_d} (X^{\otimes nD} \cdot C^{nD-1} Z \cdot X^{\otimes nD}) Q_{i'_{d'}}^\dagger \ket{\tilde{\psi}(T)}.
\end{align}
Two techniques will yield the quantities in \eqref{unitaryQuantities}. One could either use the Hadamard test circuits depicted in this section or use QPE \cite{qpe1, qpe2}. The circuit for the Hadamard test is given by Fig \ref{fig:momentcircuit}.

Here the $U$'s are gates such that $U \in \{ Q_{i_d} Q_{i'_{d'}}^\dagger, Q_{i_d} (X^{\otimes nD} \cdot C^{nD-1} Z \cdot X^{\otimes nD}) Q_{i'_{d'}}^\dagger \}$ and
\begin{align}
V := 
\begin{cases}
\mathbb{I} \quad &\mbox{for } \real \bra{\tilde{\psi}(T)} U \ket{\tilde{\psi}(T)}, \nonumber \\
S^\dagger \quad &\mbox{for } \imag \bra{\tilde{\psi}(T)} U \ket{\tilde{\psi}(T)}. \nonumber
\end{cases}
\end{align}

We shall get to the pros and cons of each technique towards the end of this section, however, it is noteworthy that the circuit has the advantage of being shallow but not efficient in the terms of the number of measurements needed.

The next step is to define the operator
\begin{align} \label{Sfwithchis}
S_{\chi[a_1,b_1][a_2,b_2]\cdots[a_D,b_D]}\ket{{\bf 0}} &= \sum_{i_1=0}^{n_m-1}\sum_{i_2=0}^{n_m-1} \cdots \sum_{i_D=0}^{n_m-1} \chi[a_1,b_1] (x_{i_1}) \nonumber \\
& \quad \quad \quad \quad \times \chi[a_2,b_2] (x_{i_2}) \cdots \chi[a_D,b_D] (x_{i_D}) \ket{{\bf i}} \nonumber \\
&=\sum_{i_1=0}^{n_m-1} \chi[a_1,b_1] (x_{i_1}) \sum_{i_2=0}^{n_m-1} \chi[a_2,b_2] (x_{i_2}) \cdots  \sum_{i_D=0}^{n_m-1} \chi[a_D,b_D] (x_{i_D}) \ket{{\bf i}}\nonumber \\
&=\sum_{x_{i_1} \in [a_1, b_1]}\sum_{x_{i_2} \in [a_2, b_2]} \cdots \sum_{x_{i_D} \in [a_D, b_D]} \ket{{\bf i}} \nonumber \\
&=\sum_{\substack{j=1,2,\cdots, D \\ x_{i_j} \in [a_j, b_j]}} \ket{{\bf i}},
\end{align}
where the interval indicator function $\chi$ is defined by
\begin{align}
\chi[a,b](x) :=
\begin{cases}
1 \quad &\mbox{if } x \in [a,b], \nonumber \\
0 \quad &\mbox{otherwise}.
\end{cases}
\end{align}
We must now connect $S_f$ coming from \eqref{Sfon0} to $S_{\chi[0,a_1][0,a_2]\cdots[0,a_D]}$ from \eqref{Sfwithchis}. Note that
\begin{align}
S_f \ket{{\bf 0}} &= \sum_{i_1=0}^{n_m-1}\sum_{i_2=0}^{n_m-1} \cdots \sum_{i_D=0}^{n_m-1} f(x_{i_1}, x_{i_2}, \cdots, x_{i_D}) \ket{{\bf i}} \nonumber \\
&= \prod_{j=1}^D \sum_{i_j=0}^{n_m-1} \sum_{k_j=0}^{d-1} \chi[\alpha_{j,k_j}](x_{i_j}) f_{j,k_j}(x_{i_1}, x_{i_2}, \cdots, x_{i_D}) \ket{{\bf i}} \nonumber \\
&= \prod_{j=1}^D \sum_{k_j=0}^{d-1} \sum_{i_j=0}^{n_m-1} \sum_{\substack{j=1,2,\cdots, D \\ x_{i_j} \in \alpha_{j,k_j}}} f_{j,k_j}(x_{i_1}, x_{i_2}, \cdots, x_{i_D}) \ket{{\bf i}}.
\end{align}
In order to perform the sums over $x_{i_j}$'s we use a multidimensional Taylor series
\begin{align} \label{Sf0D2}
S_f \ket{{\bf 0}} &= \prod_{j=1}^D \sum_{k_j=0}^{d-1} \sum_{i_j=0}^{n_m-1} \sum_{\substack{j=1,2,\cdots, D \\ x_{i_j} \in \alpha_{j,k_j}}} \sum_{|\beta| \le L_{i_j,k_j}} f_{j,k_j,\beta} {\bf x}^{\beta} \ket{{\bf i}}  \nonumber \\
&= \prod_{j=1}^D \sum_{k_j=0}^{d-1} \sum_{i_j=0}^{n_m-1} \sum_{\substack{j=1,2,\cdots, D \\ x_{i_j} \in \alpha_{j,k_j}}} \sum_{|\beta| \le L_{j,k_j}} f_{j,k_j,\beta} x_{i_1}^{\beta_1} x_{i_2}^{\beta_2} \cdots x_{i_D}^{\beta_D} \ket{{\bf i}}  \nonumber \\
&= \prod_{j=1}^D \sum_{k_j=0}^{d-1} \sum_{i_j=0}^{n_m-1}  \sum_{|\beta| \le L_{j,k_j}} f_{j,k_j,\beta} (i_1 \Delta x)^{\beta_1} (i_2 \Delta x)^{\beta_2} \cdots (i_D \Delta x)^{\beta_D} \sum_{\substack{j=1,2,\cdots, D \\ x_{i_j} \in \alpha_{j,k_j}}} \ket{{\bf i}}  \nonumber \\
&= \prod_{j=1}^D \sum_{k_j=0}^{d-1} \sum_{i_j=0}^{n_m-1}  \sum_{|\beta| \le L_{j,k_j}} f_{j,k_j,\beta} (i_1 \Delta x)^{\beta_1} (i_2 \Delta x)^{\beta_2} \cdots (i_D \Delta x)^{\beta_D} S_{\chi \prod_{j=1}^D\alpha_{j,k_j}}\ket{{\bf 0}}  \nonumber \\
&= \prod_{j=1}^D \sum_{k_j=0}^{d-1} \sum_{i_j=0}^{n_m-1} f_{j,k_j}(x_{i_1}, x_{i_2}, \cdots, x_{i_D}) S_{\chi \prod_{j=1}^D\alpha_{j,k_j}}\ket{{\bf 0}} .
\end{align} 
The operator $S_f$ is extracted from this and seen to be
\begin{align} \label{SfD}
S_f = \prod_{j=1}^D \sum_{k_j=0}^{d-1} \sum_{|\beta| \le L_{j,k_j}} f_{j,k_j,\beta} (\Delta x)^{|\beta|} (D(n))^{|\beta|} S_{\chi \prod_{j=1}^D\alpha_{j,k_j}} ,
\end{align}
where $D(n)$ was given by \eqref{D(n)}. We now note that if $D=1$, then \eqref{Sf0D2} becomes
\begin{align} \label{Sf1}
S_f \ket{0 } = \sum_{k_1=0}^{d-1} \sum_{i_1=0}^{n_m-1} f_{1,k_j}(x_{i_1}) S_{\chi \alpha_{1,k_1}}\ket{0} = \sum_{k=0}^{d-1} \sum_{i=0}^{2^n-1} f_{k}(x_i) S_{\chi \alpha_{k}}\ket{0},
\end{align}
since in this case there is only one dimension so that $f_{1,k}$ can simply be written as $f_k$, $\alpha_{1,k}$ is $\alpha_k$ and all qubits are assigned to the same dimension. Moreover for one dimension \eqref{SfD} reduces to
\begin{align} \label{SfopinD}
S_f = \sum_{k_1=0}^{d-1} \sum_{|\beta| \le L_{1,k_1}} f_{1,k_1,\beta} (\Delta x)^{|\beta|} (D(n))^{|\beta|} S_{\chi \alpha_{1,k_1}} = \sum_{k=0}^{d-1} \sum_{\beta = 0}^L f_{k,\beta} (\Delta x)^{\beta} (D(n))^{\beta} S_{\chi \alpha_{k}},
\end{align}
where we are now dealing an ordinary Taylor series of $L$th order in one dimension $f_k(x) = \sum_{\beta=0}^L \frac{f_k^{\beta}(0)}{\beta!}x^{\beta} =: \sum_{\beta=0}^L f_{k,\beta} x^{\beta}$. This means that $S_f \ket{0}\bra{0} S_f^\dagger$ is the sum of $\mathcal{O}(d^2n^{2(L+1)})$ unitaries and each $Q_i$ is composed of at most $\mathcal{O}(n^4)$ gates because $S_{\chi \alpha_k}, (D(n))^{\beta}$ and $\ket{0}\bra{0}$ are the sum of $\mathcal{O}(n), \mathcal{O}(n^\beta)$ and $\mathcal{O}(1)$ unitaries, respectively, each composed of $\mathcal{O}(n), \mathcal{O}(1)$ and $\mathcal{O}(n^2)$ gates, respectively. Equations \eqref{SfD}, \eqref{Sf1} and \eqref{SfopinD} are inaccurately given in \cite[Eqs. (39), (40) and (B24)]{stochastic}. 

Let us now return to the error term of the expectation \eqref{expectationDimensional} resulting from the Taylor expansion of the function $f$. We now assume, for simplicity of notation, that all the $L$'s in each dimension are the same. The Lagrange form of the error term of the $L$th order multidimensional Taylor series around ${\bf a} = \{hk_1, hk_2, \cdots, hk_D\}$ is given by
\begin{align}
R_{L,k}({\bf x}, {\bf a}) = \sum_{|\gamma|=L+1} \frac{\partial^{\gamma}f_k({\bf c})}{\gamma!} ({\bf x}-{\bf a})^\gamma,
\end{align}
where ${\bf c}=(c_1, c_2, \cdots, c_D)$ is such that $c_i \in [x_i,(k_i+1)h]$. For every $x_i \in {\bf x}$ and every $hk_i$ with same index $i$ we have that $x_i - hk_i = x_i - a_{k_i} \le h$. Hence we see that $R_{L,k}({\bf x}, {\bf a}) = \mathcal{O}(\prod_{i=1}^D h^{L+1}) = \mathcal{O}(h^{D(L+1)})$. Let $\mathbb{E}_L$ be the expectation that would result from approximating $f$ with the $L$th order Taylor series, then
\begin{align} \label{EEL}
|\mathbb{E}-\mathbb{E}_L| &= \bigg|\sum_{k_1=0}^{d-1} \cdots \sum_{k_D=0}^{d-1} \int_{k_1 h}^{(k_1+1)h} \cdots \int_{k_D h}^{(k_D+1)h} f({\bf x})p_{{\bf X}}({\bf x}) d {\bf x} \nonumber \\
&\quad - \sum_{k_1=0}^{d-1} \cdots \sum_{k_D=0}^{d-1} \int_{k_1 h}^{(k_1+1)h} \cdots \int_{k_D h}^{(k_D+1)h} f_L({\bf x})p_{{\bf X}}({\bf x}) d {\bf x}\bigg| \nonumber \\
&= \bigg|\sum_{k_1=0}^{d-1} \cdots \sum_{k_D=0}^{d-1} \int_{k_1 h}^{(k_1+1)h} \cdots \int_{k_D h}^{(k_D+1)h} R_{L,k}({\bf x}, {\bf a}) p_{{\bf X}}({\bf x}) d {\bf x} \bigg| \nonumber \\
&\le \bigg(\max_{\bf k} \max_{i=1,\cdots,D: \; hk_i \le x_i \le h(k_i+1)} R_{L,k}({\bf x}, {\bf a})\bigg) \nonumber \\
& \quad \times \bigg|\sum_{k_1=0}^{d-1} \cdots \sum_{k_D=0}^{d-1} \int_{k_1 h}^{(k_1+1)h} \cdots \int_{k_D h}^{(k_D+1)h} p_{{\bf X}}({\bf x}) d {\bf x} \bigg| \nonumber \\
&= \bigg(\max_{\bf k} \max_{i=1,\cdots,D: \; hk_i \le x_i \le h(k_i+1)} R_{L,k}({\bf x}, {\bf a})\bigg) \nonumber \\
&= \mathcal{O}(h^{D(L+1)}),
\end{align}
where $p_{{\bf X}}({\bf x})$ denotes the continuous joint probability distribution function. In order to secure the error below a desired threshold $\varepsilon$, we need to take $d > x_{\max}\varepsilon^{-\tfrac{1}{D(L+1)}}$.

The final issue we need to resolve is the construction of the operator $S_f$ and its complexity. This construction is explained in \cite{stochastic} for one dimension, and we now give the $D$ dimensional generalization needed for the rest of the bounds on the complexity of the algorithm. For any given $i=1,2,\cdots,D$ we have that $a_i \in [0, x_{i, \max}]$ and hence there exists an integer $k_{a_i}$ with $0 < k_{a_i} \le \lfloor n/D \rfloor$ such that $\Delta x 2^{k_{a_i}-1} \le a_i \le \Delta x 2^{k_{a_i}}$. The binary expansion of $a_i / \Delta x$ is $a_i/\Delta x = \sum_{j=0}^{k_{a_i}-1} s_{i,j} 2^j$ with $s_{i,j} \in \{0,1\}$. Next, we define the matrix of elements $\ell$
\begin{align}
{\bf L} = \left( 
\begin{array}{*{20}{c}}
     \ell_{1,1} & \ell_{1,2} & \cdots & \ell_{1,k_{a_1}-1}  \\
     \ell_{2,1} & \ell_{2,2} & \cdots & \ell_{2,k_{a_2}-1}  \\
     \vdots & \vdots & \ddots & \vdots  \\
     \ell_{D,1} & \ell_{D,2} & \cdots & \ell_{D,k_{a_D}-1} 
\end{array} \right)
\end{align}
satisfying $s_{i,\ell}=1$ in ascending order for each $i$, and we define the collection of intervals
\begin{align}
\chi_{\ell_{i,j}}^{a_i} = \bigg[2^{\ell_{i,k_{a_i}-1}} + \sum_{k=0}^{\ell_{i,j}-1} s_{i,k}2^k + 1,  2^{\ell_{i,k_{a_i}-1}} + \sum_{k=0}^{\ell_{i,j}} s_{i,k}2^k\bigg]
\end{align}
for $\ell_{i,j}$ a member of the matrix ${\bf L}$. These intervals allow us to divide a dimension interval $[0, a_i / \Delta x]$ into disjoint intervals as
\begin{align}
[0, a_i / \Delta x] = [0, 2^{k_{a_i}-1}] \cup \bigcup_{m=1}^{k_{a_i}-1} \chi_{\ell_{i,m}}^{a_i}.
\end{align}
The effect of the operator $S_{\chi_{\ell_{i,j}}^{a_i}}$ on the interval $S_{\chi_{i,j}}^{a_i}$ is defined as
\begin{align}
S_{\chi_{\ell_{i,j}}^{a_i}} \ket{0} = 2^{\ell_{i,j}/2} \mathbb{I}^{\otimes k_{a_i}-1} \otimes X \otimes \bigotimes_{k=0}^{\lfloor n/D \rfloor-k_{a_i}-1-\ell_{i,j}-1} \mathbb{X}_{i, s_{\lfloor n/D \rfloor} - k_{a_i}-m} \otimes H^{\otimes \ell_{i,j}} \ket{0}
\end{align}
where the operator $\mathbb{X}$ takes two different values according to
\begin{align}
\mathbb{X}_s =
\begin{cases}
X, \quad &\mbox{if } s=1, \nonumber \\
\mathbb{I}, \quad &\mbox{if } s=0.
\end{cases}
\end{align}
For the last interval on the right, the action of the operator is defined as 
\begin{align}
S_{\chi_{[0,2^{k_{a_i}-1}]}} \ket{0} = 2^{(k_{a_i}-1)/2} \mathbb{I}^{\otimes \lfloor n/D \rfloor - k_{a_i}+1} H^{\otimes k_{a_i}-1} \ket{0}.
\end{align}
Thus the interval $S_{\chi[0,a_i]}$ is constructed by summing over $S_{\chi_{\ell_{i,j}}^{a_i}}$ and $S_{\chi_{[0,2^{k_{a_i}-1}]}}$. Lastly, we obtain $S_{\chi \alpha_{i,k}} = S_{\chi[0,a_{i,k+1}]} - S_{\chi[0,a_{i,k}]}$, which is a sum of at most $\mathcal{O}(n)$ unitaries composed of $\mathcal{O}(n)$ gates.

The last part is the bound on the computational complexity in several dimensions. In order to upper bound the error $\varepsilon$ coming from $E = \sqrt{\bra{\tilde \psi} S_f \ket{{\bf 0}} \bra{{\bf 0}} S_f^\dagger \ket{\tilde \psi}}$ we shall upper bound the error $\varepsilon'$ of each of the terms in \eqref{Sfbf0bf0Sf} and then find the number of measurements and gate complexity needed to achieve this error. Following \cite[$\mathsection$A]{stochastic} we assume that $S_f \ket{{\bf 0}} \bra{{\bf 0}} S_f^\dagger$ can be written as a linear combination of $N_{U_j}$ unitaries for each dimension $j=1,2,\cdots, D$, i.e.
\begin{align}
S_f \ket{{\bf 0}} \bra{{\bf 0}} S_f^\dagger = \bigotimes_{j=1}^D \sum_{i_j=1}^{N_{U_j}} \beta_{i_j} U_{i_j}, \quad \textnormal{where $U_{i_j}$ are unitary operators}.
\end{align}
The normalized state is
\begin{align} \label{normalizedunnormalized}
\ket{\psi} &= \bigg(\sum_{j_1=0}^{n_m-1} \sum_{j_2=0}^{n_m-1} \cdots \sum_{j_D=0}^{n_m-1} \mathbb{P}^2 [X_1(t) = x_{j_1}, X_2(t) = x_{j_2}, \cdots, X_D(t) = x_{j_D}]\bigg)^{-1/2} \ket{\tilde \psi} \nonumber \\
&= \bigg(\sum_{j_1=0}^{n_m-1} \sum_{j_2=0}^{n_m-1} \cdots \sum_{j_D=0}^{n_m-1} \mathbb{P}^2_{j_1,j_2,\cdots,j_D}\bigg)^{-1/2} \ket{\tilde \psi}.
\end{align}
The error $\varepsilon'$ is defined as the error of the expectation value of each term in a state $\ket{\psi}$. This implies that the estimated expectation value of each term $\tilde{u}_{i_1, i_2, \cdots, i_D}$ satisfies
\begin{align} \label{tensorbound}
\bigg |\tilde{u}_{i_1, i_2, \cdots, i_D} - \bra{\psi} \bigotimes_{j=1}^D U_{i_j}\ket{\psi} \bigg| \le \varepsilon'
\end{align}
The error in the linear combination of expectation values is determined using \eqref{tensorbound}
\begin{align} \label{boundstensorepsilon'}
\bigg |\prod_{j=1}^D \sum_{i_j=1}^{N_{U_j}} \beta_{i_j}  \tilde{u}_{i_1, i_2, \cdots, i_D} - \bra{\psi} \bigotimes_{j=1}^D \sum_{i=1}^{N_{U_j}} \beta_{i_j} U_{i_j} \ket{\psi}\bigg| &= \bigg| \prod_{j=1}^D \sum_{i_j=1}^{N_{U_j}} \beta_{i_j} (\tilde{u}_{i_1, i_2, \cdots, i_D} - \bra{\psi} \bigotimes_{j=1}^D U_{i_j} \ket{\psi})\bigg| \nonumber \\
&\le  \prod_{j=1}^D \sum_{i_j=1}^{N_{U_j}} \bigg|\beta_{i_j} (\tilde{u}_{i_1, i_2, \cdots, i_D} - \bra{\psi} \bigotimes_{j=1}^D U_{i_j} \ket{\psi})\bigg| \nonumber \\
&\le  \varepsilon' \prod_{j=1}^D \sum_{i_j=1}^{N_{U_j}} |\beta_{i_j} | .
\end{align}
If we write $\tilde E$ for the estimation of the expectation $E$, then by the use of \eqref{normalizedunnormalized} and \eqref{boundstensorepsilon'} we arrive at
\begin{align}
|\tilde E - E| &\le \bigg(\sum_{j_1=0}^{n_m-1} \sum_{j_2=0}^{n_m-1} \cdots \sum_{j_D=0}^{n_m-1} \mathbb{P}^2_{j_1,j_2,\cdots,j_D}\bigg) \frac{\varepsilon'}{\tilde E + E} \prod_{j=1}^D \sum_{i_j=1}^{N_{U_j}} |\beta_{i_j} | \nonumber \\
&\sim \bigg(\sum_{j_1=0}^{n_m-1} \sum_{j_2=0}^{n_m-1} \cdots \sum_{j_D=0}^{n_m-1} \mathbb{P}^2_{j_1,j_2,\cdots,j_D}\bigg) \frac{\varepsilon'}{2E} \prod_{j=1}^D \sum_{i_j=1}^{N_{U_j}} |\beta_{i_j} |.
\end{align}
Therefore, in order to obtain an upper bound for $\varepsilon$ in the expectation $E$, the error $\varepsilon'$ must be such that
\begin{align} \label{gamma}
\varepsilon' \le  \frac{2E}{(\sum_{j_1=0}^{n_m-1} \sum_{j_2=0}^{n_m-1} \cdots \sum_{j_D=0}^{n_m-1} \mathbb{P}^2_{j_1,j_2,\cdots,j_D}) \prod_{j=1}^D \sum_{i_j=1}^{N_{U_j}} |\beta_{i_j}|} \varepsilon =: \gamma \varepsilon.
\end{align}

In terms of the Hadamard test circuit above, one requires $\mathcal{O}(1/\varepsilon'^2)$ measurements to bound the error in the expectation value to $\varepsilon'$. Up to the preparation of the state, the depth of the unitaries in the circuit is $\mathcal{O}(1)$. The trade-off with QPE is that the same accuracy in the expectation is $\mathcal{O}(\log 1/\varepsilon')$ but the depth is now $\mathcal{O}(1/\varepsilon')$. With the $L$th order Taylor approximation, one has that $N_{U_j} = \mathcal{O}(d^2 n^{2L+2})$ for $j=1,2,\cdots, D$, where $d$ was the number of intervals. Moreover, the total number of measurements is the number of measurements for each term affected by a multiplication by $N_{U_j}$. Effectively, this yields that the total number of measurements by Hadamard test is $\mathcal{O}((\prod_{j=1}^D d^2 n^{2L+2})/(\gamma \varepsilon^2))$, whereas using QPE, one would need $\mathcal{O}((\prod_{j=1}^D d^2 n^{2L+2}) \log(1/\gamma \varepsilon))$, but QPE needs $\mathcal{O}(1/\gamma \varepsilon)$ many $U$ gates.

Lastly, we need the bound for $\gamma$ from \eqref{gamma}. Since $\sum_{j_1=0}^{n_m-1} \sum_{j_2=0}^{n_m-1} \cdots \sum_{j_D=0}^{n_m-1} \mathbb{P}^2_{j_1,j_2,\cdots,j_D} \le 1$, we see that $\gamma \ge 2E / (\prod_{j=1}^D \sum_{i_j=1}^{N_{U_j}} |\beta_{i_j}|)$. Next, using \eqref{SfDDef} we get
\begin{align}
\prod_{j=1}^D \sum_{i_j=1}^{N_{U_j}} |\beta_{i_j}| = 2^D \prod_{j=1}^D \sum_{\ell_j, \ell'_j} |\xi_{\ell_j} \xi_{\ell'_j}^*| = 2^D \prod_{j=1}^D \sum_{\ell_j, \ell'_j} |\xi_{\ell_j}| |\xi_{\ell'_j}|.
\end{align}
Moreover, from \eqref{D(n)} we see that $(D(n))^{|\beta|} = \mathcal{O}(2^{n |\beta|}) = \mathcal{O}(x_{\max}^{|\beta|})$. For an interval in a given dimension $S_{\chi \alpha_k}$ is a linear combination of operators $S_{\chi_\ell^\alpha}$, therefore $|S_{\chi \alpha_k}| = \mathcal{O}(2^{(k_{a_i}-1)/2}) = \mathcal{O}(x_{\max}^{1/2})$ since $k_{a_i}$ was such that $0 < k_{a_i} \le \lfloor n/D \rfloor$. This implies that the largest $|\xi_{\ell_j}|$ is as large $[\max_{k_j} \max_{|\beta|} |f_{j, k_j, \beta}| x_{\max}^{|\beta|} ] \mathcal{O}(x_{\max}^{1/2})$. Therefore, we get
\begin{align}
\prod_{j=1}^D \sum_{i_j=1}^{N_{U_j}} |\beta_{i_j}| = \mathcal{O} \bigg(\prod_{j=1}^D[\max_{k_j} \max_{|\beta|} |f_{j,k_j, |\beta|}| x_{\max}^{|\beta|}]^2  d^2 n^{2L+2} x_{\max}\bigg),
\end{align}
which is the last bound we needed for $\gamma$.

The operator $S_f \ket{{\bf 0}} \bra{{\bf 0}} S_f^\dagger$ will be the sum of $\mathcal{O}(d^{2D}n^{2D(L+1)})$ unitaries and each $Q_k$ will be composed of $\mathcal{O}(n^4)$ gates. The number of gates will grow exponentially with the dimensions and this is a universal problem in many algorithms, such as the use of QAE when computing moments when the arguments of the function are multidimensional. In such cases one needs exponentially many resources to build a multidimensional oracle.

As $L$ grows, $\varepsilon^{-\tfrac{2}{D(L+1)}}$ from \eqref{EEL} becomes smaller while $d^{2D} n^{D(2L+2)}$ grows larger. This means that the number of unitaries is not monotonic with respect to $L$ and thus there could be an optimal $L$ for the desired accuracy. In \cite{stochastic}, the authors point out that the expectation values of those unitaries could be computed in parallel with a second quantum device.

\section{\label{sec:futurework}Future work}
As discussed in Section \ref{sec:conclusion}, some of the following areas of research might have to be pursued using a combination of real and imaginary time evolutions.

\subsection{Stochastic volatility and stochastic interest rates}
A key weakness in the Black-Scholes is that the volatility term $\sigma$ is constant. One can go around this problem by defining two SDEs: geometric Brownian motion for the underlying stock $S$
\begin{align} \label{geometric}
dS_t = \mu S_t dt + \sigma_t S_t dW_{1,t},
\end{align}
where $dW_1$ is standard Brownian motion, as well as another SDE for $\sigma$
\begin{align} \label{stocvol}
d\sigma_t = p(S, \sigma, t)dt + q(S,\sigma,t)dW_{2,t},
\end{align}
where $dW_2$ is another Brownian motion whose correlation with $dW_1$ is given $\rho$, i.e. $dW_1 dW_2 = \rho dt$. Let $V(S,\sigma,t)$ be the value of an option on the underlying $S$ at time $t$ and with stochastic volatility $\sigma$. Portfolio replication coupled with It\^{o}'s lemma and non-arbitrage arguments yields the partial differential equation
\begin{align} \label{merton}
\frac{\partial V}{\partial t} + rS\frac{\partial V}{\partial S}  + \frac{1}{2}\sigma^2 S^2 \frac{\partial^2 V}{\partial S^2} + (p-\lambda)q \frac{\partial V}{\partial \sigma} + \frac{1}{2} q^2 \frac{\partial^2 V}{\partial \sigma^2} + \rho \sigma S q \frac{\partial^2 V}{\partial S \partial \sigma} - rV = 0.
\end{align}
Here $\lambda(S,\sigma,t)$ is a universal function that is used in portfolio replication to ensure that options are only dependent on the variables $S, \sigma$ and $t$. The function $\lambda$ is called the market price of volatility risk. Comparing  \eqref{merton} with Black-Scholes
\begin{align} \label{bs}
\frac{\partial V}{\partial t} + rS\frac{\partial V}{\partial S}  + \frac{1}{2}\sigma^2 S^2 \frac{\partial^2 V}{\partial S^2} - rV = 0,
\end{align}
we see the appearance of the additional linear and second order partial derivatives with respect to $\sigma$ coming from \eqref{stocvol} as well as the presence of the mixed term in $S$ and $\sigma$ due to the correlation $\rho$ of the two Brownian motions $dW_1$ and $dW_2$. Note that if the volatility were to be constant, then \eqref{merton} would reduce to \eqref{bs}. 
Different choices of $p$ and $q$ lead to different named models, e.g. 
\begin{itemize}
\item Hull and White: $d(\sigma^2) = a(b-\sigma^2) dt + c \sigma^2 dW_2$,
\item Square-root model / Heston: $d(\sigma^2) = (a-b\sigma^2)dt + c \sqrt{\sigma^2} dW_2$,
\item 3/2 model: $d(\sigma^2) = (a\sigma^2-b\sigma^4)dt + c \sigma^{3/2} dW_2$,
\item Ornstein–Uhlenbeck: $d(\log \sigma^2) = (a-b \log \sigma^2)dt + cdW_2$.
\end{itemize}
see \cite[p. 861]{wilmott}. Similar equations also exist for stochastic interest rates \cite{wilmott}.
\subsection{American and decision-embedded options}
The American option is also an involved one to price. Unlike in the European set-up, the holder of an American option may exercise at any time prior and up to expiration. The details of this set up can be found in \cite[$\mathsection$2]{vecer}. We are dealing with our usual geometric Brownian motion \eqref{geometric} as well as the additional SDE
\begin{align} \label{dXtq}
dX_t^q = q_t dS_t + \mu(X_t^q -q_t S_t)dt, \quad X_0^q = X_0,
\end{align}
where $q_t$ is the number of shares held at time $t$. The strategy of $q_t$ is subject to the contractual constraint $q_t \in [\alpha_t, \beta_t]$, where $\alpha_t \le \beta_t$. Financial arguments lead us to the conditional expectation
\begin{align}
V^{[\alpha, \beta]}(t, S_t, X_t^q) = \max_{q_u \in [\alpha, \beta]} e^{-r(T-t)} \mathbb{E}[\max(x_T^q,0) \;|\; \mathcal{F}_t], \quad t \in [0,T]. \nonumber
\end{align}
These expectations arise naturally in stochastic optimal control and are characterized by the corresponding Hamilton-Jacobi-Bellman equation
\begin{align}
-rV + \frac{\partial V}{\partial t} &+ rs \frac{\partial V}{\partial s} \nonumber \\
&+ \max_{q \in [\alpha,\beta]} \bigg[(\mu x + qs(r-u)) \frac{\partial V}{\partial x} + \frac{1}{2}\sigma^2 s^2 \bigg(\frac{\partial^2 V}{\partial s^2} + 2q \frac{\partial^2 V}{\partial s \partial x} +q^2 \frac{\partial^2 V}{\partial x^2} \bigg) \bigg] = 0, \nonumber
\end{align}
with associated boundary condition $V(T,s,x)=\max(x,0)$. Note the typo in \cite[Eq. (2.4)]{vecer}. To model the American option using \eqref{dXtq}, one sets $\mu=0$ and one allows only one switch in $q_t$, either from 1 to 0 for the American call, or from $-1$ to 0 for the American put.

\subsection{The Poisson and wave equations}
The heat equation is not the only candidate for this algorithm. In one of its simplest forms the Poisson equation reads as follows
\begin{align}
\nabla^2 u(x,y,z) = f(x,y,z), \nonumber
\end{align}
where $\nabla^2$ is the Cartesian Laplacian in three dimensions. Note the similarity with the heat equation, except for the absence of the time derivative. It\^{o}'s lemma can be used \cite[$\mathsection$4.1.1]{nolen} to transform the conditional expectation
\begin{align}
w({\bf x}) = \mathbb{E} \bigg[\int_0^\infty \exp\bigg(-\int_0^s c({\bf X}_\tau)d\tau\bigg) \; f({\bf X}_s) \; | \; {\bf X}_t = {\bf x}  \bigg] \nonumber
\end{align}
arising from the vectorized SDE
\begin{align}
d{\bf X}_t = b({\bf X}_t)dt + \sum_j \sigma_{ij}({\bf X}_t) dW_t^j \nonumber
\end{align}
into the solution of the PDE
\begin{align}
\sum_{i,j}\frac{1}{2}\Sigma_{ij}(x) \frac{\partial^2 w}{\partial x_i \partial x_j}+\sum_j b_j \frac{\partial w}{\partial x_j}-c({\bf x})w = f({\bf x}). \nonumber
\end{align}
Here $\Sigma_{ij}=\sum_k \sigma_{ik}\sigma_{kj}$. Note the absence of $t$ in $w$, as $w$ is now only a function of space ${\bf x}$.

Another venue would be to consider the wave or Klein-Gordon equations (or in general, other type of hyperbolic differential equations). In its simplest form the wave equation is written as
\begin{align} \label{waveequation}
\frac{\partial^2 u}{\partial t^2} = c^2 \nabla^2 u(x,y,z), \quad c>0.
\end{align}
The time derivative is now of second order, unlike in the heat equation, which was of first order. In these cases, other types of probability distributions and Brownian motions need to be added, specially by considering jumps and Poisson processes, see \cite{dalang}.

In these situations, different variational principles will be needed to account for the different orders of the time derivative that appears in the PDE. For example, when it comes to the wave equation \eqref{waveequation}, we will need a varitational principle of the type
\begin{align}
\delta \; \bigg| \bigg| \frac{d^2}{dt^2} \ket{\tilde{v}(\boldsymbol{\theta}(t))} - \mathcal{O}(t) \ket{\tilde{v}(\boldsymbol{\theta}(t))} \bigg| \bigg| = 0, \nonumber
\end{align}
as opposed to \eqref{mclachlan}. This procedure, along with more generalizations of the time derivative and related differential operators, will be carried out in subsequent research.

\end{document}